\documentclass[twocolumn,twocolappendix]{aastex631}
\pdfoutput=1

\usepackage{gensymb}
\usepackage{xcolor}
\usepackage{soul} 

\shorttitle{Luminosity History of NGC 5972}
\shortauthors{Finlez et al.}

\graphicspath{{./}{figures/}}

\begin{document}

\title{Detailed accretion history of the supermassive black hole in NGC 5972 over the past $\gtrsim$10$^4$ years through the extended emission line region}

\correspondingauthor{Finlez, C.}
\email{cfinlez@astro.puc.cl}

\author[ 0000-0003-1778-1061]{Finlez, C.}
\affiliation{Instituto de Astrof\'isica, Facultad de F\'isica, Pontificia Universidad Cat\'olica de Chile, Casilla 306, Santiago 22, Chile}

\author[0000-0001-7568-6412]{Treister, E.}
\affiliation{Instituto de Astrof\'isica, Facultad de F\'isica, Pontificia Universidad Cat\'olica de Chile, Casilla 306, Santiago 22, Chile}

\author{Bauer, F.}
\affiliation{Instituto de Astrof\'isica, Facultad de F\'isica, Pontificia Universidad Cat\'olica de Chile, Casilla 306, Santiago 22, Chile}
\affiliation{Millennium Institute of Astrophysics, Nuncio Monse\~nor S\'otero Sanz 100, Of 104, Providencia, Santiago, Chile}
\affiliation{Space Science Institute, 4750 Walnut Street, Suite 205, Boulder, Colorado 80301}

\author[0000-0002-6131-9539]{Keel, W.}
\affiliation{Department of Physics and Astronomy, University of Alabama, Box 870324, Tuscaloosa, AL 35487, USA}

\author[0000-0002-7998-9581]{Koss, M.}
\affiliation{Eureka Scientific, 2452 Delmer Street Suite 100, Oakland, CA 94602-3017, USA}
\affiliation{Space Science Institute, 4750 Walnut Street, Suite 205, Boulder, Colorado 80301}

\author[0000-0001-6920-662X]{Nagar, N.}
\affiliation{Astronomy Department, Universidad de Concepci\'on, Casilla 160-C, Concepci\'on, Chile}

\author{Sartori, L.}
\affiliation{ETH Z\"urich, Institute for Particle Physics and Astrophysics, Wolfgang-Pauli-Strasse 27, CH-8093 Z\"urich, Switzerland}

\author[0000-0002-2203-7889]{Maksym, W.P.}
\affiliation{Center for Astrophysics, Harvard \& Smithsonian, 60 Garden Street, Cambridge, MA 02138, USA}

\author[ 0000-0001-8349-3055]{Venturi, G.}
\affiliation{Instituto de Astrof\'isica, Facultad de F\'isica, Pontificia Universidad Cat\'olica de Chile, Casilla 306, Santiago 22, Chile}
\affiliation{INAF-Osservatorio Astrofisico di Arcetri, Largo E. Fermi 5, 50125 Firenze, Italy}

\author[ 0000-0002-2688-7960]{Tub\'in, D.}
\affiliation{Leibniz-Institut f\"ur Astrophysik Potsdam, An der Sternwarte 16, D-14482 Potsdam, Germany}

\author[ 0000-0002-4130-636X]{Harvey, T.}
\affiliation{Center for Astrophysics, Harvard \& Smithsonian, 60 Garden Street, Cambridge, MA 02138, USA}

\begin{abstract}
We present integral field spectroscopic observations of NGC 5972 obtained with the Multi Unit Spectroscopic Explorer (MUSE) at VLT. NGC 5972 is a nearby galaxy containing both an active galactic nucleus (AGN), and an extended emission line region (EELR) reaching out to $\sim 17$ kpc from the nucleus. We analyze the physical conditions of the EELR using spatially-resolved spectra, focusing on the radial dependence of ionization state together with the light travel time distance to probe the variability of the AGN on  $\gtrsim 10^{4}$ yr timescales.
The kinematic analysis  suggests multiple components: (a) a faint component following the rotation of the large scale disk; (b) a component associated with the EELR suggestive of extraplanar gas connected to tidal tails; (c) a kinematically decoupled nuclear disk. Both the kinematics and the observed tidal tails suggest a major past interaction event.
Emission line diagnostics along the EELR arms typically evidence Seyfert-like emission, implying that the EELR was primarily ionized by the AGN. We generate a set of photoionization models and fit these to different regions along the EELR. This allows us to estimate the bolometric luminosity required at different radii to excite the gas to the observed state. Our results suggests that NGC 5972 is a fading quasar, showing a steady gradual decrease in intrinsic AGN luminosity, and hence the accretion rate onto the SMBH, by a factor $\sim 100$ over the past $5 \times 10^{4}$ yr. 
\end{abstract}


\section{Introduction}
\label{sec:introduction}

Active galactic nuclei (AGN) can have a significant impact on the interstellar medium (ISM) of their host galaxies, through the mechanical input of radio jets, AGN wind-driven, and photoionization of the gas \cite[e.g.,][]{morganti+2017}. The energy injected into the host galaxy is thought to play a critical regulatory role in both galaxy and supermassive black hole (SMBH) evolution \cite[e.g.,][]{gebhardt+2000,kormendy+2013}.

The AGN can photoionize gas out to $\gtrsim$ 1 kpc, which is considered the so-called narrow line region (NLR), as the kinematics of this gas are typically $< 1000$ km/s. However, observations have shown that AGN-ionized gas can, and often does, extend well beyond this limit. These extended emission line regions (EELRs) can extend through the entire host galaxy and reach tens of kpcs \cite[e.g.,][]{liu+2013,harrison+2014}. EELRs can exhibit complex morphologies, and be spatially and kinematically distinct from the NLR. Some EELRs show no direct morphological relation to the host galaxy, and have been connected to tidal tails from galaxy interactions \cite[e.g.,][]{keel+2012a} or large-scale outflows \cite[e.g.,][]{harrison+2015}. While other EELRs show conical or biconical shapes, with their apex near the active nucleus, and can be considered as large-scale extensions of the NLR. The ionized gas within an EELR can show large velocities, ranging between several hundreds to $>$ 1000 km s$^{-1}$ with respect to the systemic velocity of the host, while also showing low velocity dispersion \cite[e.g.,][]{fu+2009,husemann+2013}. 

While early studies connected EELRs with the presence of radio-jets \cite[][]{stockton+2006}, they have also been observed in active galaxies with no appreciable radio emission \cite[][]{husemann+2013}. Some EELRs show narrow line widths and modest electron temperatures, indicating that they must be ionized by radiation from the nucleus, rather than direct interaction with either a radio jet or an outflow, for which the linewidth and electron temperature would be much higher due the presence of shocks \citep[e.g.][]{knese+2020}. 
EELRs can potentially offer a unique way to probe the effect of the AGN on the galaxy due to their physical extension, which connects the large scales of the host with the active nucleus \cite[e.g.,][]{harrison+2015,sun+2017}. Furthermore, by considering the light-time travel to the clouds and then to us, we can obtain a view into the past of the AGN, on longer timescales than typical AGN variability \cite[e.g.,][]{dadina+2010,lintott+2009,keel+2012a}.

The suggestion that EELRs can be light echoes from a former high-luminosity AGN was originally explored in a nearby, highly ionized, extended cloud near the active galaxy IC 2497 \cite[][]{lintott+2009,sartori+2016}, this object and others alike have been since called Voorwerpjes. The current AGN luminosity failed to account for the ionization observed in the cloud, thus suggesting that the AGN decreased by factor $\sim100$ in luminosity over a $\sim10^{5}$ yrs timescale \citep{lintott+2009,keel+2012a,schawinski+2010}. Further studies of nearby Voorwerpjes have found comparable variability amplitudes in similar timescales \citep{keel+2012a,keel+2017,sartori+2018}.\\

AGN variability has been observed and inferred to occur on virtually all timescales, ranging from seconds to days, to decades, to $> 10^4$ yr. These different timescales indicate that there are likely different physical processes at play at different spatial scales of the system,from the SMBH vicinity to galaxy-wide scale. Days to years variability, observed in the optical and UV and probed by ensemble analysis, may arise from accretion disk instabilities \citep[e.g.,][]{sesar+2006,dexter+2019,caplar+2017}. While possible explanations for years-to-decades timescales, observed in the so-called 'changing look AGN', include changes in accretion rate or accretion disk structure \citep[e.g.,][]{lamassa+2015,macleod+2019,graham+2020}. Variability on longer timescales ($> 10^{4}$ yr), probed through AGN-photoionized EELRs \citep[e.g.,][]{schawinski+2010,lintott+2009, keel+2012a}, are possibly linked to dramatic changes in accretion rate. Simulations indicate mergers, bar-induced instabilities and clumpy accretion as possible mechanisms behind these changes \citep[e.g.,][]{hopkins+2010,bournaud+2011}.

A typical AGN duty cycle is estimated to last $10^7 - 10^9$ yr \citep[][]{marconi+2004}. This however does not constrain whether the total mass growth is achieved during a single active accretion phase or is broken up in shorter phases.
Simulations and theoretical models have addressed accretion variability \citep[e.g.,][]{king+2015,gabor+2013,schawinski+2015,sartori+2018,sartori+2019}, suggesting a typical timescale for AGN phases of $10^{5}$ yr, implying that nearby AGN switch "on" rapidly during $\sim 10^4$ yr and stay on for $\sim 10^5$ yr before switching "off". The AGN continues "flickering" on these $10^5$yr cycles resulting in a total $10^7 - 10^9$ yr "on" lifetime over the course of the host. 

Quantifying the effect the AGN has on the host galaxy evolution requires a better understanding of how the nuclear activity evolves and varies during the galaxy lifetime. Following the analysis of IC 2497, efforts to find similar objects in the nearby Universe were made, based on SDSS DR X optical imaging. In order to accomplish this, both targeted and serendipitous searches were carried out at z $< 0.1$ by Galaxy Zoo volunteers during a six-week period \citep{keel+2012a}. This search retrieved 19 objects, including NGC 5972, a nearby (z = 0.02974, where 1\arcsec\ corresponds to 0.593 kpc) Seyfert 2 galaxy \citep{veron+2006}.

NGC 5972 was previously known for the presence of powerful ($10^{23.9}$ WHz$^{-1}$ at 4850 MHz) extended (9\farcm4, corresponding to 0.3 Mpc) double radio lobes \citep{condon+1988,veron+1995} that extend along a position angle (PA) of 100\degree, almost perpendicular to the major axis of the optical emission.\\
The galaxy has been morphologically classified as a S0/a. Further modeling of the I-band luminosity profile shows complex residuals that indicate a possible merger or galaxy interaction event \citep{veron+1995}.
The EELRs extends to a radius of $20\arcsec$ from the center, which corresponds to $\sim$12 kpc,forming a double-helix shape which shows a highly ionized complex filamentary  structure, \cite[narrow and medium band HST imaging][]{keel+2015}. 

In this paper we present results from VLT/MUSE observations of NGC 5972. We study the long-term variability of the source by analysing the changes in luminosity required to ionize the EELR to its current state as a function of radius.

This paper is organized as follows. In Sect. \ref{sec:observations} we describe the observations and the data reduction process. In Sect. \ref{sec:results} we present the results of our analysis. The results are discussed and summarized in Sect. \ref{sec:discussion} and Sect. \ref{sec:summary}, respectively.

\section{Observations and Data Reduction}
\label{sec:observations}

The Multi-Unit Spectroscopic Explorer \citep[MUSE,][]{bacon+2010} is an integral field spectrograph installed on the Very Large Telescope (VLT). Its field of view (FOV) in wide field mode (WFM) covers 1\arcmin $\times$ 1\arcmin, with a pixel scale of 0\farcs2. The wavelength range spans $\sim$4600-9300 \AA\ and the resolving power is R$=$1770-3590.

NGC 5972 was observed with MUSE in WFM on the night of March 10th, 2019 (program ID 0102.B-0107, PI L. Sartori), in two observing blocks (OBs). Each OB consisted of 3 on-target observations with an exposure of 950 seconds each, with 1\arcsec dithers, and a seeing constraint of 1\farcs3. The last exposure of the second OB was finished 8 minutes into twilight, causing a small increase in the blue background. The exposure was included in the analysis.  
The data were reduced with the ESO VLT/MUSE pipeline (v2.8) under the ESO Reflex environment \citep{freudling2013}.
Briefly, this pipeline generates a master bias and flat. This is followed by the wavelength calibration, employing the arc-lamp exposures, and after the flux calibration. Both of these are applied to the raw science exposures. 

A sky model is created from selected pixels free of source emission and is subtracted from the science exposures. The coordinate offsets are calculated for each FOV image to align the 6 exposures. Finally, we combined all the exposures by resampling the overlapping pixels to obtain the final data cube. 

The final data cube has a mean seeing-limited spatial resolution of $\sim$ 0\farcs78, as estimated from a point-source located in the southern portion of the FOV. The datacube is rotated 35\degree\ so North is up and East to the left. This results into our final data cube containing $432 \times 431$ spaxels, corresponding to 86\farcs4 and 86\farcs2. At the redshift of NGC 5972, the physical area observed is 50$\times$50 kpc$^{2}$, sufficient to cover the entire galaxy and its EELR, as can be seen in Fig. \ref{fig:compositecolor}.

\section{Results}
\label{sec:results}

In this section we present the results from our analysis of the MUSE observations of NGC 5972. The results are organized as follows. We first present the fitting to the stellar component (Sect. \ref{sec:stellarcomp}). In Sect. \ref{sec:emlines} we present an analysis of the ionized gas component by studying the moment maps from the emission lines, the dust extinction and the dominant ionization mechanism. In Sect. \ref{sec:kinematics} we present a kinematic analysis of the stellar and ionized gas components. Finally, in Sect. \ref{sec:lum_hist} we present our analysis to the extended emission line region, comparing the observed emission line ratios to ionization models in order to constrain the required luminosity to ionize the gas to its observed state.

\begin{figure}
	\includegraphics[width=\columnwidth]{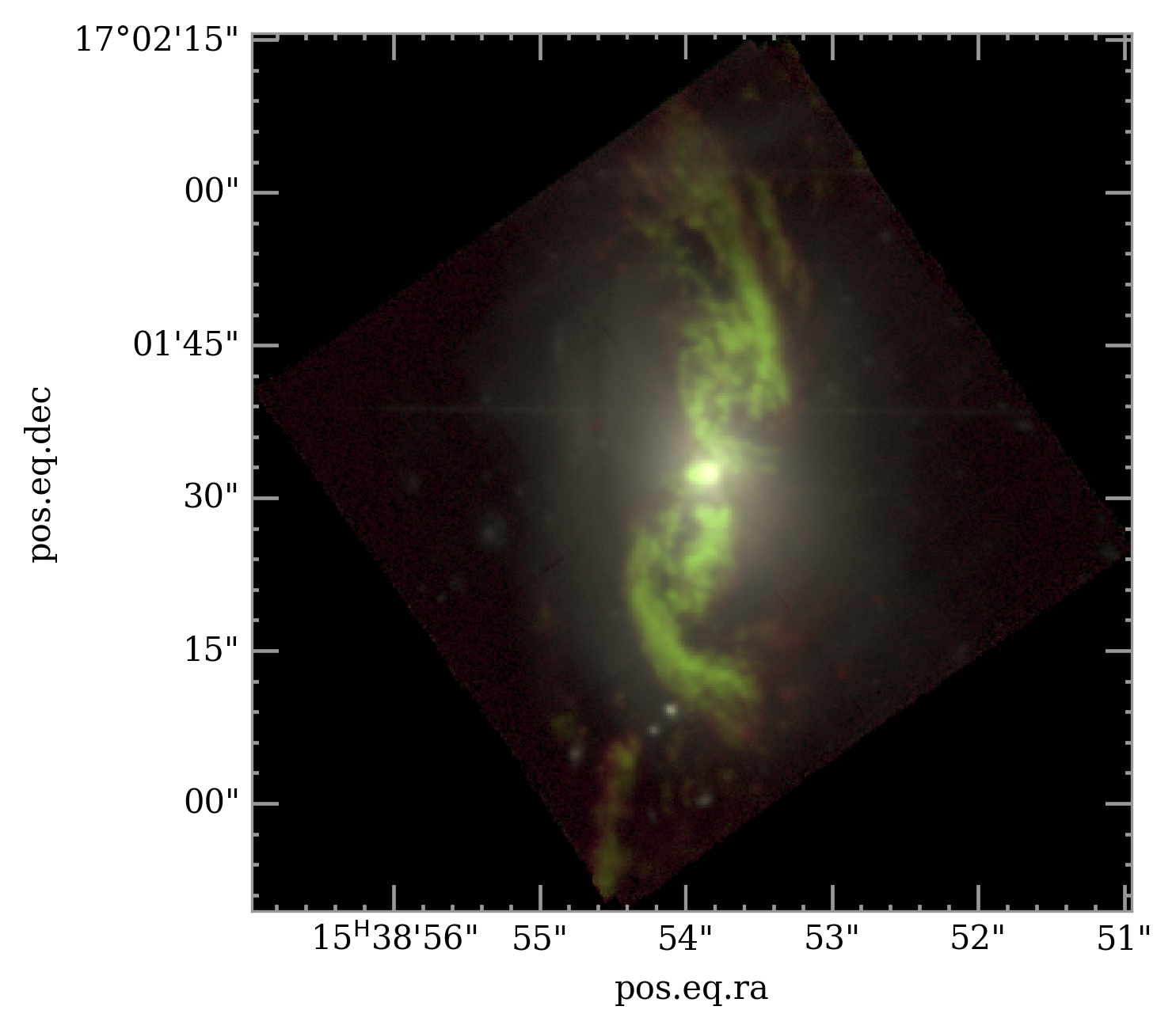}
    \caption{Composite RGB color image using red, green and blue pseudo-colors, collapsing three bands in the data cube (See Sect. \ref{sec:results} for details). In this image North is up and East to the left. This image shows the intricate filamentary distribution of the ionized gas.}
    \label{fig:compositecolor}
\end{figure}

We create a color composite image (Fig. \ref{fig:compositecolor}) by collapsing three separate sections of the data cube, one that encompasses the [OIII] emission line, the second one for the H$\alpha$ emission line, and a third one collapsing the range 7000-9000 \AA\ to represent the stellar continuum. These three images are represented by green, red and white pseudo-colors, respectively. Each color image was stretched with an logarithmic scale before being combined. 
This composite image shows the large scale morphology of the ionized gas, as well as the filamentary complex structure that extends from the center to the N and S. The emission line gas forms a double helix shape to the S and an arm that seems to twist on itself towards the W is observed in the N. Low-luminosity tidal tails are observed beyond these bright arms both to the NE and SE. 

Narrow band imaging obtained with the 2.1 m telescope at Kitt Peak in Arizona, where V-band continuum has been subtracted from the narrow filter centered at redshifted [OIII], show ionized structure reaching 70\arcsec\ as a lower limit in the S, and up to 81\arcsec\ in the N, this corresponds to 41 and 48 kpc, respectively (Keel, W; priv. comm.). When considering fainter features that are revealed with smoothing this limits can extend up to 76\arcsec\ and 91\arcsec\ for the S and N regions.

\subsection{Stellar Component fitting}
\label{sec:stellarcomp}

To analyze the stellar component of NGC 5972 we use the python implementation of the penalised PiXel-Fitting software 
\citep[pPXF;][]{cappellari+2004,cappellari+2017}, together with the MILES single stellar population (SSP) models \citep{vazdekis+2015} as spectral templates for the stellar continuum. The templates have an intrinsic spectral resolution of 2.51 \AA, and were broadened to the wavelength-dependent MUSE resolution before any fits with pPXF were performed. We further adopt the line-spread function (LSF) model of the MUSE spectra as described in \cite{bacon+2017}.

The first step in the analysis is to apply the adaptive Voronoi tessellation routine of \cite{cappellari+2003}, to guarantee a minimum signal-to-noise ratio (SNR) across the entire field-of-view (FOV), in order to ensure reliability of the measurement for the stellar component.
The SNR is measured using the der\_SNR algorithm \citep{stoehr+2008} in the wavelength range 6400 $\leq \lambda $ (\AA) $\leq $ 6500, as there are no strong emission lines present in this spectral window. With this we achieve a minimum SNR of 50, on average per spectral bin, in 1407 spatial bins, in contrast to the 180,000 original pixels. Furthermore, we masked the FOV to exclude spaxels with SNR $< 4$.

\begin{figure*}
	\includegraphics[width=\textwidth]{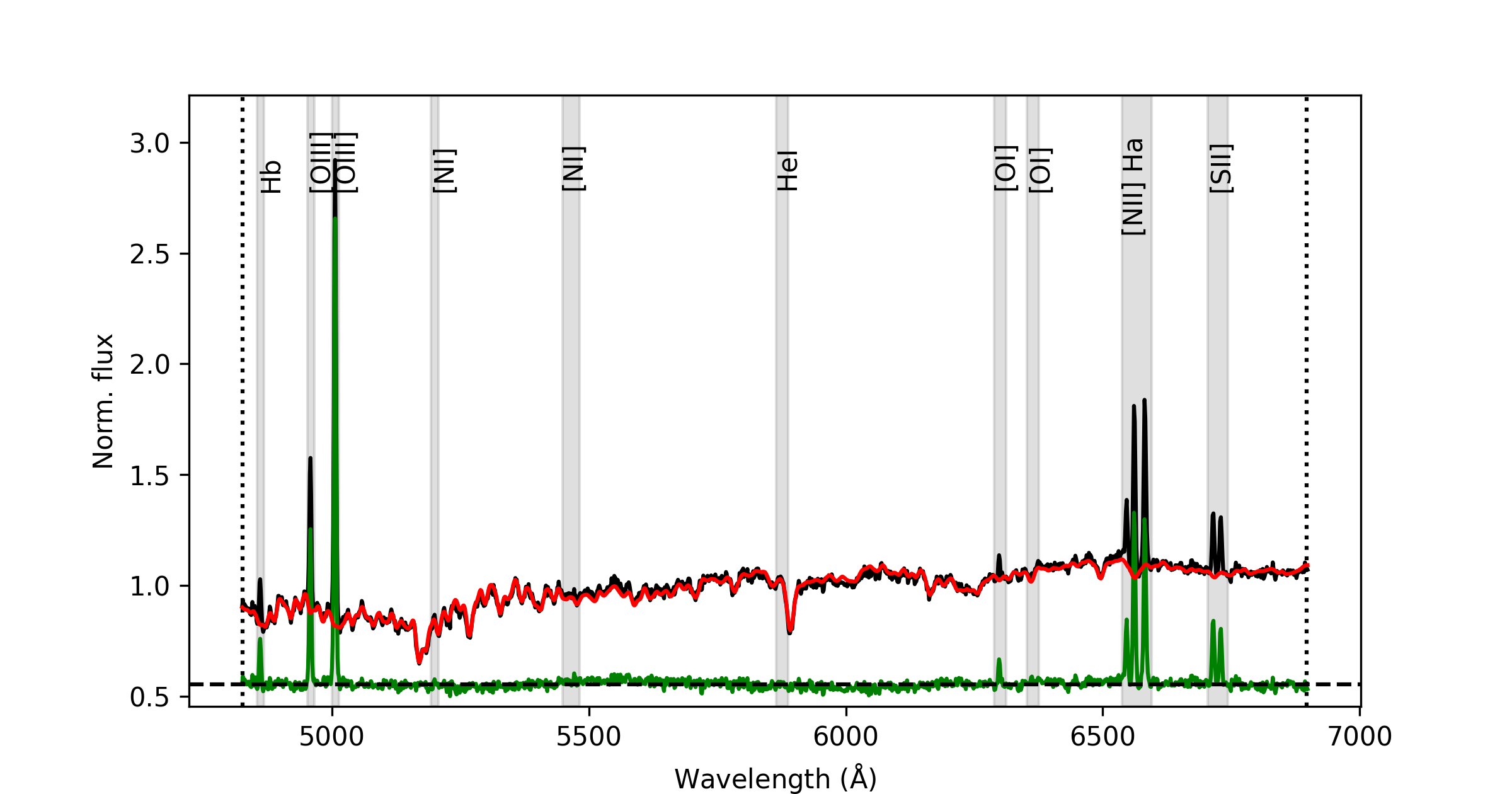}
    \caption{An example of a pPXF fit to one of the central bins. The observed spectra, the best-fit stellar continuum model and the residuals are represented by the black, red and green lines, respectively. The colored grey areas represent the masked spectral regions around the emission lines observed. }
    \label{fig:ppxf_example_fit}
\end{figure*}

The first run of pPXF was unregularised, fitting the wavelength range 4800-7000 \AA. We masked the following emission lines present in this range, using a width of 1800 km/s: H$\beta \lambda 4861$ \AA, [OIII]$\lambda 4958$ \AA, [OIII]$\lambda 5007$ \AA, [NI]$\lambda 5197$ \AA, [NI]$\lambda 5200$ \AA, HeI$\lambda 5875$ \AA, [OI]$\lambda 6300$ \AA, [OI]$\lambda 6363$ \AA, [NII]$\lambda 6547$ \AA, H$\alpha \lambda 6562$ \AA, [NII]$\lambda 6583$ \AA, [SII]$\lambda 6716$ \AA, [SII]$\lambda 6730$ \AA.
We use a fourth-order multiplicative Legendre polynomial to match the overall spectral shape of the data. The purpose of this first fit is to derive the stellar component kinematics. In Fig. \ref{fig:ppxf_example_fit} we show an example of the pPXF fitting to one of the central Voronoi bins.  

\begin{figure*}
	\includegraphics[width=\textwidth]{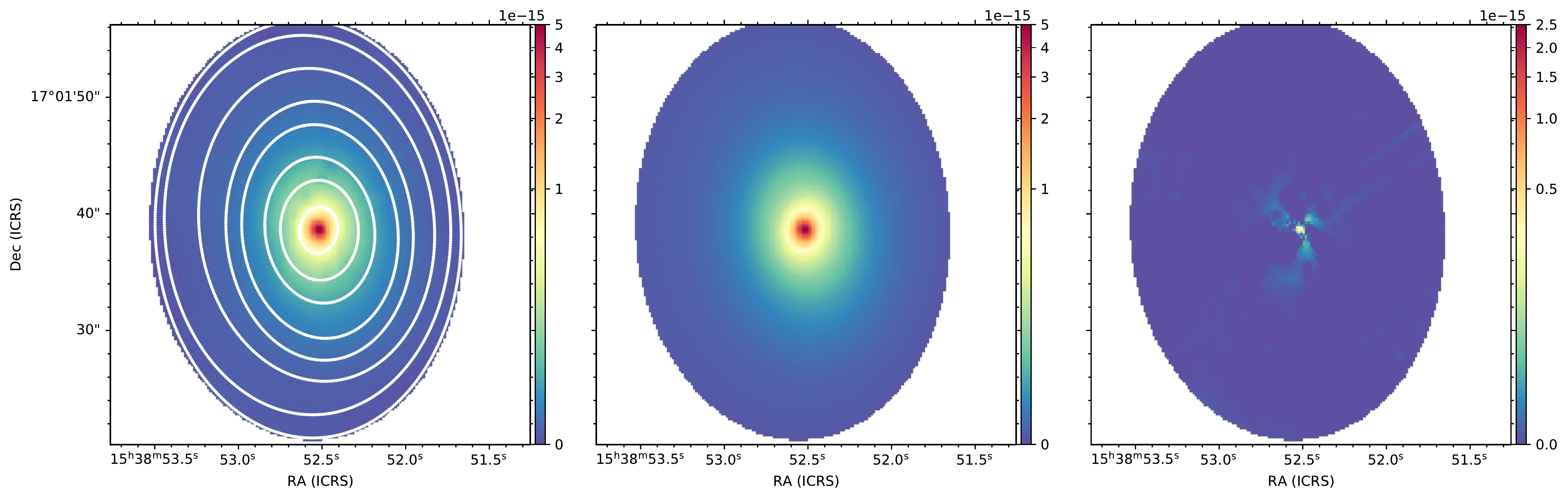}
    \caption{Isophotal ellipse fits to the flux distribution of the stellar component. {\it Left panel:} Flux distribution of the stellar component, obtained from collapsing a 200 \AA\  slab of the data cube near 9000 \AA\  is shown in color, with some example best-fit ellipses overlaid in white contours. {\it Center panel:} Model reconstructed from the best fit to each ellipse. {\it Right panel:} Residual map (observed minus best fit model), showing remaining structure in the central region. The color bar represents the flux in $5.5 \times 10^{-5}$ erg/s/cm$^{2}$/Angstrom.}
    \label{fig:ellipse_fit}
\end{figure*}

\begin{figure}
	\includegraphics[width=\columnwidth]{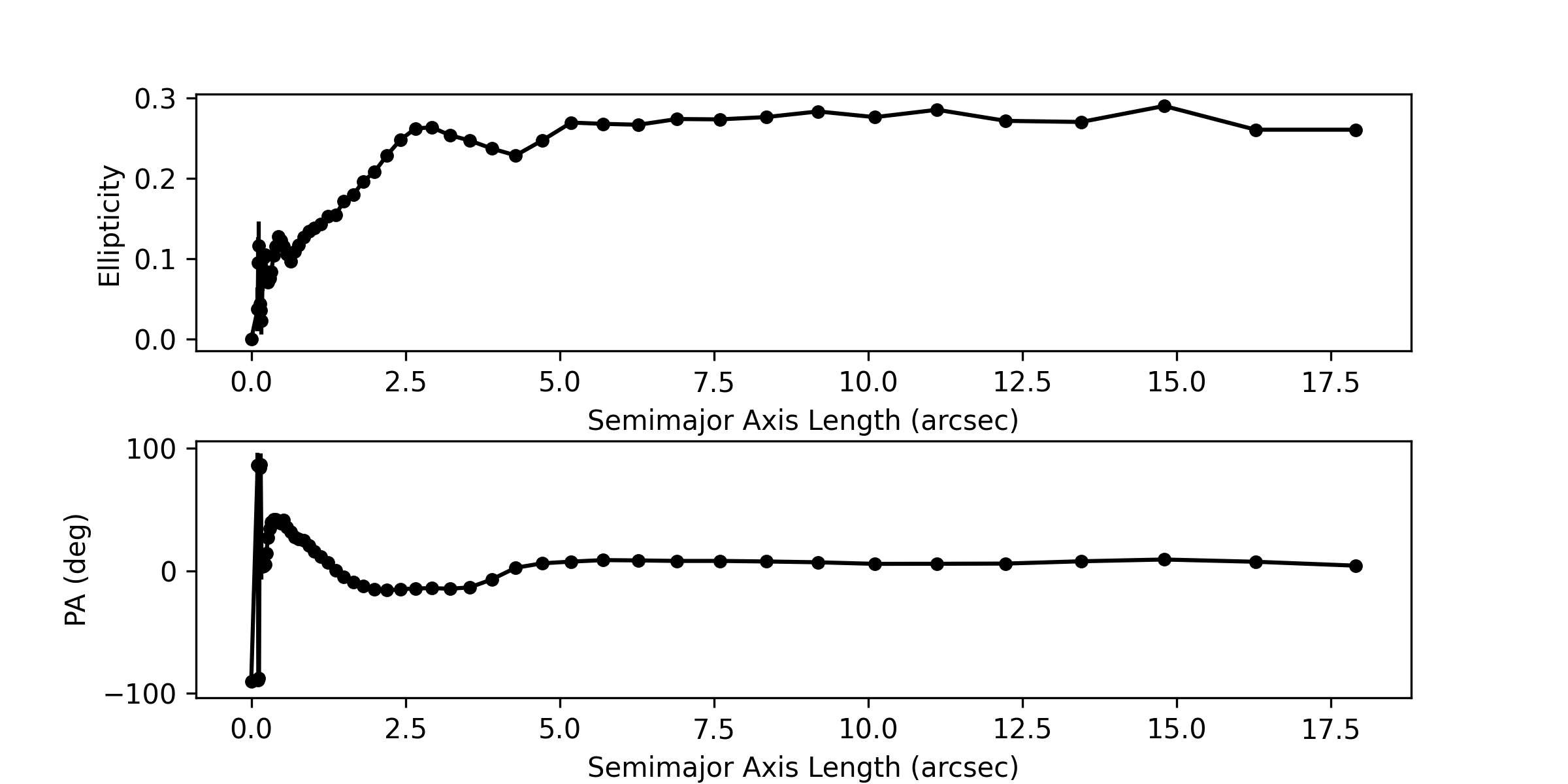}
    \caption{Best-fitted ellipticity (top panel) and PA (bottom panel) for every fitted ellipse as a function of the radius in arcseconds. Different values of PA and ellipticity are observed in the inner $\sim$ 2\farcs5, while the values remain more or less constant outside this radius. }
    \label{fig:ellipse_fit_rad}
\end{figure}

A fit of elliptical isophotes (Fig. \ref{fig:ellipse_fit}) to the stellar moment 0 map, as obtained from the pPXF fit, shows that the inner region ($\lesssim$ 2\arcsec, Fig. \ref{fig:ellipse_fit_rad}) is best fitted with a different PA ($\sim 15$\degree) and ellipticity ($\epsilon \sim 0.1 $) than the rest of the disk (PA $\sim 3$\degree, $\epsilon \sim 0.25 $). 

\begin{figure}
	\includegraphics[width=\columnwidth]{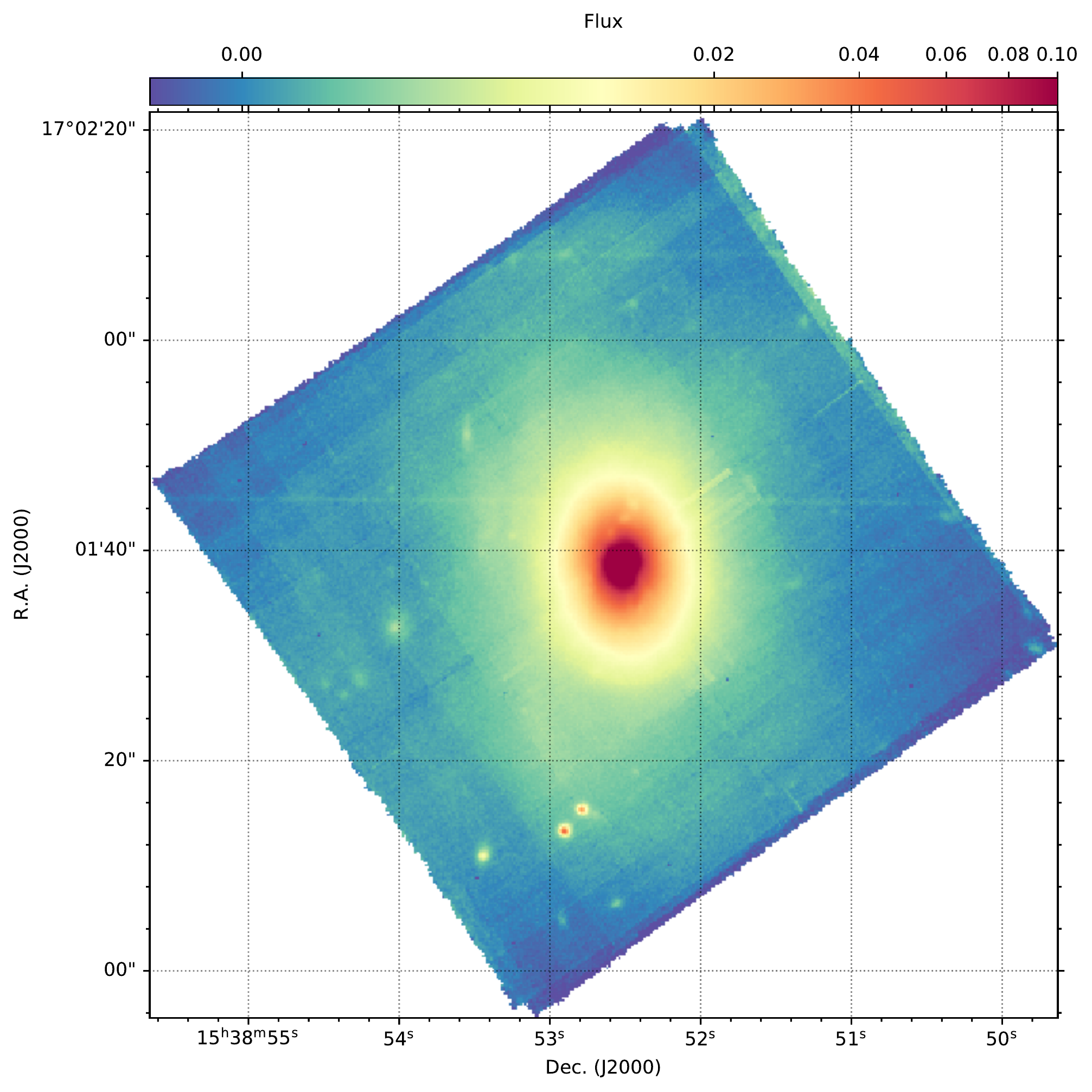}
    \caption{Flux distribution map from collapsing the cube around 9000~\AA. }
    \label{fig:stellar_flux}
\end{figure}

\begin{figure}
	\includegraphics[width=\columnwidth]{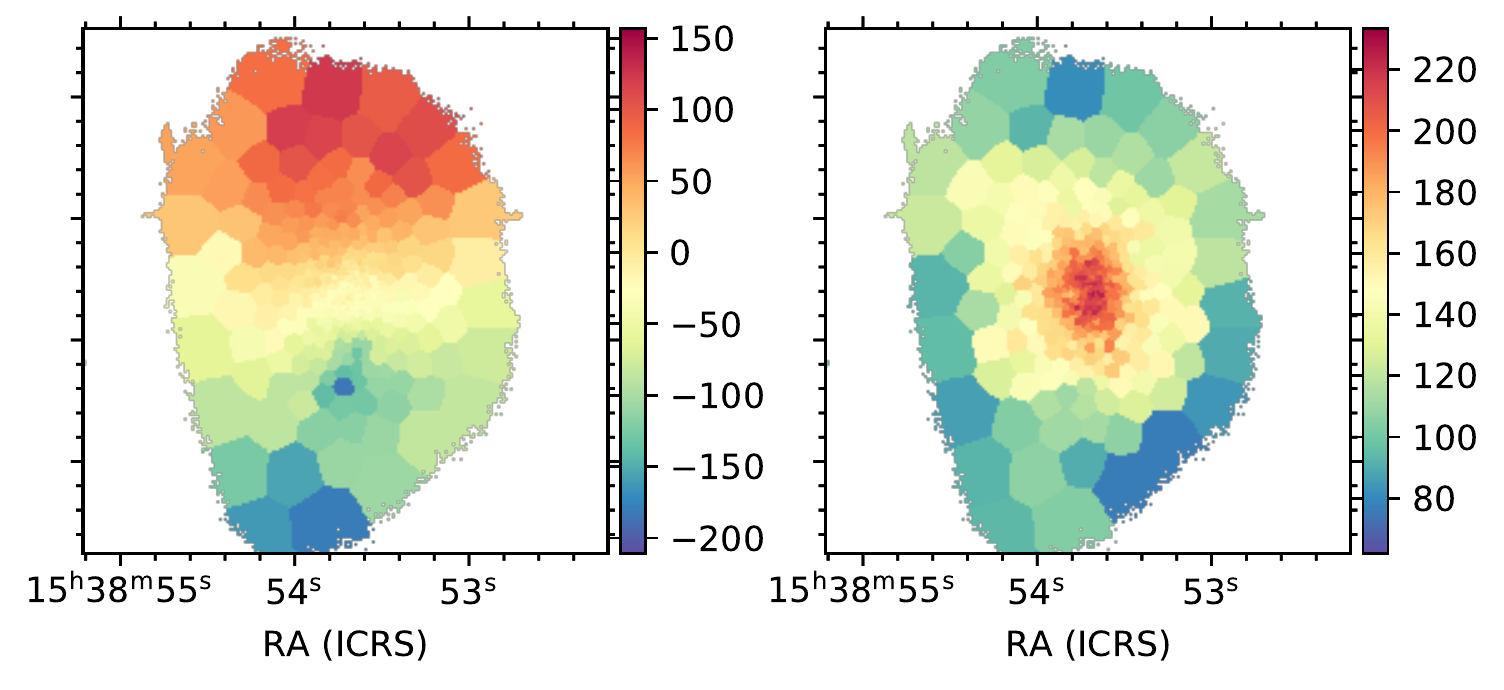}
    \caption{Moment maps from the stellar components. {\it Left panel:} Moment 1 map, showing kinematics dominated by disk rotation with perturbations observed in the N and S directions, color bar units in km/s. {\it Right panel:} Moment 2 map, color bar units in km/s.}
    \label{fig:stellar_moments}
\end{figure}

In Fig. \ref{fig:stellar_flux} we show the integrated flux distribution (moment 0) obtained from collapsing the cube near 9000 \AA\ with a 200 \AA\ range. On Fig. \ref{fig:stellar_moments} we show the stellar component velocity field (moment 1) and velocity dispersion (moment 2) maps from the pPXF fit.
The velocity distribution shows a fairly regular rotating system that reaches velocities of $\pm 150$ km/s along a PA $\sim 10$\degree. However a slightly bent zero-velocity contour and some  blueshifted (redshifted) features in the southern (northern) regions indicate the presence of a perturbation to the rotation. While the velocity dispersion map shows a distribution peaking in the center, as expected for a rotating disk, the values of up to $\sim 220$ km/s (slightly higher than the rotation velocities) suggests some turbulence present in the system. 

\begin{figure}
	\includegraphics[width=\linewidth]{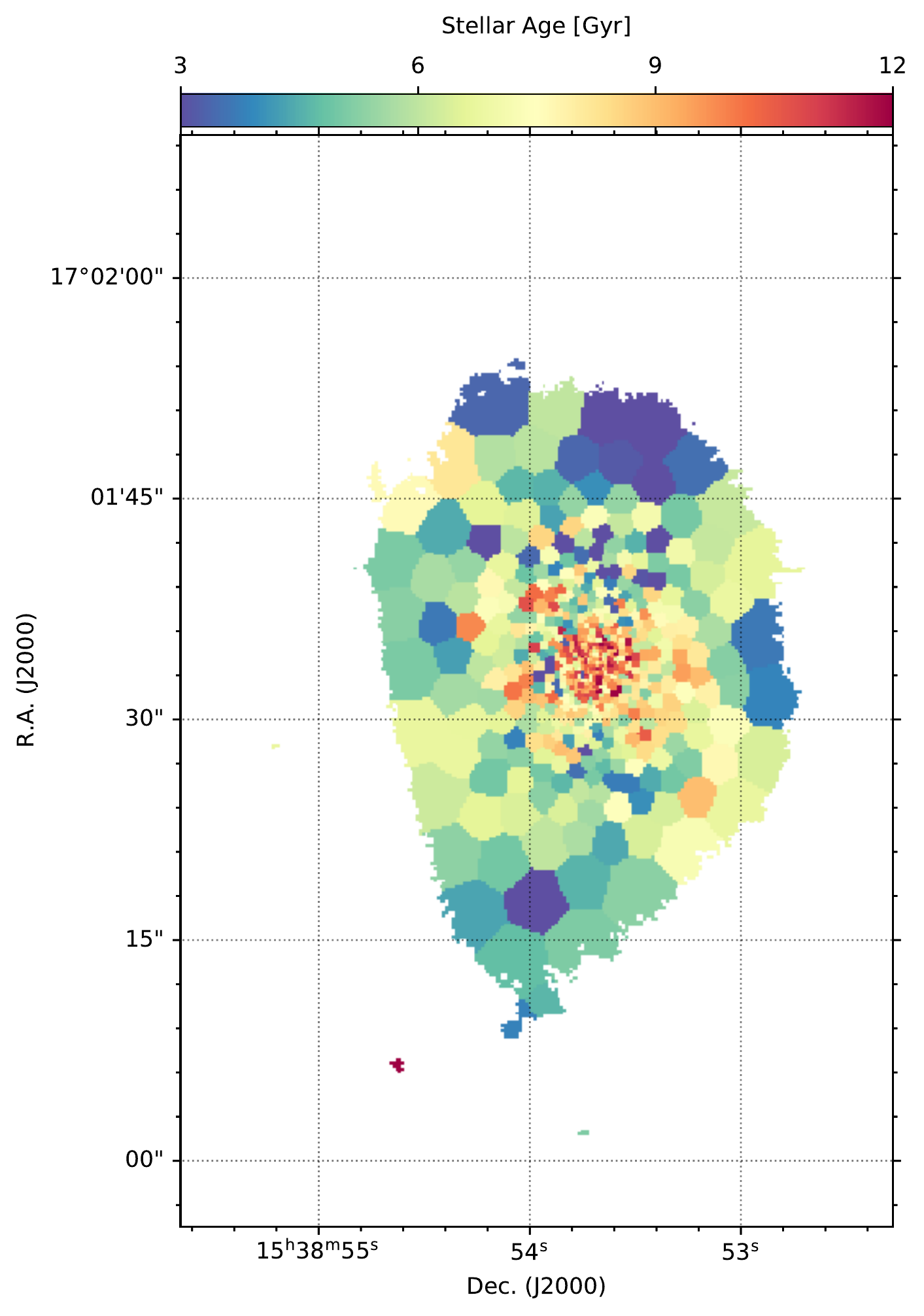}
    \caption{Light-weighted stellar age map, color bar in units of Gyr. The distribution of stellar age shows older populations in the central region, while younger populations are observed at larger radii.}
    \label{fig:metalage}
\end{figure}

The second pPXF run is regularised to impose a smoothness constraint on the solution, applying the 'REGUL' option on pPXF. For this fit we mask the emission lines and apply an eight-order multiplicative Legendre polynomial, fixing the stellar kinematics to those of the first run to avoid degeneracies between stellar velocity dispersion and metallicity. Weights are applied to every template, which are regularly sampled on an age and metallicity grid that covers 0.03 $<$ Age (Gyr) $<$ 14 and -2.27 $<$ Metallicity (dex) $<$ 0.40. The regularisation allows templates with similar age and metallicity to have smoothly varying weights. 
The stellar ages map is shown in Fig. \ref{fig:metalage} where we can observe a stellar population distribution with older populations at the center and younger populations at larger radii.

\subsection{Emission Lines}
\label{sec:emlines}

\subsubsection{Moment maps}

We run one further pPXF fit to the complete unbinned FOV with the purpose of subtracting the stellar continuum from the data cube. We use the unbinned data cube in order to recover the fine spatial structure of the gas component. This is possible due to the high SNR of the emission lines in contrast to the absorption lines. Following the same procedure as above, we masked the emission lines. 
A scaled version of the best-fitted stellar continuum is subtracted to every spaxel of the unbinned cube associated to a given bin. We therefore obtain a continuum-free data cube, from which we can recover the emission lines. The total flux of the emission lines are model-dependent estimates based on this pPXF fitting. 

\begin{figure*}
\gridline{\fig{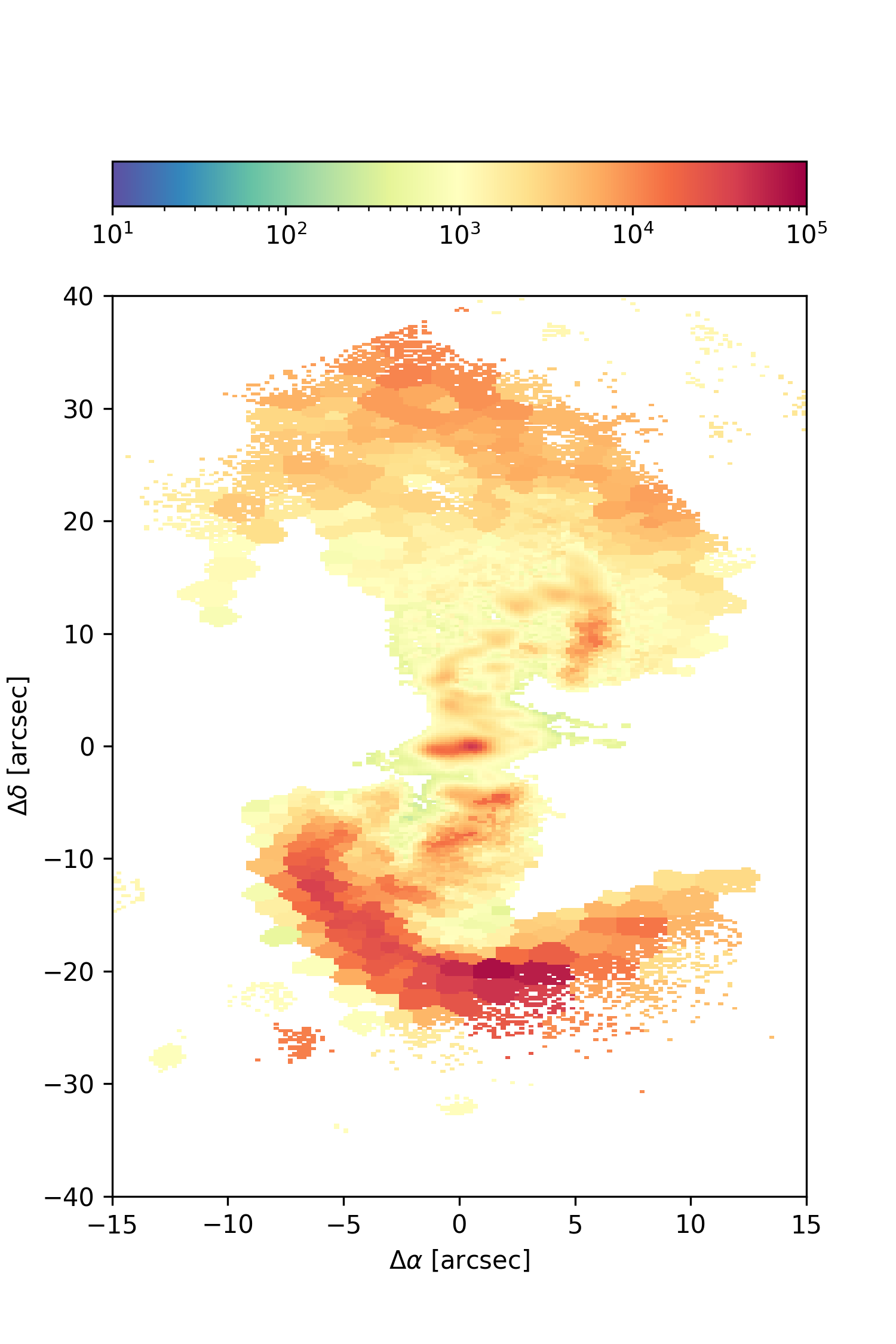}{0.33\textwidth}{}
          \fig{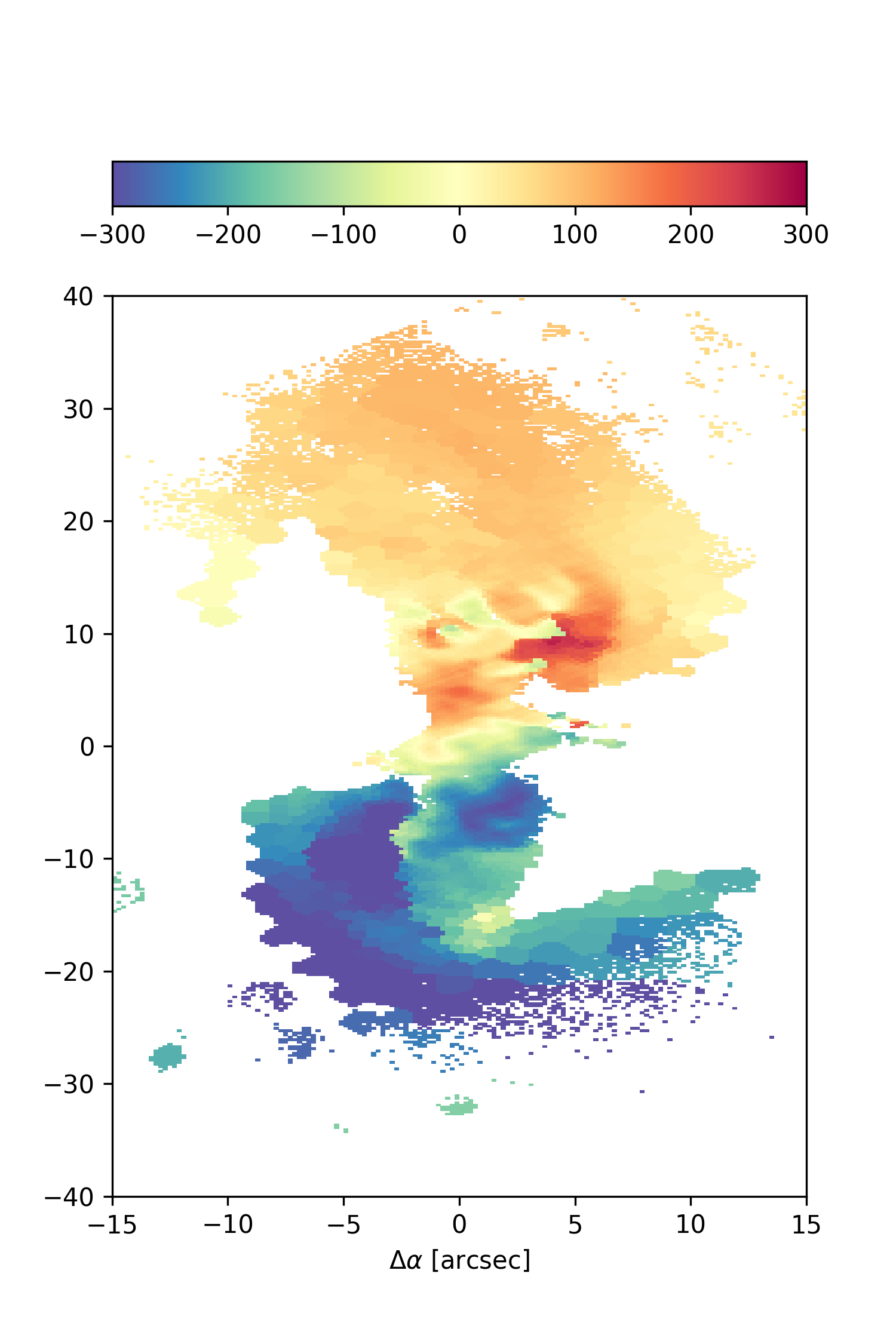}{0.33\textwidth}{}
          \fig{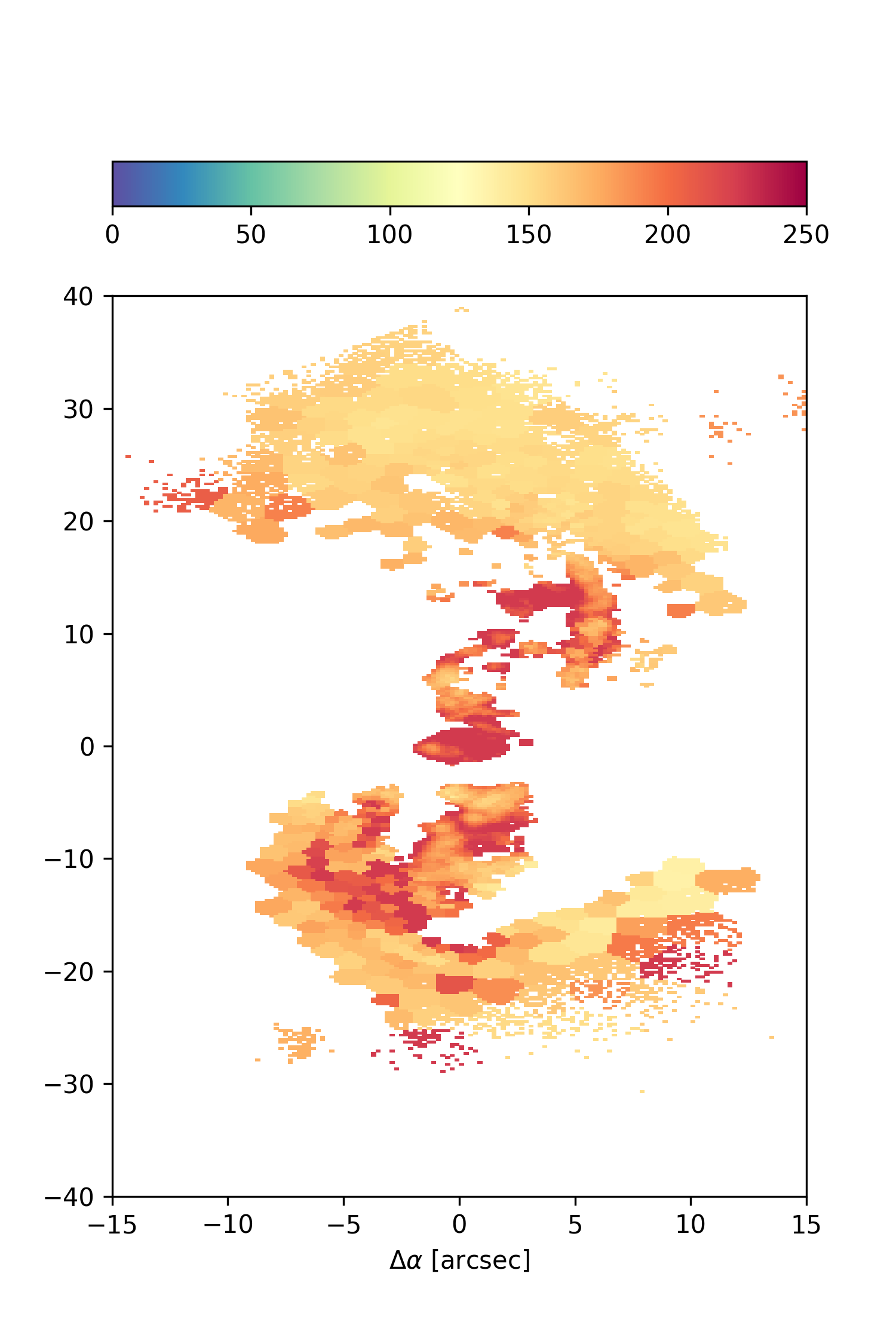}{0.33\textwidth}{}
          }
\gridline{\fig{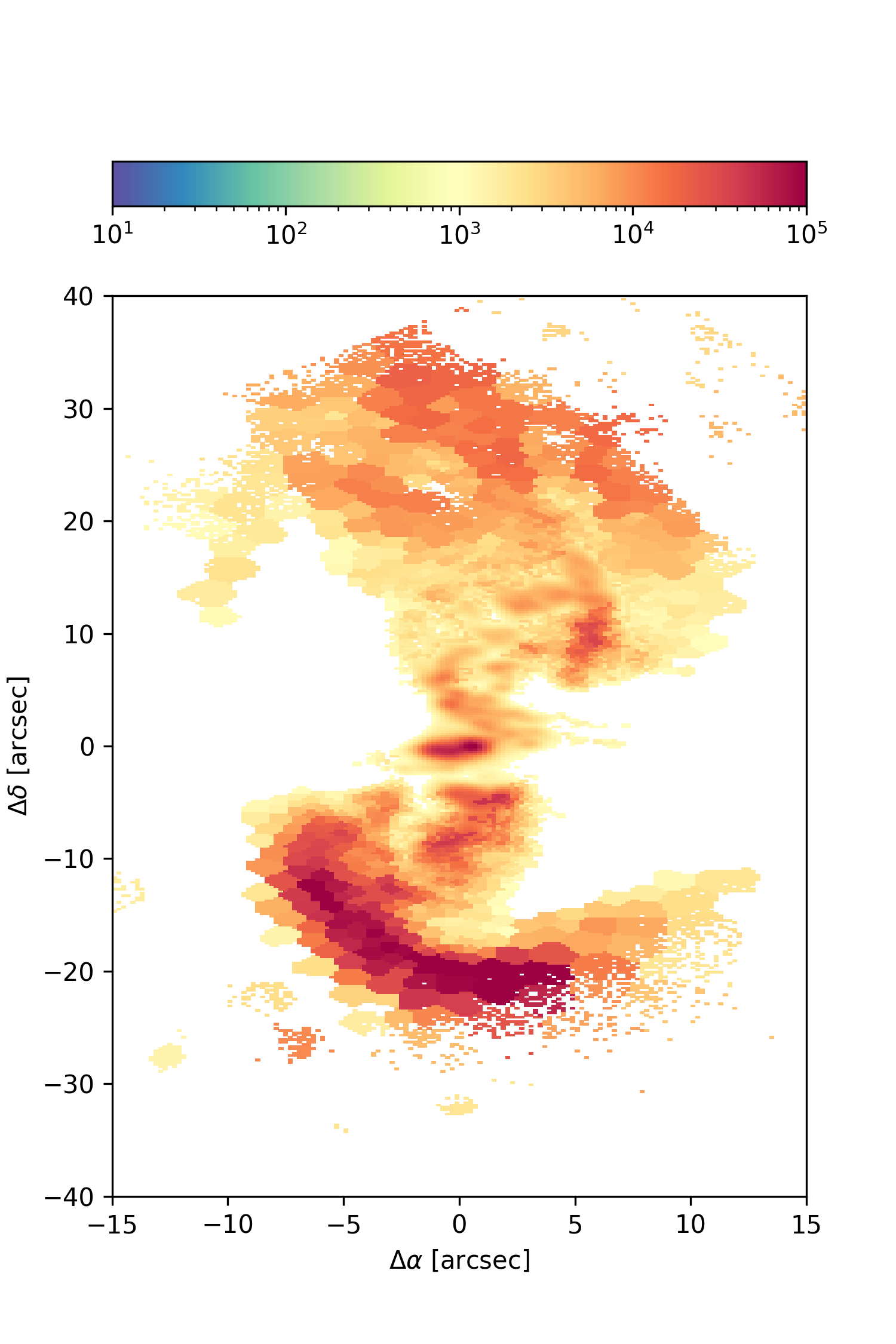}{0.33\textwidth}{}
          \fig{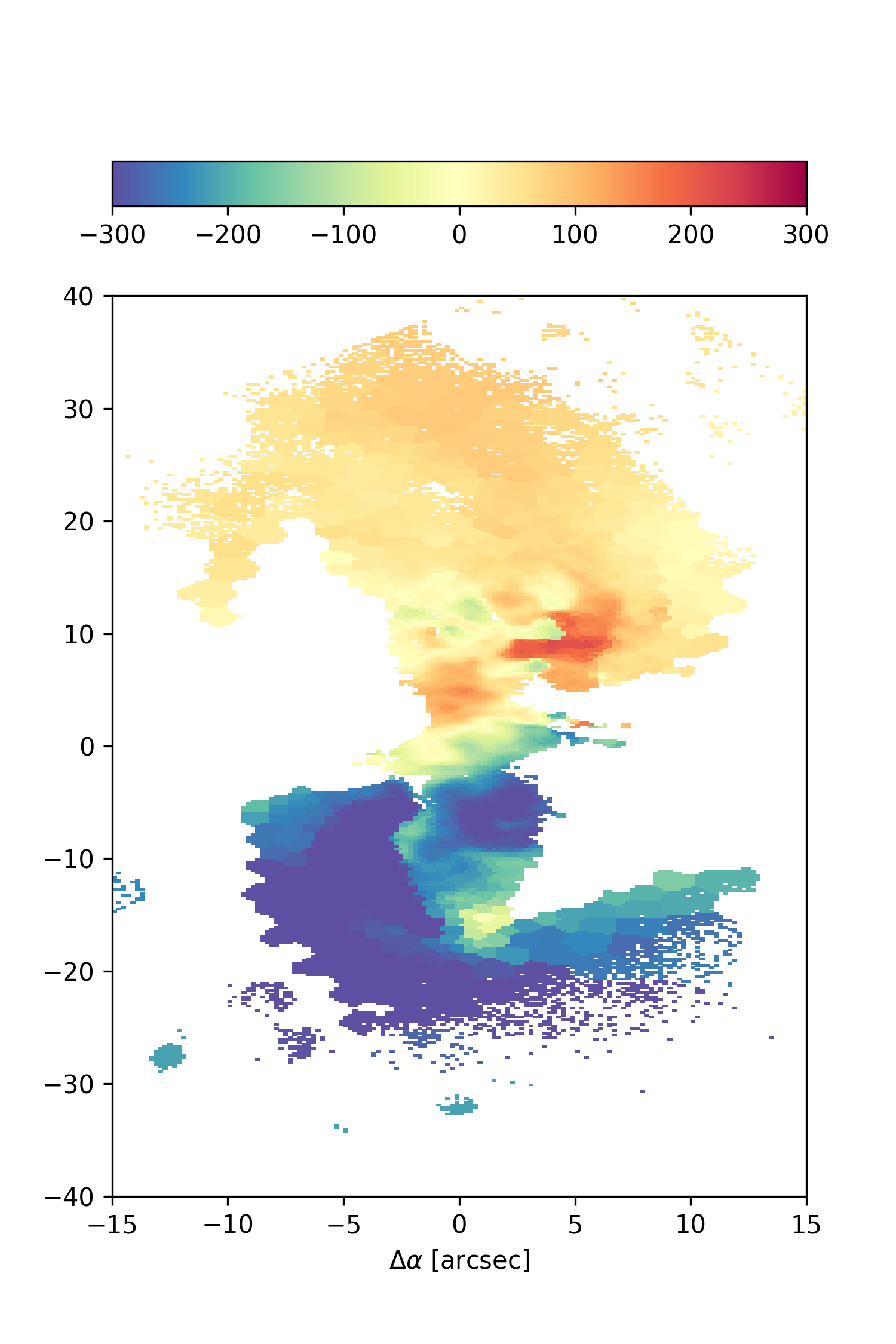}{0.33\textwidth}{}
          \fig{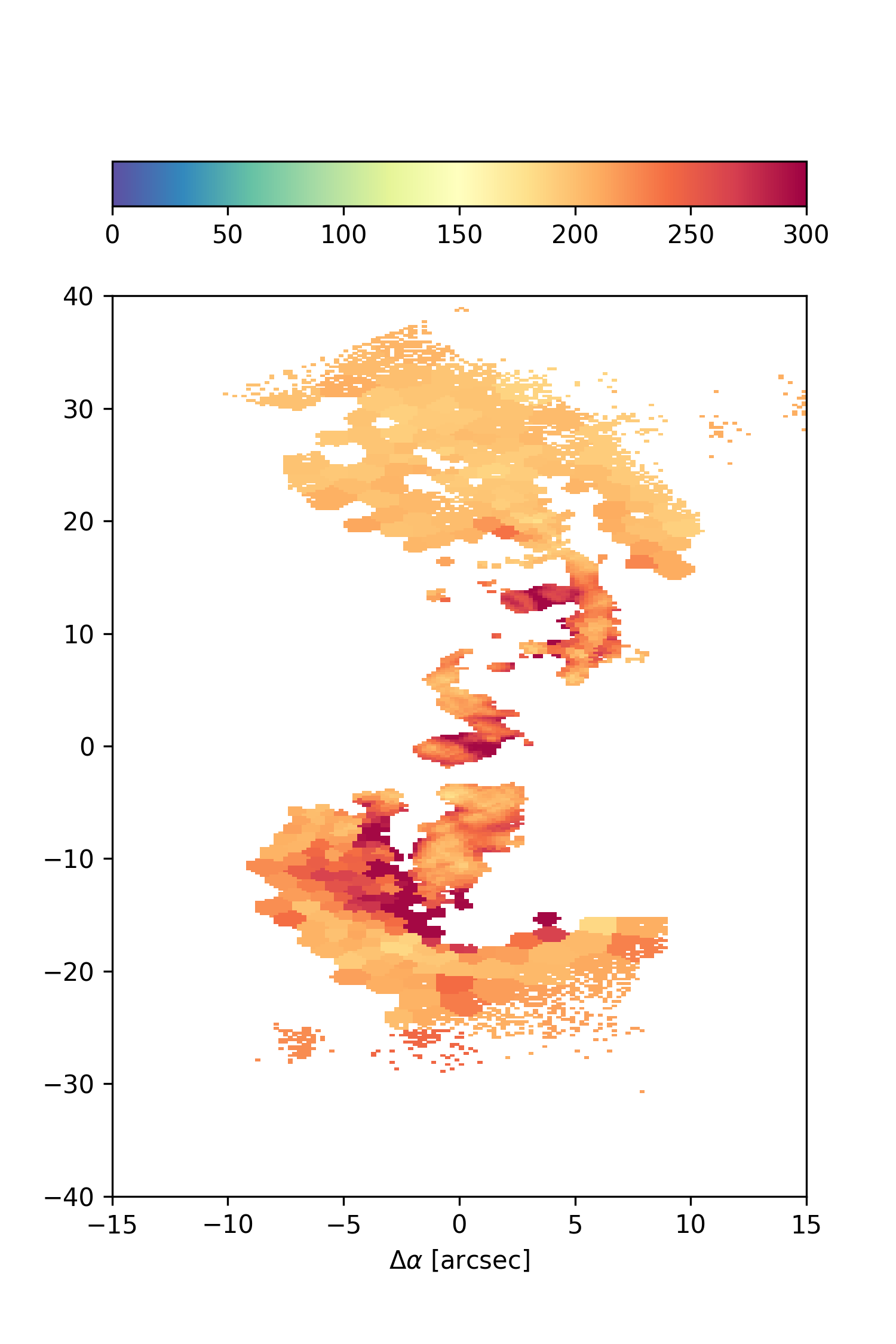}{0.33\textwidth}{}
          }
\caption{Moment maps from a one component Gaussian fit to the emission lines H$\alpha$ ({\it top panels}) and [OIII] ({\it bottom panels}). {\it Left panels:} Moment 0 maps (flux distribution). Color bar in $10^{-17}$ erg s$^{-1}$cm$^{-2}$\AA$^{-1}$. {\it Middle panels:} Moment 1 maps (velocity distribution). Color bar in km/s. {\it Right panels:} Moment 2 maps (velocity dispersion). Color bar in km/s. All the maps are clipped at 3$\sigma$.
\label{fig:moments_onecomp}}
\end{figure*}

The emission lines, in every bin, were then fitted with Gaussian profiles, tying the [OIII]$\lambda4958,5007$ \AA\ and [NII]$\lambda$6548,6562 \AA\ flux ratios, H{$\beta$} and H{$\alpha$} line widths, the relative positions of the emission lines, and the forbidden emission lines width to the [OIII]$\lambda$5007 \AA\ line width, given its considerably higher SNR. This approach does not leave considerable residuals in the other emission lines. This fit was made for all spaxels with SNR $> 4$, where the SNR was calculated in a wavelength range that includes the [OIII]$\lambda$5007 \AA\ forbidden emission line. From this one Gaussian component fit we extract the flux and kinematics for the emission lines as shown in Fig. \ref{fig:moments_onecomp}.
The flux distribution of the gas component shows a complex spatial distribution. The ionized gas extends further out than the stellar component, forming a "helix"-shaped pattern that displays very fine filamentary structure. High SNR ionized gas extends $\sim$25\arcsec\ from the center, while fainter structure is observed beyond this radius, up to 32\arcsec. 

The northern trail of gas appears to extend from the center and then curve back towards the nucleus, forming at the same time fine gas filaments towards the NE. In the Southern region the ionized gas appears to extend from the nucleus in two arms. One curves slightly towards the E, and shows higher flux. The other extends from the E of the nucleus, and curves towards the W. 
The gas velocity field shows complex kinematics that, while mainly redshifted in the N region and blueshifted in the S, do not appear to be a primarily rotating disk, in contrast to the stellar component. This is supported by the distribution of the velocity dispersion map, which shows a complex distribution that does not resemble the expectation for a purely rotating disk.

\subsubsection{Dust Extinction}

\begin{figure}
	\includegraphics[width=\columnwidth]{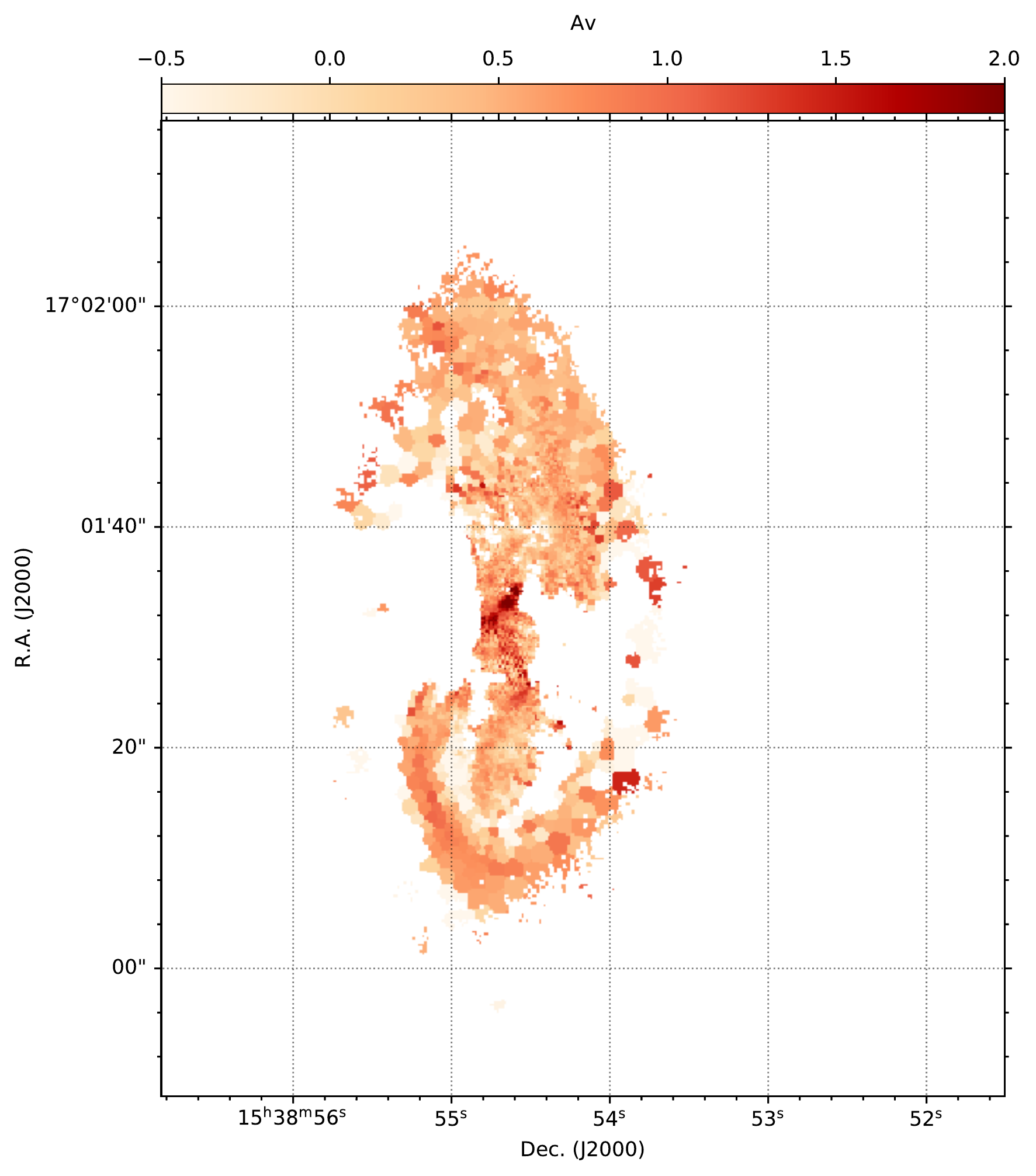}
    \caption{V-band extinction map obtained from the H$\alpha$ to H$\beta$ flux ratio. This map shows the presence of a high dust obscuration, in the central region and along the southern arms.}
    \label{fig:extinction}
\end{figure}

To obtain a map of the extinction by dust in the line of sight we calculate the Balmer decrement (using the H$\alpha$/H$\beta$ emission line ratio). From this, we estimate the total extinction (A$_{V}$), in the V-band, following \cite{dominguez+2013}. We use the average Galactic extinction curve from \cite{osterbrock+2006} assuming an intrinsic value of H$\alpha$/H$\beta = 2.86$ which corresponds to a temperature T$= 10^4$ K and electron density n$_{e} = 10^{2}$ cm$^{-3}$ for case B recombination. The extinction map (Fig. \ref{fig:extinction}) shows the presence of dust in  an upside-down V shape with its apex close to the center, and some dust extending along the SE and SW arms. It has been suggested \citep{keel+2015} that these dust lanes were formed by a differentially precessing warped disk, as modeled by \cite{steiman+1988}. We use the derived extinction values to correct the emission line fluxes used in the analysis of Section \ref{sec:lum_hist}.\\

A starlight attenuation map was already created for this galaxy based on HST WFC3 imaging, as presented in Fig. 7 of \citet{keel+2015}. Assuming it represents pure continuum, which is then divided by a smooth model. This map shows almost no starlight attenuation in the SE arm, which suggests that it is on the far side of the system, and only absorbs a small fraction of the starlight behind it.
The three dots observed in Fig. \ref{fig:extinction} in the NW arm match well with the features observed in the starlight attenuation map, implying the filament in that region is in front of most of the starlight. Finaly, a dust feature is observed aligned N to S from 3.5\arcsec\ to 5.6\arcsec, this feature is S of the nucleus and it does not show a counterpart in the Balmer-decrement map, indicating that it may be decoupled from the ionized structure.

\subsubsection{Origin of the ionized gas}
\label{sec:origin_ionization}
\begin{figure}
	\includegraphics[width=\columnwidth]{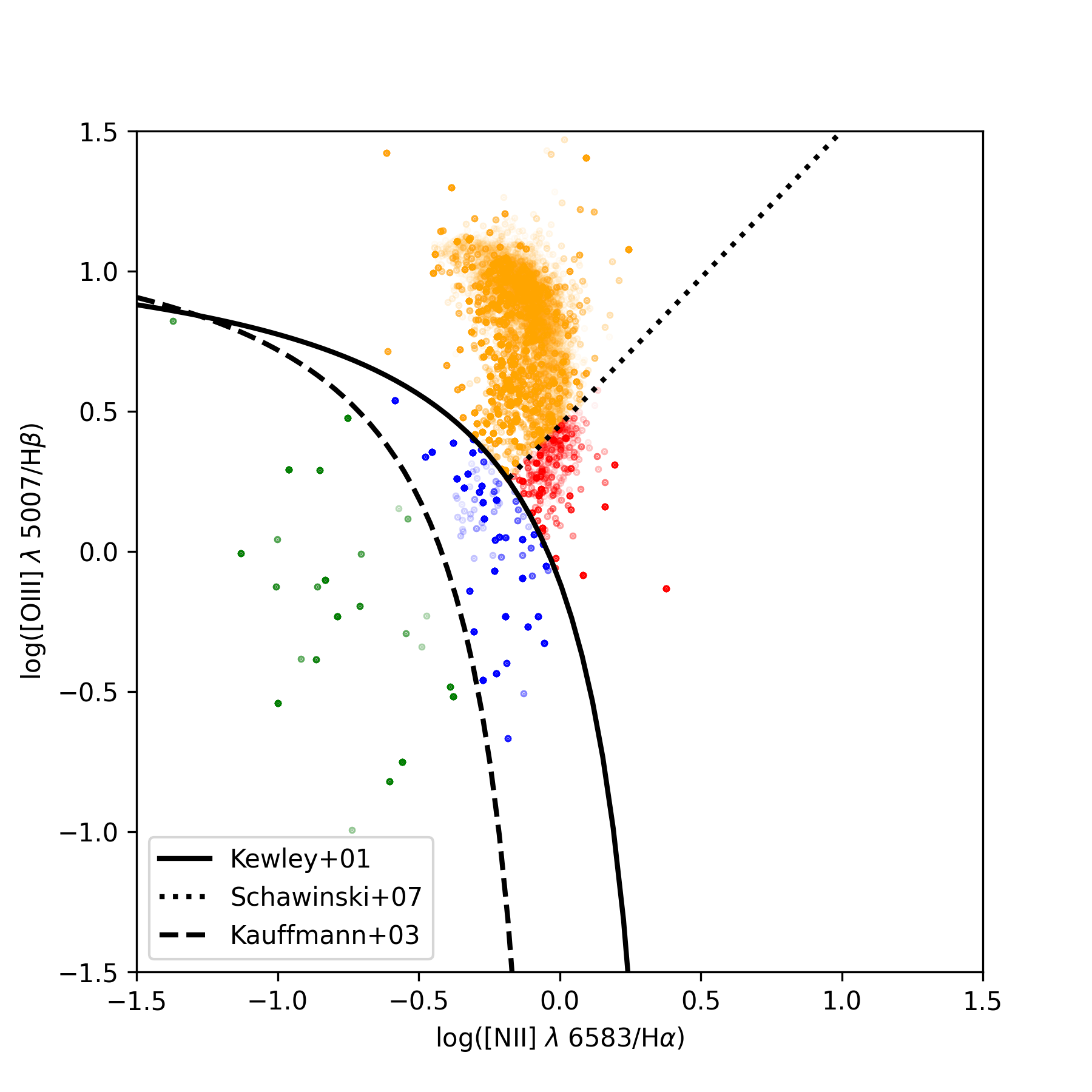}
    \caption{BPT diagram for the resolved NGC 5972 map as presented in Fig. \ref{fig:bpt_2d}. The dashed line marks the separation between Composite and Star-Forming, as proposed by \cite{kauffman+2003}. The solid line marks the transition between Composite and AGN as proposed by \cite{schawinski+2007}, and the dotted line marks the separation between Seyfert and LINER, as proposed by \cite{kewley+2001}. The colored circles represent every spaxel in Fig. \ref{fig:bpt_2d} with Star-Forming emission shown in green, Composite in blue, LINER in red, and Seyfert in orange.}
    \label{fig:bpt_1d}
\end{figure}

\begin{figure}
	\includegraphics[width=\columnwidth]{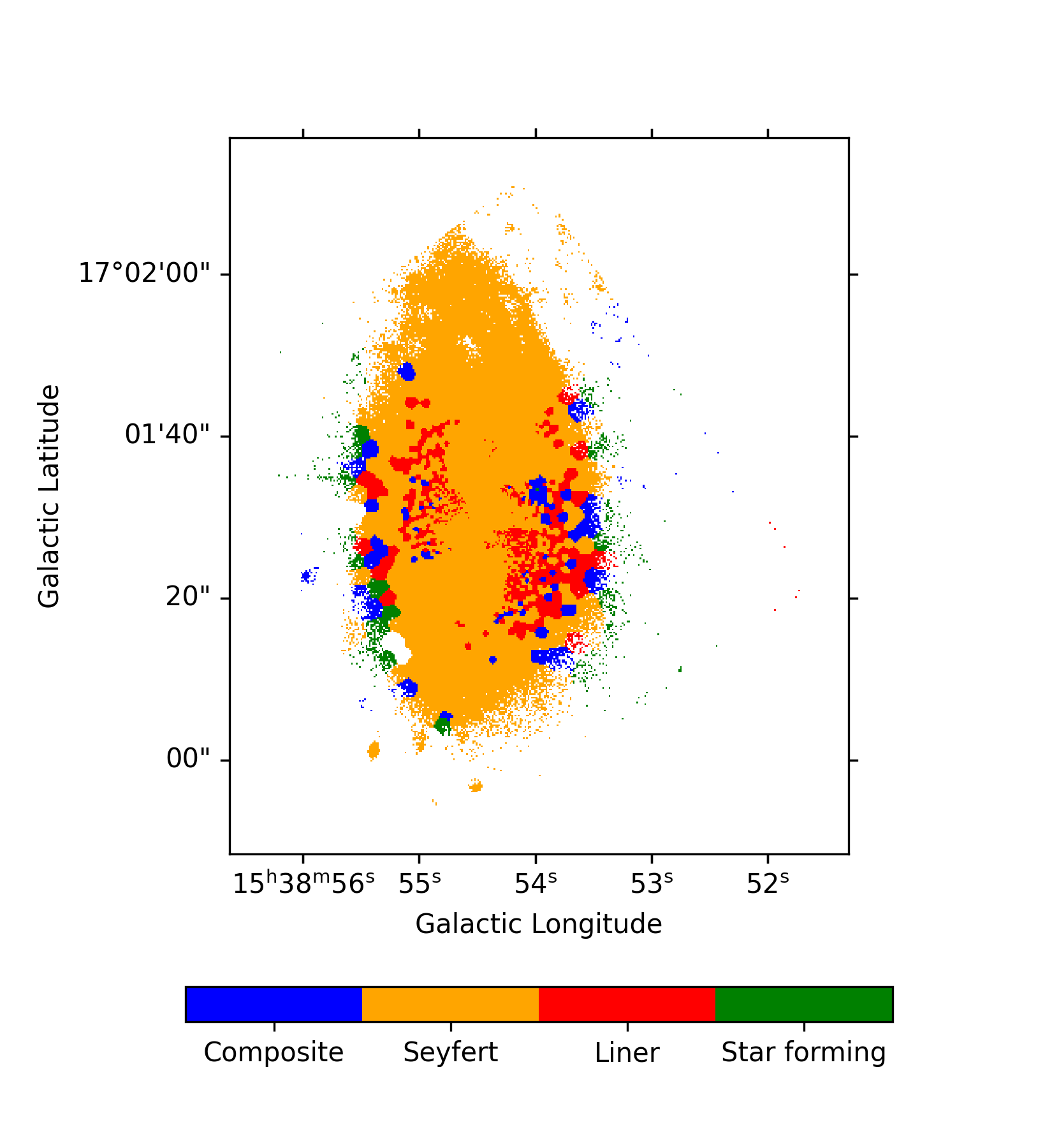}
    \caption{Resolved BPT diagram of NGC 5972, colors represent different sections of the BPT diagram as described in Fig. \ref{fig:bpt_1d}. This map shows that the arms are dominated by Seyfert-like emission, and are surrounded by a mixture of LINER-like and Composite emission. }
    \label{fig:bpt_2d}
\end{figure}

\begin{figure}
	\includegraphics[width=\columnwidth]{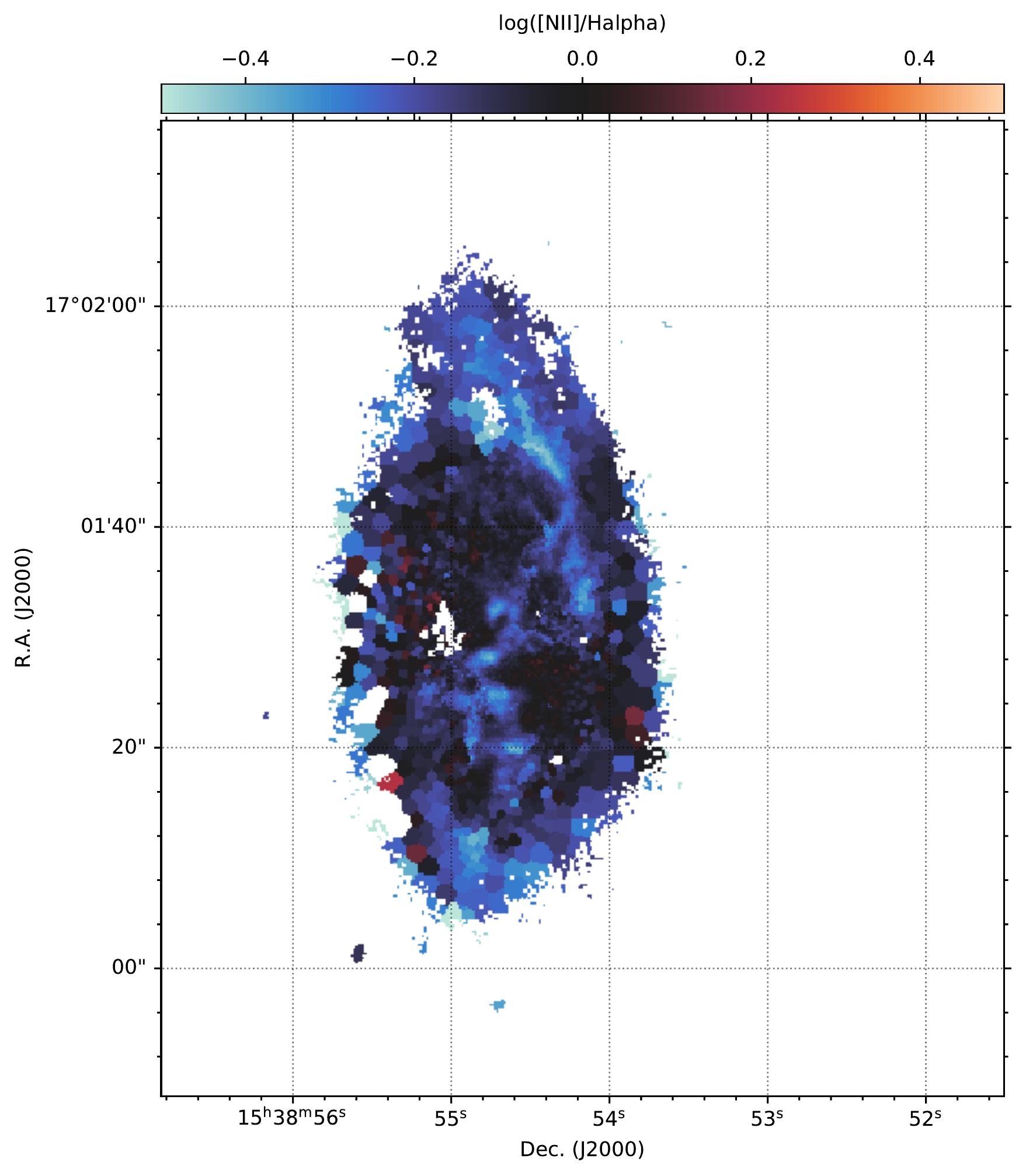}
    \caption{Resolved [NII]/H$\alpha$ flux ratio map. In this map we observe an excess of H$\alpha$ over [NII] by a factor of $\sim 2$ in particular knots over the arms (light-blue areas). This may be an indication of star formation in these areas. }
    \label{fig:bpts_gradient}
\end{figure}

We compute emission line ratios from the fluxes obtained from the one-component Gaussian fit, in order to analyse the distribution and origin of the ionized gas. We use the [OIII]/H$\beta$ and [NII]/H$\alpha$ emission line ratios to create a Baldwin, Phillips \& Telervich \citep[BPT;][]{baldwin+1981} diagram, where every spaxel is classified between Star-Forming, Composite, LINER or Seyfert. This color-coded classification is presented in Figures \ref{fig:bpt_1d}-\ref{fig:bpt_2d}, and shows the presence of Seyfert-like ionization along the arms that form the helix pattern, which is surrounded by a mixture of LINER-like and Composite ionization. The [OIII]/H$\beta$ map (Fig. \ref{fig:bpts_gradient}) shows that [OIII] dominates over H$\beta$ in the arms, with higher [OIII] to H$\beta$ ratio inside the arms, falling to a lower ratio quickly outside the filamentary structure. The [NII]/H$\alpha$ map shows the structure inside the arms where H$\alpha$ emission is higher than the [NII]. Some nodes with higher H$\alpha$ emission are observed near the nucleus and in the filamentary structure of the arms.\\
The resolved BPT reveals that AGN ionized gas extends uninterrupted from the central region up to radius $\sim$22\arcsec, and that AGN photoionization is the main ionization mechanism observed in the galactic disk, with Seyfert-like ionization in the arms and mostly LINER-like ionization in the disk outside the arms. 
To evaluate the possibility that the ionized region extends over a bicone centered in the nucleus, tracing a large-scale extended NLR, we assume the observed ionization is bounded by the availability of radiation rather than gas. We estimate a required full opening angle for the ionization cones of $\sim 75$\degree\ to encompass the entire EELR. This angle is estimated from the projected angular width of each half of a notional bicone that encompasses the observed ionized regions. Given the projection effects this angle is an upper limit. Making the ionization cones narrower than the observed $\sim 75$\degree\ on each side would require the axis to be closer to our line-of-sight, putting all the EELR features farther from the AGN. In the context of the analysis conducted on Sect. \ref{sec:lum_hist} this would imply at even larger mismatch between the present-day AGN luminosity and that of the ionized clouds.

\subsection{Kinematics}
\label{sec:kinematics}

\subsubsection{Stellar component}

\begin{figure*}
	\includegraphics[width=\linewidth]{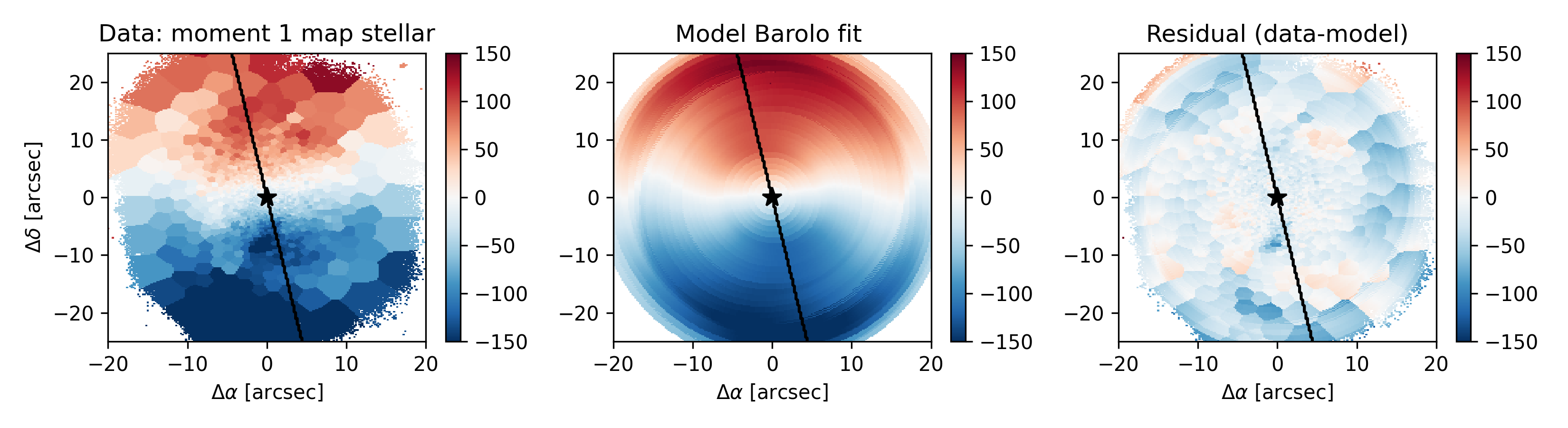}
    \caption{{\it Left panel:} Moment 1 for the stellar component obtained from the unbinned pPXF run.
	{\it Middle panel:} Kinematic rotation model for the stellar component obtained from the {\it 2DFIT} task inside the routine \texttt{$ ^{3D}$Barolo}.
	{\it Right panel:} Residual map obtained from subtracting the kinematic model from the moment 1 map.
	All color bars units are in km/s. The black star shows the kinematic center and the black line marks the average PA of 2\degree.
	}
    \label{fig:stellar_model}
\end{figure*}

We modelled the kinematics of the stellar component using the \texttt{2DFIT} task included in the package \texttt{$ ^{3D}$Barolo} \citep{diteodoro+2015}, which fits tilted-rings of a chosen thickness, we use a width of 0\farcs4, fixing the center (as the peak of the continuum) and the systemic velocity and varying the rotation velocity, the PA and the inclination for every ring. The best-fit model and the residuals are shown in Fig. \ref{fig:stellar_model}. The residuals, obtained by subtracting the model of our stellar velocity map from the measured one, show a small blueshifted excess S and SE of the nucleus and redshifted E and SW of the nucleus. Some of these features coincide with the region where the arms observed in the gas component are present. The model reaches $\pm 150$ km/s and it has a curved zero-velocity contour, indicative of some perturbation in the disk. Along the different radii, the PA remains close to 6\degree\ varying to $\sim 353$\degree\ at radius $\sim 15$\arcsec and then returning to 6\degree. The inclination remains closer to 41\degree in the inner $\sim 15$\arcsec and increases to $\sim 50$\degree at larger radii. The rotation curve reaches $\sim 150$ km/s at radius 7\arcsec and remains more or less constant until radius 15\arcsec from where it increases steadily up to 220 km/s.

\subsubsection{Gas component}
\begin{figure*}
	\includegraphics[width=\linewidth]{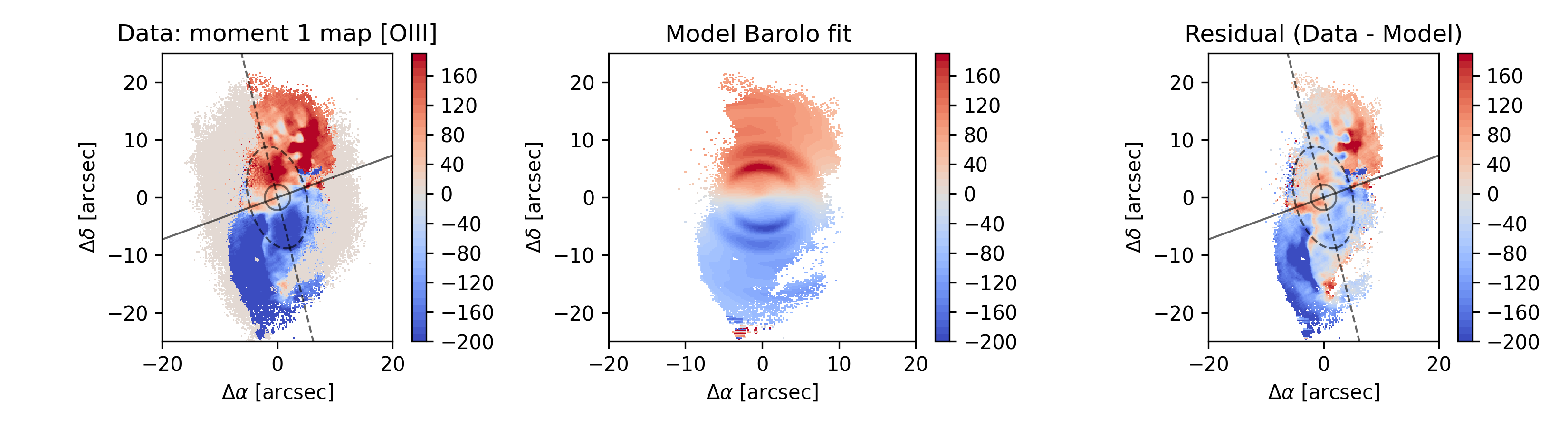}
    \caption{{\it Left panel:} Moment 1 map for the [OIII] emission line. 
	{\it Central panel:} Kinematic model fitted to a data cube trimmed to contain only the [OIII] emission line using the \texttt{$ ^{3D}$Barolo} routine. The dashed line indicate the mean PA (14\degree) of the fit, while the dashed ellipse shows the extension of the inner disk. The solid line indicates de PA of the central feature, with the solid black circle indicating the extension of the nuclear feature. {\it Right panel:} Residual map obtained from subtracting the fitted-model to the moment 1 map.
	}\label{fig:3dfit_moments}
\end{figure*}

\begin{figure}
	\includegraphics[width=\linewidth]{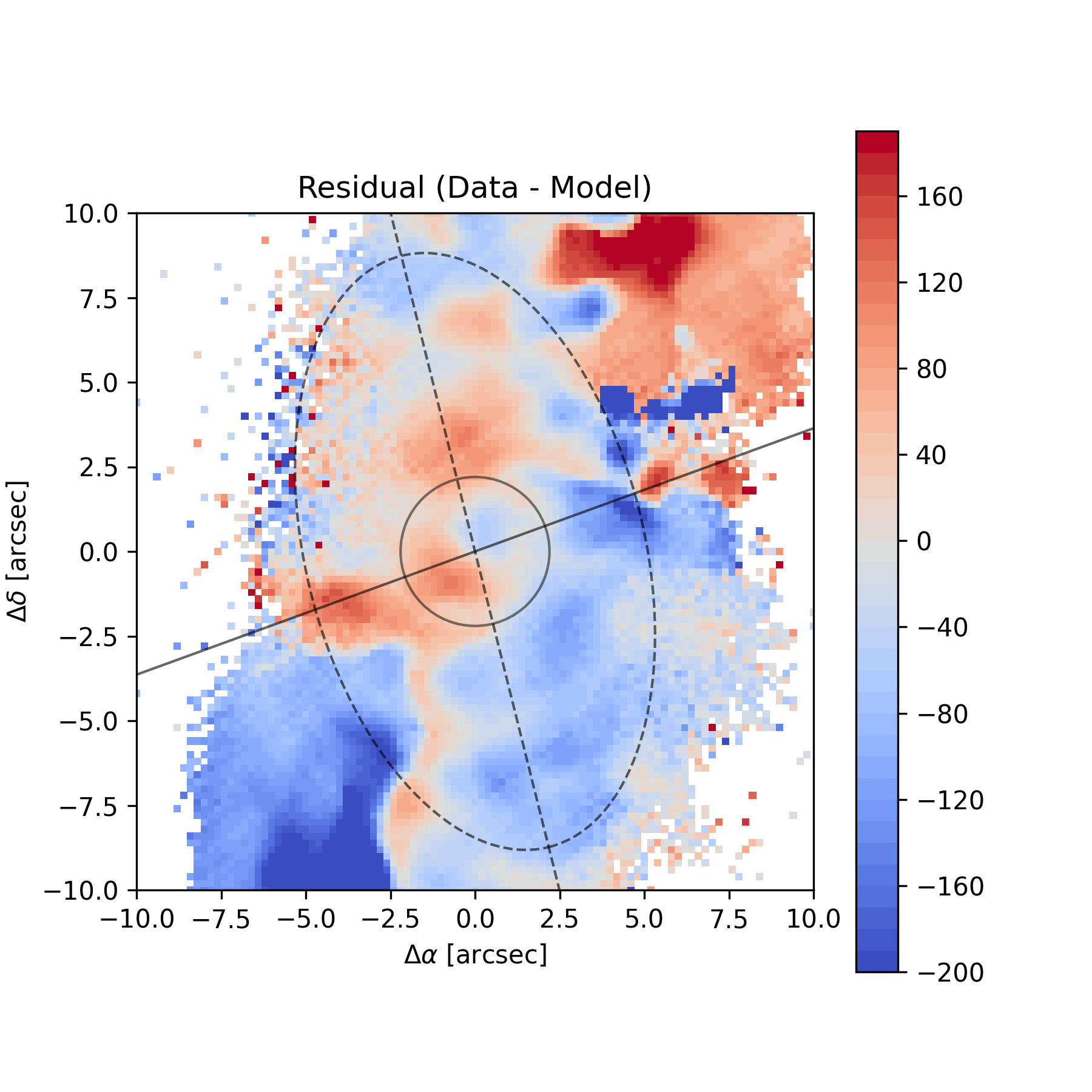}
    \caption{ Same as right panel of Fig. \ref{fig:3dfit_moments}, zoomed in at the central region.}\label{fig:oiii_resd_zoom}
\end{figure}

\begin{figure}
\label{fig:3dfit_pvds}
\gridline{\fig{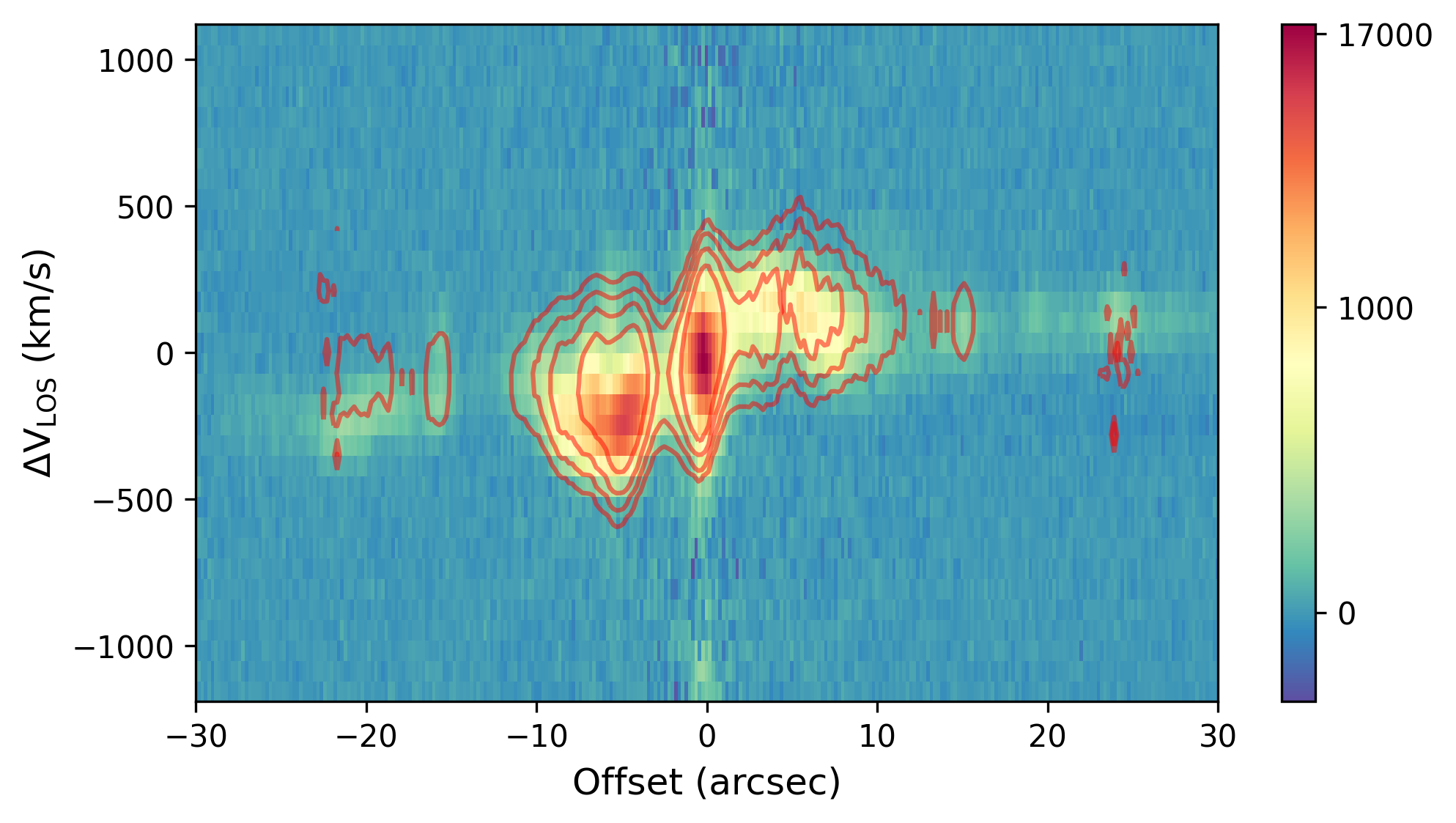}{\columnwidth}{}
          }
\gridline{\fig{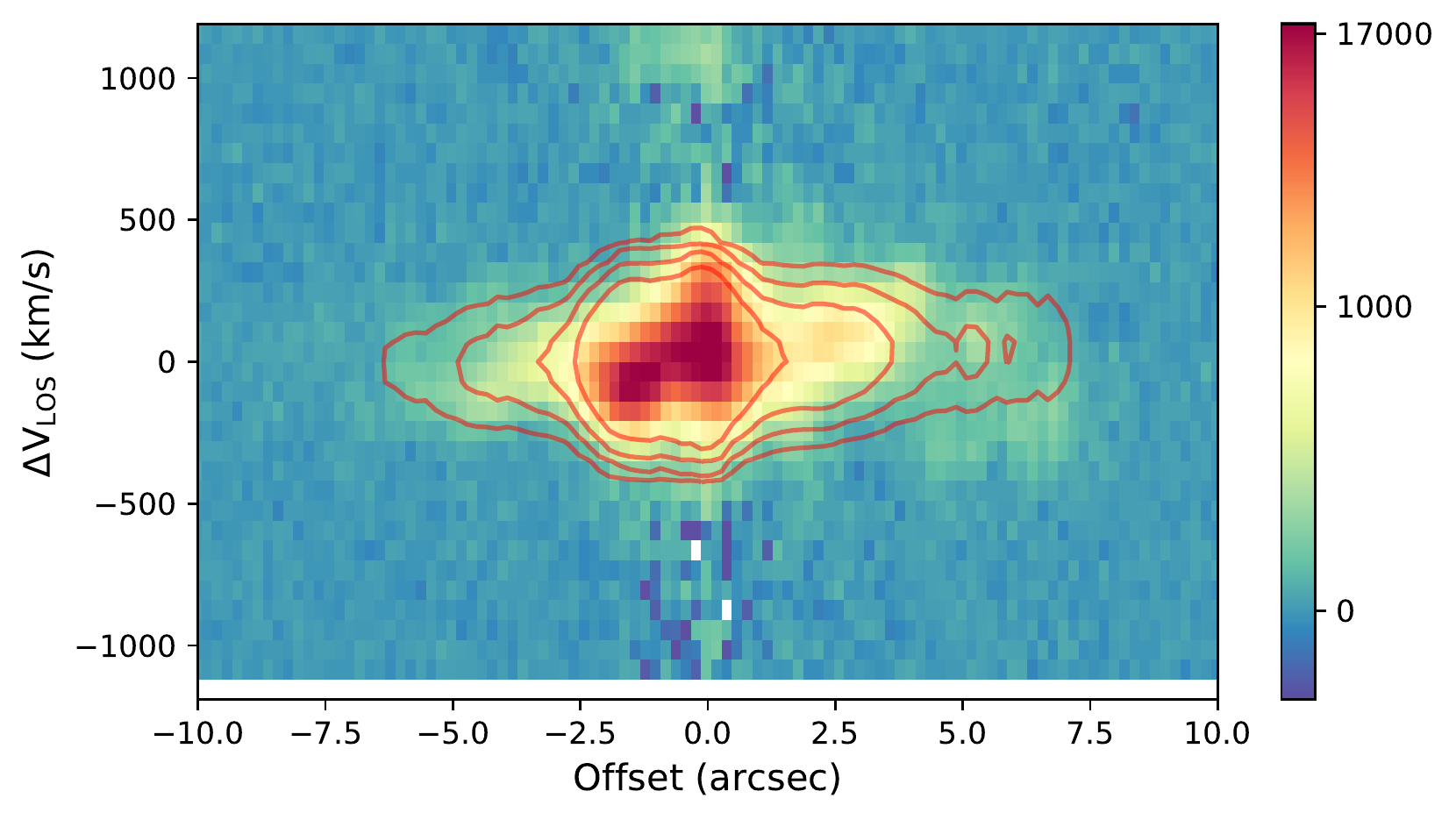}{\columnwidth}{}
          }
\caption{Position velocity diagram extracted from the continuum subtracted data cube along the major (top panel) and minor (bottom panel) axes, centered on the [OIII] emission line. Red contours show the PV-diagram from the best-fitted 3D model obtained with the the \texttt{$ ^{3D}$Barolo} routine.  }
\end{figure}

The moment 1 for the [OIII] emission line reveals complex kinematics that seem to be, at least partially, rotating on a disk. To model this kinematics we utilize the \texttt{3DFIT} task in the \texttt{$ ^{3D}$Barolo} routine, to perform a 3D-fitting to a data cube trimmed on the wavelength-axis to contain only the [OIII] emission line. The fit is done in two stages. For the first stage we leave as free parameters the inclination, major axis PA, circular velocity and velocity dispersion. From this fit we obtain values for the inclination and the PA on each radius. The mean of these parameters over all the rings are 14\degree and 35\degree, respectively. The values from ring to ring do not deviate considerable from the mean values.  For the second stage we fix the center to the peak of the continuum, and the inclination (35\degree) is fixed to the value obtained in stage one. We leave the PA as a free parameter but we use the mean value from stage one as the initial guess for the second fit. The radial velocity is fixed at zero at this point, assuming only a rotational component. 
The code fits a pure circular rotation model to the data cube, while the non-circular motions can be observed in a residual map (Fig. \ref{fig:3dfit_moments}, and a zoomed-in version in Fig. \ref{fig:oiii_resd_zoom}), in which very complex structures can be observed. The fitted-model shows a large-scale rotation disk that reaches $\sim 120$ km/s. In the inner radius (6\arcsec), a higher velocity gradient is observed reaching $\sim 200$ km/s. This feature can correspond to an inner disk that has been disrupted, given the complex distribution observed in the velocity field. In the nuclear 1\arcsec, a feature can be observed along PA 100\degree, with velocities of $\sim 50$ km/s.  
To confirm that these features are not model-dependent we create a position-velocity diagram (PVD), which consists on extracting the flux along a slit from the datacube. The positions are taken relative from the center, and the wavelength is transformed into a velocity offset centered at systemic velocity. In the PVD extracted along the minor axis, two loci can also be observed (Fig. \ref{fig:3dfit_pvds}). Considering the possibility of the near side corresponding to the W side of the disk, as the extinction map indicates, this feature could be explained by a nuclear outflow, we further explore this nuclear outflow by fitting the spectra with two Gaussian components on Appendix \ref{sec:outflow}. 

\begin{figure}
	\includegraphics[width=\linewidth]{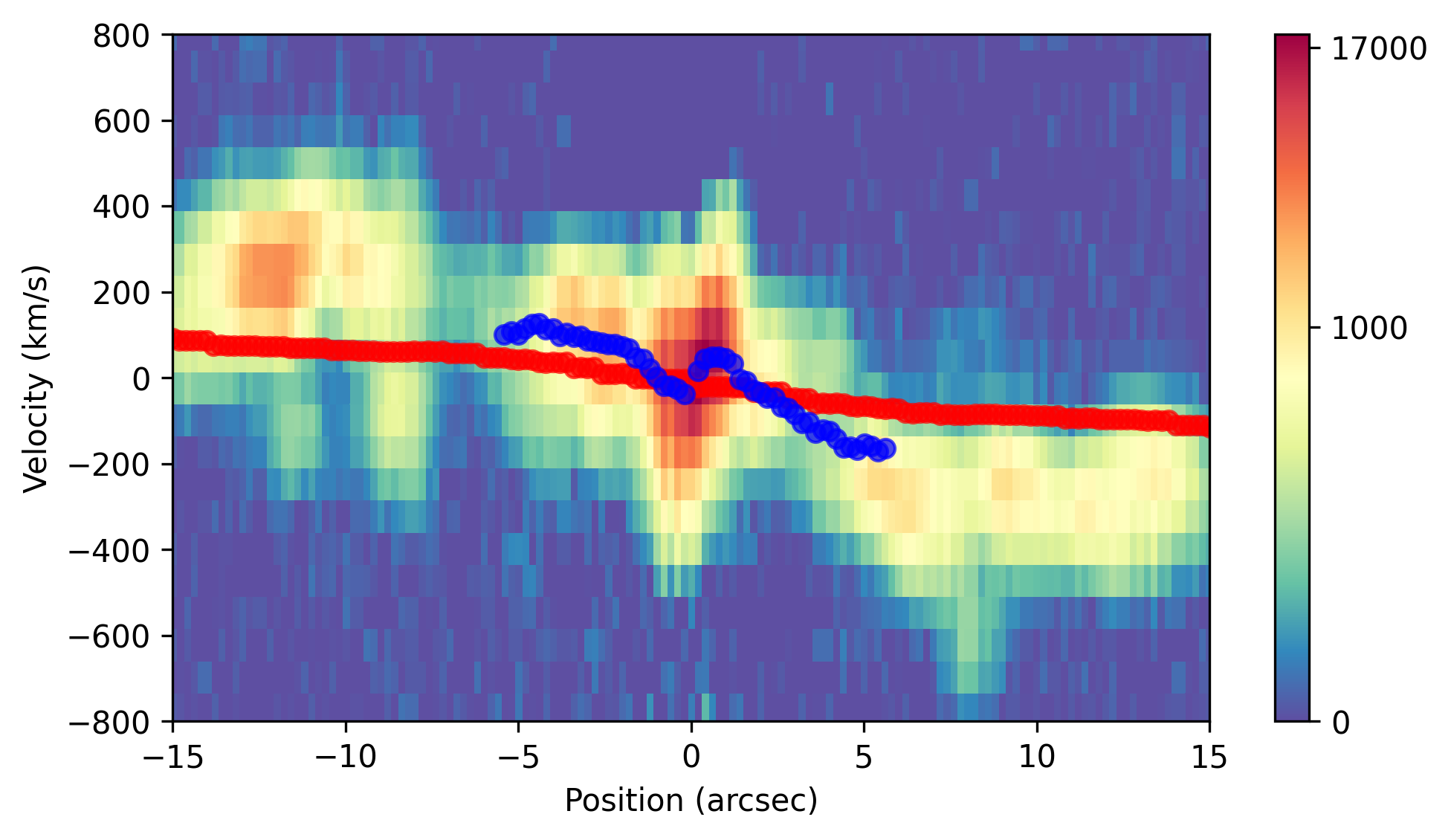}
    \caption{Position-velocity diagram extracted along PA 150\degree, this angle crosses along the NW and SE arms. The red dots mark the best fit model for the stellar component, and the blue dots the best fit model for the ionized gas component as fitted to a data cube centered in the [OIII] emission line.}
    \label{fig:pvd_150}
\end{figure}

A final feature is observed along the NW and SE arms, which show excess redshift in the approaching side of the galaxy and blueshift in the receeding side, with velocities reaching $\sim 300$ km/s (Fig. \ref{fig:pvd_150}). If we consider that the near side corresponds to the W side of the disk, this feature would correspond to an inflow in the plane of the galaxy disk. However, given the high velocities reached for an inflow, it is possible that, given the uncertainty about the three dimensional distribution of the gas, that it corresponds to a secondary component in the line-of-sight, caused by extraplanar gas related to tidal debris.\\

\begin{figure}
    \includegraphics[width=\linewidth]{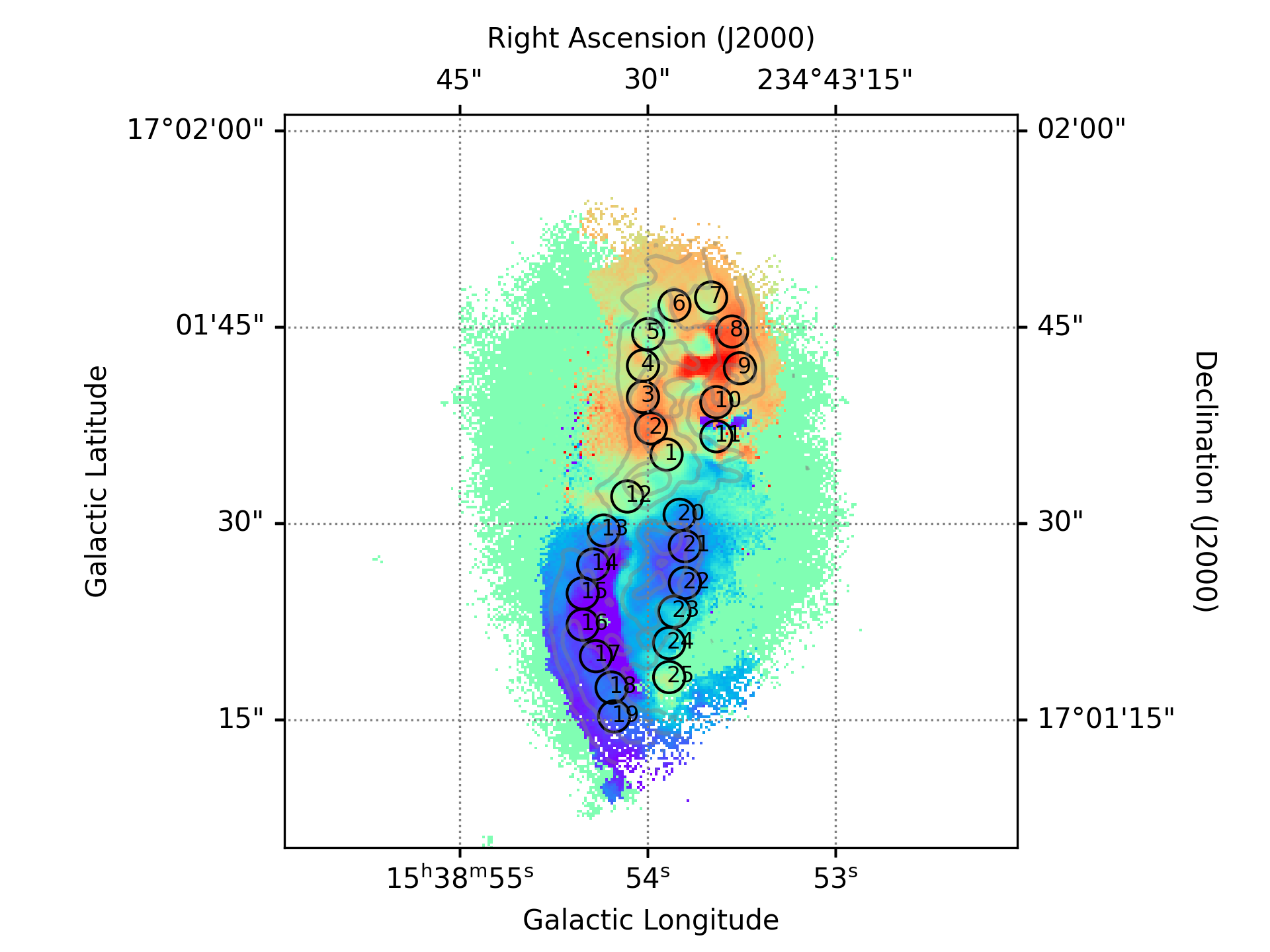}
	\caption{ Moment 1 map for the [OIII] emission line, overlayed in black are the apertures used to extract the spectra. Gray contours show the moment 0 distribution for the [OIII] emission line.}
    \label{fig:apertures_1}
\end{figure}

\begin{figure*}
    \includegraphics[width=\textwidth]{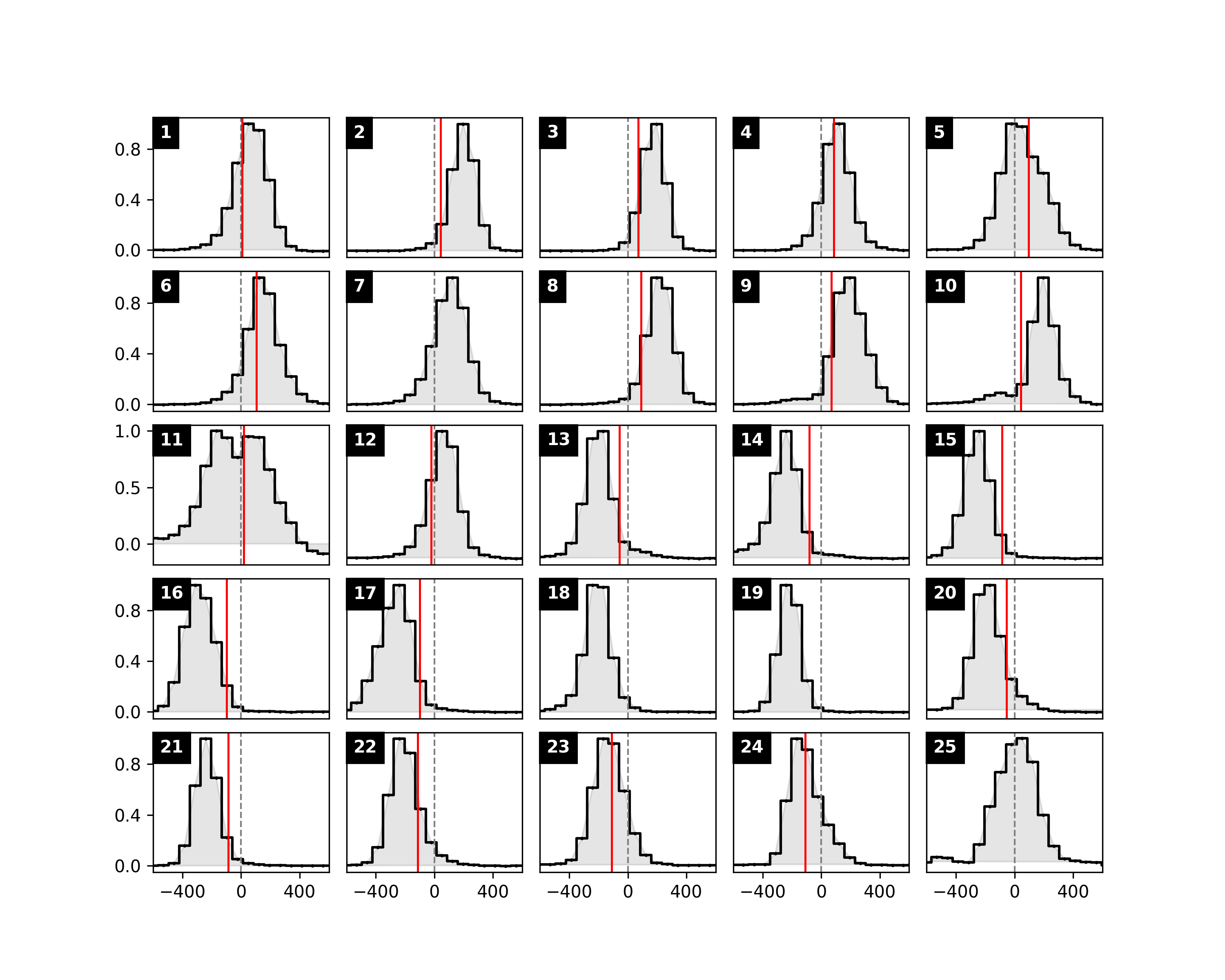}
	\caption{Extracted spectra from the apertures shown in top panel, spectral window centered on [OIII] emission line. X-axis correspond to velocity in km/s and the y-axis to flux in $10^{-20}$erg s$^{-1}$cm$^{-2}$\AA. The red vertical line corresponds to the stellar best-fitted Barolo model, and the dashed vertical line corresponds to the receeding velocity.}
    \label{fig:apertures_2}
\end{figure*}

To further understand the kinematics of the gas in NGC 5972, we extract the spectra from 6\arcsec apertures in different regions of the disk, along the arms (Figures \ref{fig:apertures_1} and \ref{fig:apertures_2}). The spectral windows, centered on the [OIII] emission line show that the ionized gas along the arms is more redshifted (blueshifted) in the receeding (approaching) side than the stellar rotation, while in apertures outside the arms, the gas seems to coincide with the stellar rotation (apertures 23-24 and 4-6). This can also be observed by comparing with the position-velocity diagrams (PVDs) extracted from slits along the major and minor axis (Fig. \ref{fig:3dfit_pvds}), and along PA $\sim 150$\degree, which crosses the SE and NW arms (Fig. \ref{fig:pvd_150}). The arms seem to be dominated by a component that has velocities deviating from the model by about 150 km/s in both the redshifted and blueshifted sides. The spectra of these NW and SE arms (apertures 8-10 and 14-17) show that a secondary component, appearing as an asymmetric wing (or tail) in the line profiles, follows the rotation, indicating that the emission in these apertures is dominated by a secondary kinematic component, which may be emission from extraplanar gas in the line-of-sight.

\subsection{Extended emission line region}
\label{sec:lum_hist}

\begin{figure}
	\includegraphics[width=\linewidth]{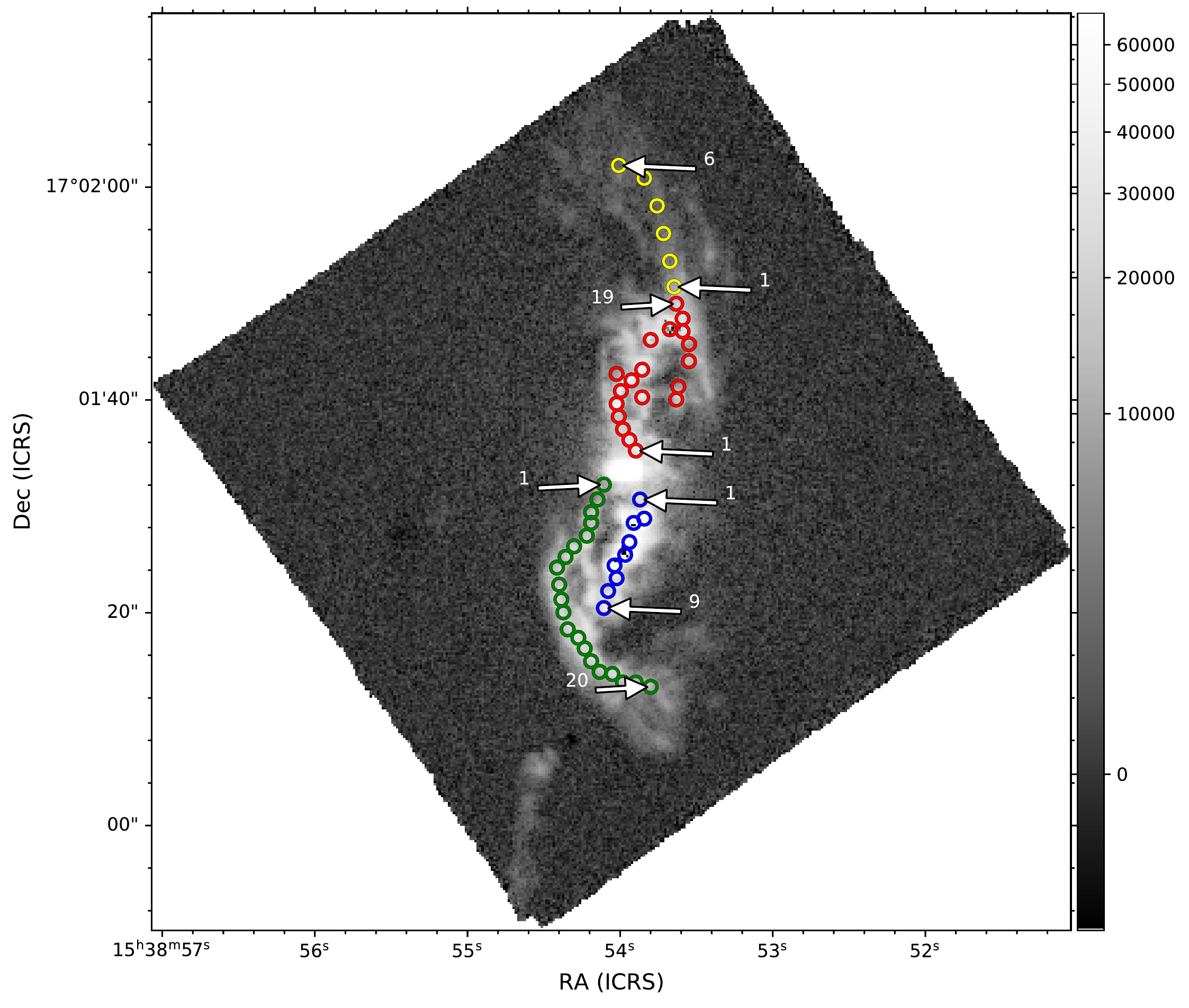}
    \caption{Moment 0 map for the [SII] emission line, overplotted are the apertures on each arm, in color red for arm ' A', yellow for 'AT', green for 'B', and blue for 'C'. The color bar is in $10^{-20}$ erg s$^{-1}$cm$^{-2}$\AA$^{-1}$ units, with an asinh scale to highlight the faint tidal tails.}
    \label{fig:arms_apers}
\end{figure}

Having established that the gas from the arms has been ionized primarily by the AGN (See Sect. \ref{sec:origin_ionization}) and given the extension of the ionized clouds, which is large enough to be substantially influenced by light-travel time effects, we can trace the AGN luminosity from the nucleus to $\sim 17$ kpc and assess possible luminosity changes over time. 
For this analysis we compare our observed emission line ratios to the spectra obtained from photoionization models to derive a best-fit ionization parameter (defined as the dimensionless ratio of hydrogen-ionizing photon to total-hydrogen densities), and use it to estimate the AGN luminosity required to produce the EELR ionization.
In Fig. \ref{fig:arms_apers} we present the apertures (radius 0\farcs6) where the analysis was carried out. We separated the regions in the arms 'A' (N arm), 'B' (SE arm) and 'C' (SW arm), each covering distances of 11, 12 and 8 kpc to the inner kpc respectively. Furthermore, we add apertures over the northern tidal tail of the 'A' arm, labeled as 'AT', which covers between 11 to 17 kpc. Given the low flux in this region we use apertures of 1.2\arcsec.\ The furthest distance corresponds to $5.5 \times 10^{4}$ ly, located in the A arm. For every aperture we obtain the integrated spectrum, which is fitted with Gaussian profiles to obtain the fluxes for the emission lines H$\beta \lambda 4861$ \AA, [OIII]$\lambda 5007$ \AA, HeI$\lambda 5875$\AA, [OI]$\lambda 6300$ \AA, [NII]$\lambda 6547,6583$ \AA, $H\alpha \lambda 6563$ \AA, [SII]$\lambda 6716, 6731$ \AA. 

After creating a continuous sequence of apertures from the center covering the extension of every arm, we fill the vicinity of each aperture with 30 new apertures of the same size. From these new apertures we choose the one with largest SNR, to the corresponding spectra we fitted a two-component Gaussian profile to separate the multiple kinematic components. An example of this can be observed in Fig.~\ref{fig:fit_example}. 

To model the emission line ratios observed in our data, we use the photoionization code \texttt{Cloudy} \cite[version 17.02 last described in][]{ferland+2017}, following the method outlined in \cite{treister+2018}. \texttt{Cloudy} resolves the equations of thermal and statistical equilibrium on a plane-parallel slab of gas being illuminated and ionized by a central source. For the central source we consider an spectrum consistent with the observed spectra of local AGN \citep[e.g.,][]{elvis+1994}. This spectrum is described by a broken power law ($L_{\nu} \propto \nu^{\alpha}$), where $\alpha = -0.5$ for E $<$ 13.6 eV, $\alpha = -1.5$ for 13.6 eV $<$ E $<$ 0.5 keV and $\alpha = -0.8$ for E $>$ 0.5 keV. 

The ionization parameter is defined as the dimensionless ratio of number of ionizing photons to hydrogen atoms at the face of the gas cloud:
$$U = \frac{Q(H)}{4\pi r_{0}^{2}n(H)c}$$ 
where $r_{0}$ is the distance between source and the ionized cloud, n$_{H}$ is the hydrogen number density (cm$^{-3}$), and {\it c} is the speed of light. \textit{Q(H)} is the number of ionizing photons per unit of time (s$^{-1}$), which is defined by 
$$Q(H) = \int_{\nu_{0}}^{\infty} \frac{L_{\nu}}{h\nu}d\nu$$
where $L_{\nu}$ is the AGN luminosity as a function of frequency, $h$ is the Planck constant, and $\nu_{0} = 13.6$ eV$/h$ is the frequency corresponding to the ionization potential of hydrogen \citep{osterbrock+2006}. 

The first run of simulations assumes solar metallicity with a grid covering the ionization parameter ($-3.5 <$ U $< -2.0$) and the hydrogen density (1.0 $<$ n$_{H}$ $<$ 5.0). However these simulations do not cover the entire range of measurements on a BPT-diagram. Thus, following \cite{bennert+2006}, we run a series of models changing the metallicity from 1.0 to 4.0 times the solar metallicity, and a third calculation changing only the nitrogen and sulphur metallicity from 0.1 to 4.0 $Z_{\sun}$. From these runs, we find that changing the N and S metallicity to $1.5 \times Z_{\sun}$ and $1.5 \times Z_{\sun}$, respectively,while maintaining the other elements at their solar values, best covers the parameter space of our data (see Fig. \ref{fig:cloudy_sims}) for the arms 'A', 'B', and 'C'.
This metallicity, however, does not fit well the fluxes observed for arm 'AT'. Therefore we maintain a solar metallicity. More details on the metallicity choice can be found on Appendix \ref{sec:metal}.

\begin{figure}
\gridline{\fig{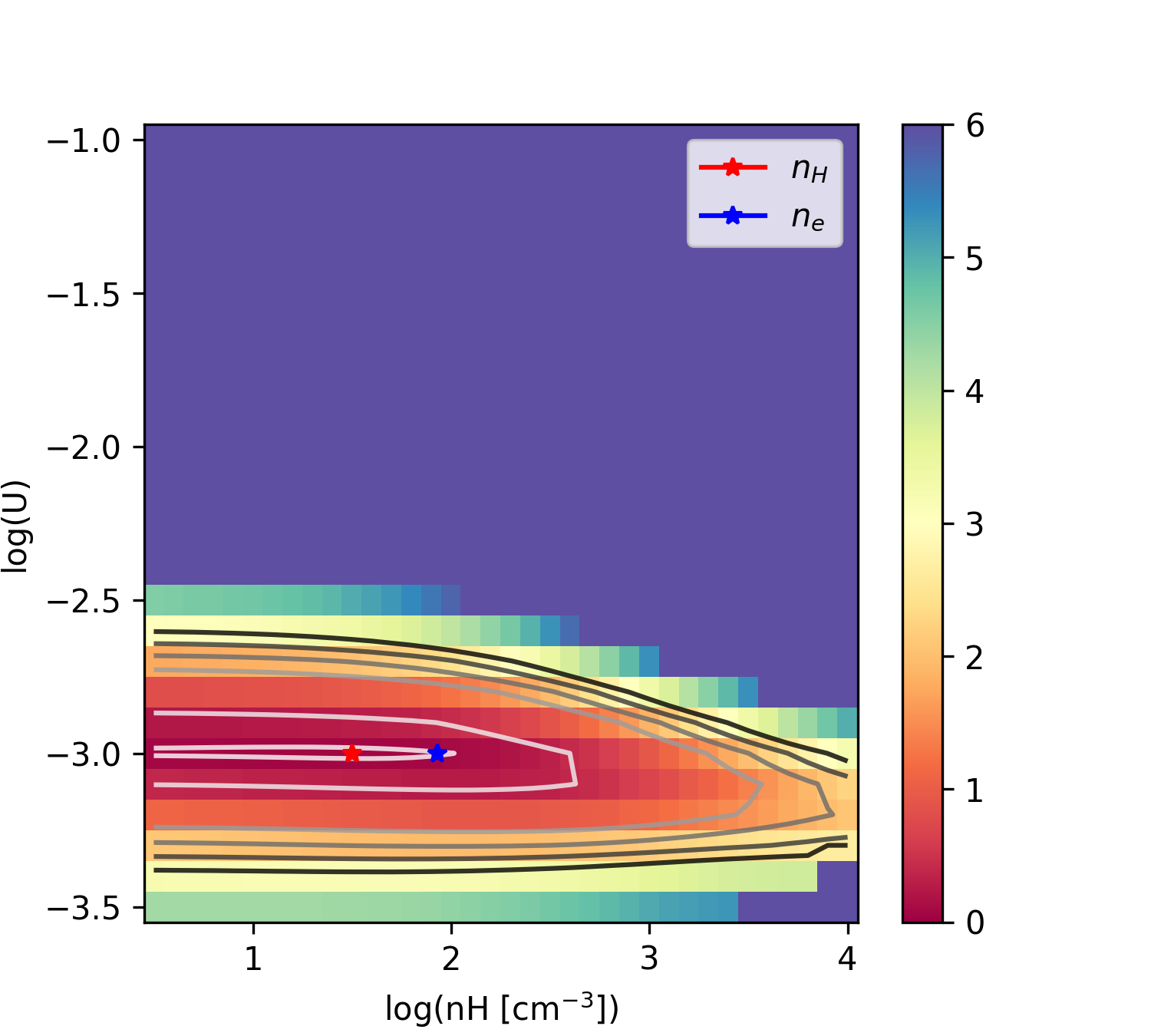}{0.5\columnwidth}{}
          \fig{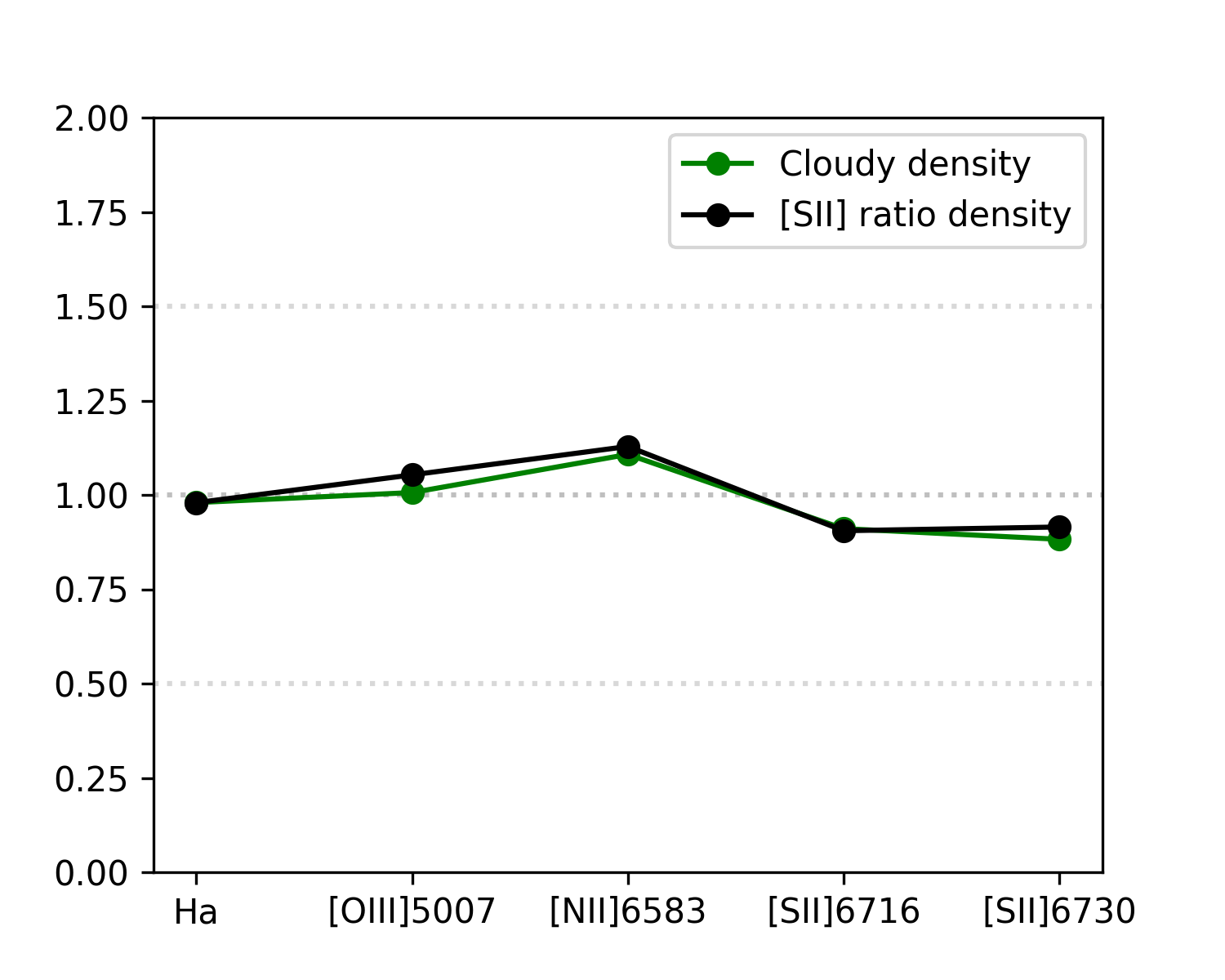}{0.5\columnwidth}{}}
\gridline{\fig{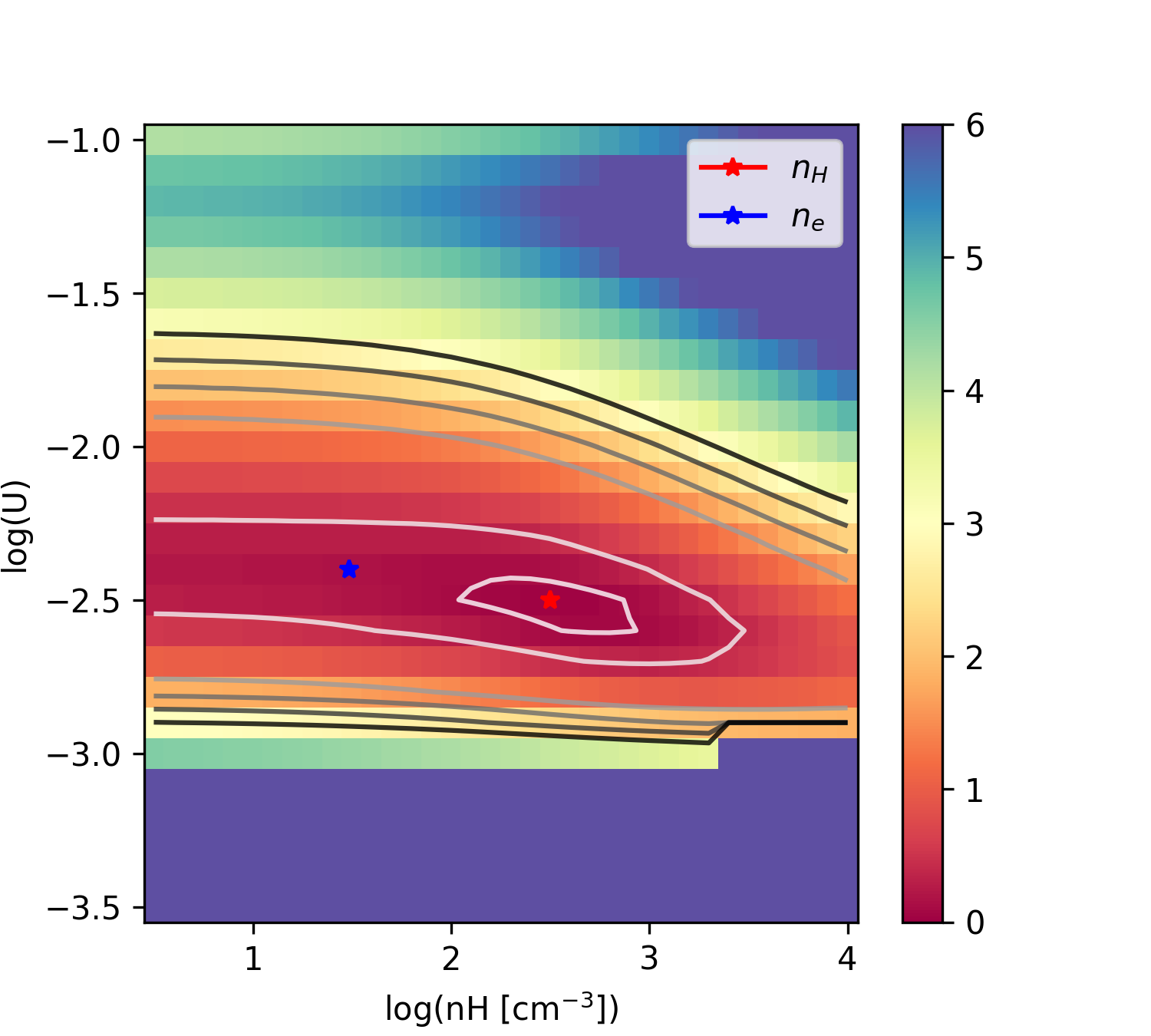}{0.5\columnwidth}{}
          \fig{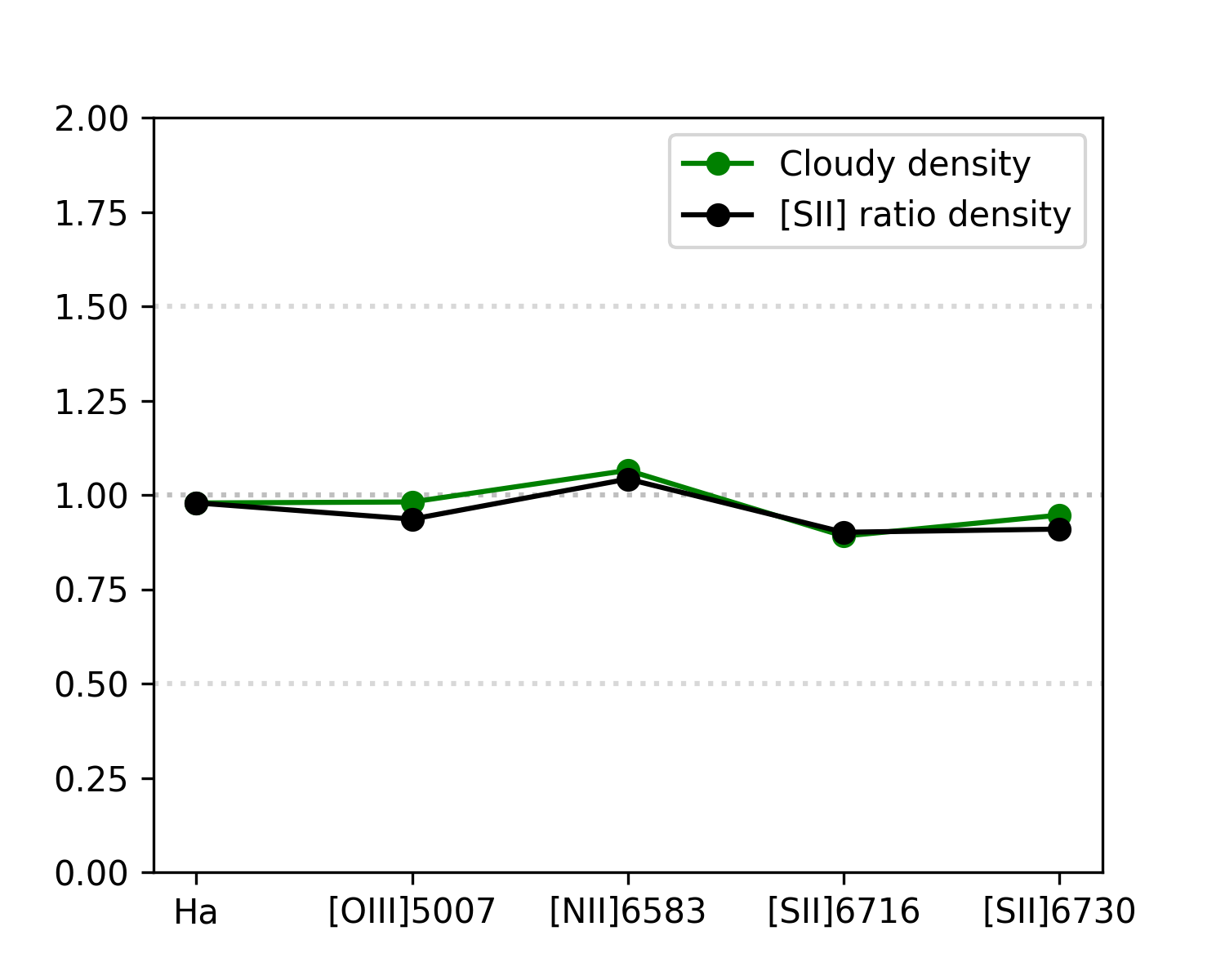}{0.5\columnwidth}{}}          
\gridline{\fig{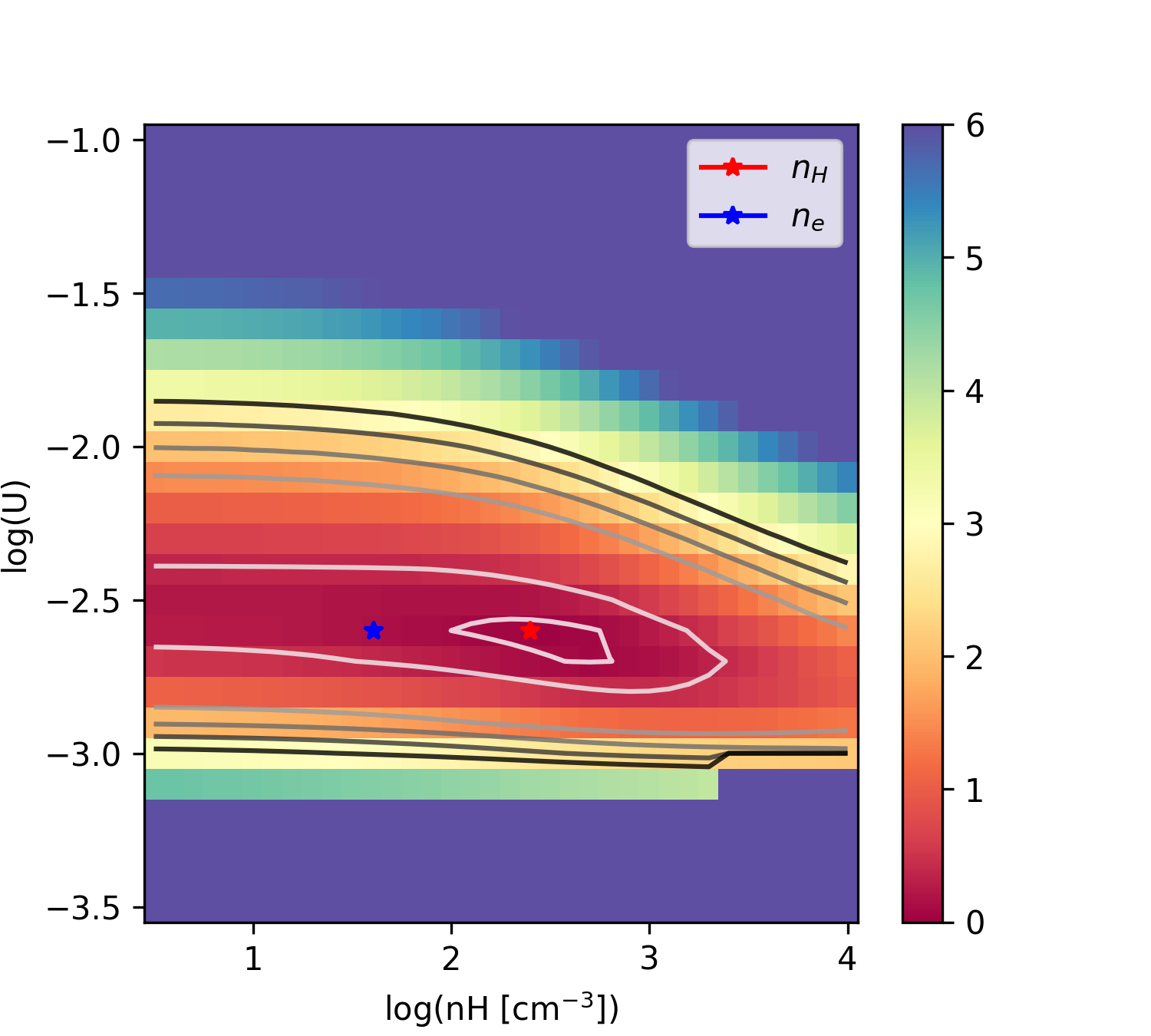}{0.5\columnwidth}{}
          \fig{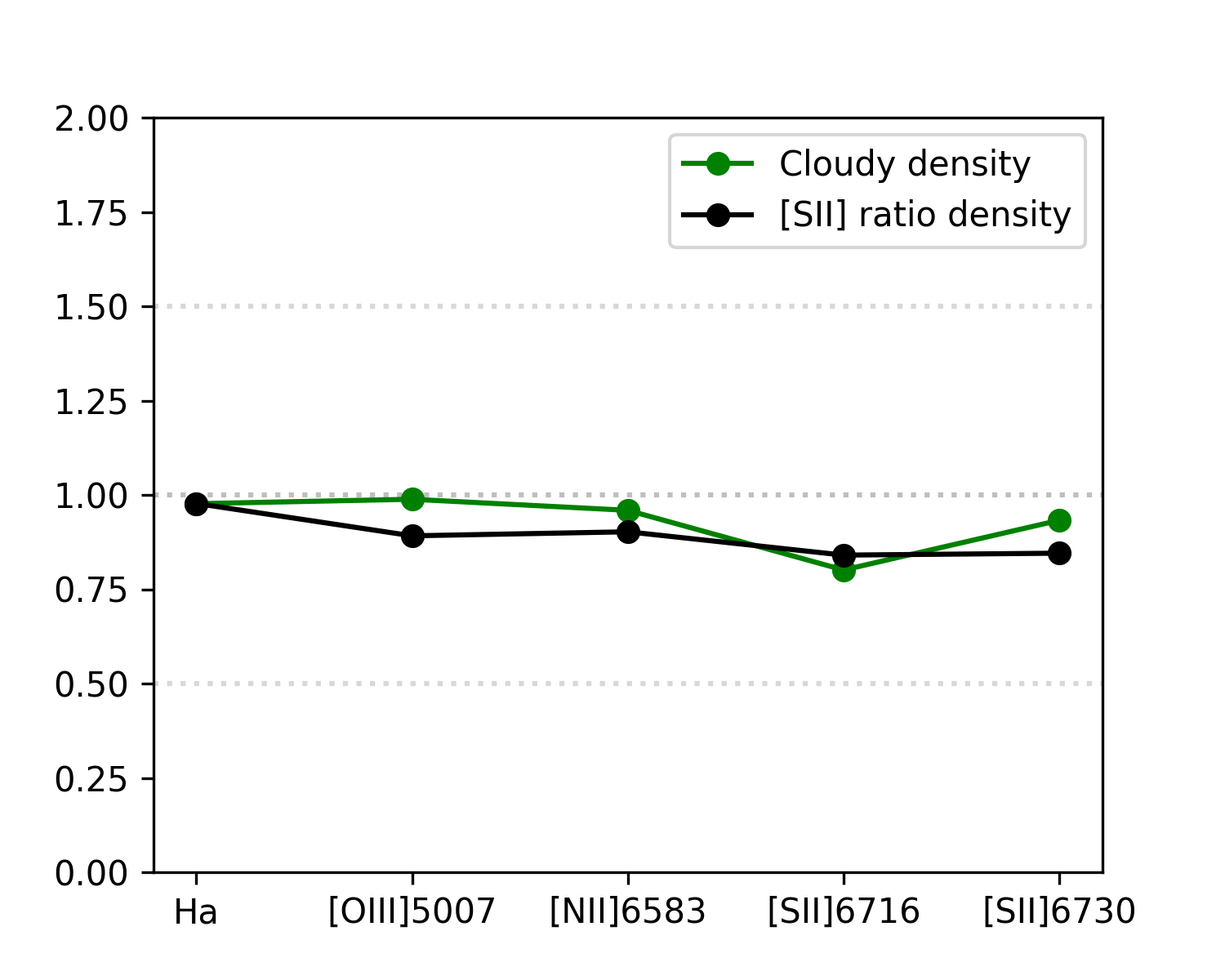}{0.5\columnwidth}{}}
\caption{ {\it Left panels:} Examples of reduced $\chi^{2}$ for arms 'A', 'B' and 'C' from top to bottom. The reduced $\chi^{2}$ distribution is shown in color, and is obtained from the fitting of the simulations grid over U and n$_{H}$. The blue stars shows the best values of (U,n$_{H}$) from model 1 and the red for model 2. {\it Right panels:} Examples of the difference between emission line ratios from the model versus the CLOUDY simulations, in the same order as the left panels. }
    \label{fig:cloudy_chisqr}
\end{figure}

\begin{figure*}
\gridline{\fig{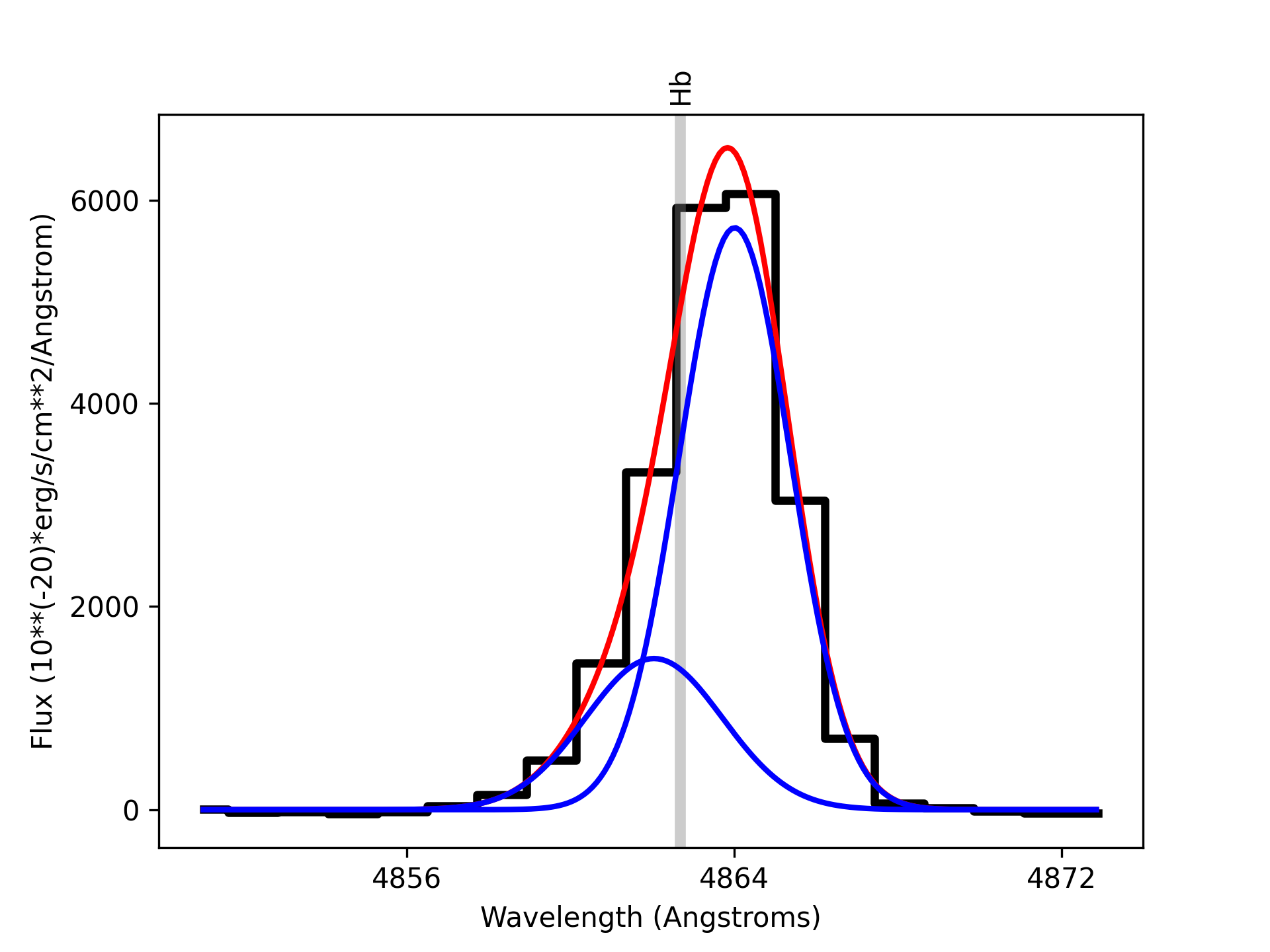}{0.25\textwidth}{}
          \fig{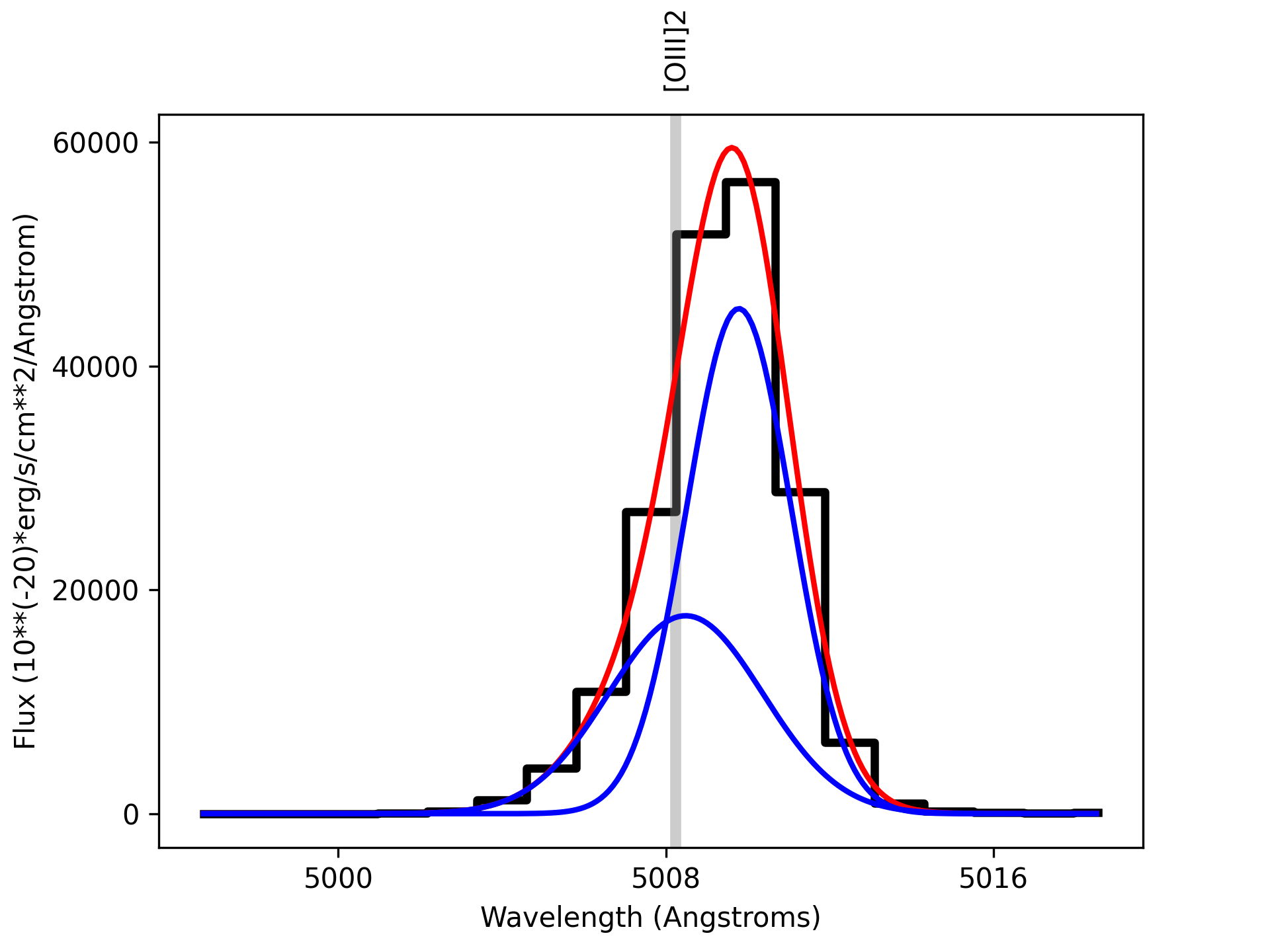}{0.25\textwidth}{}
          \fig{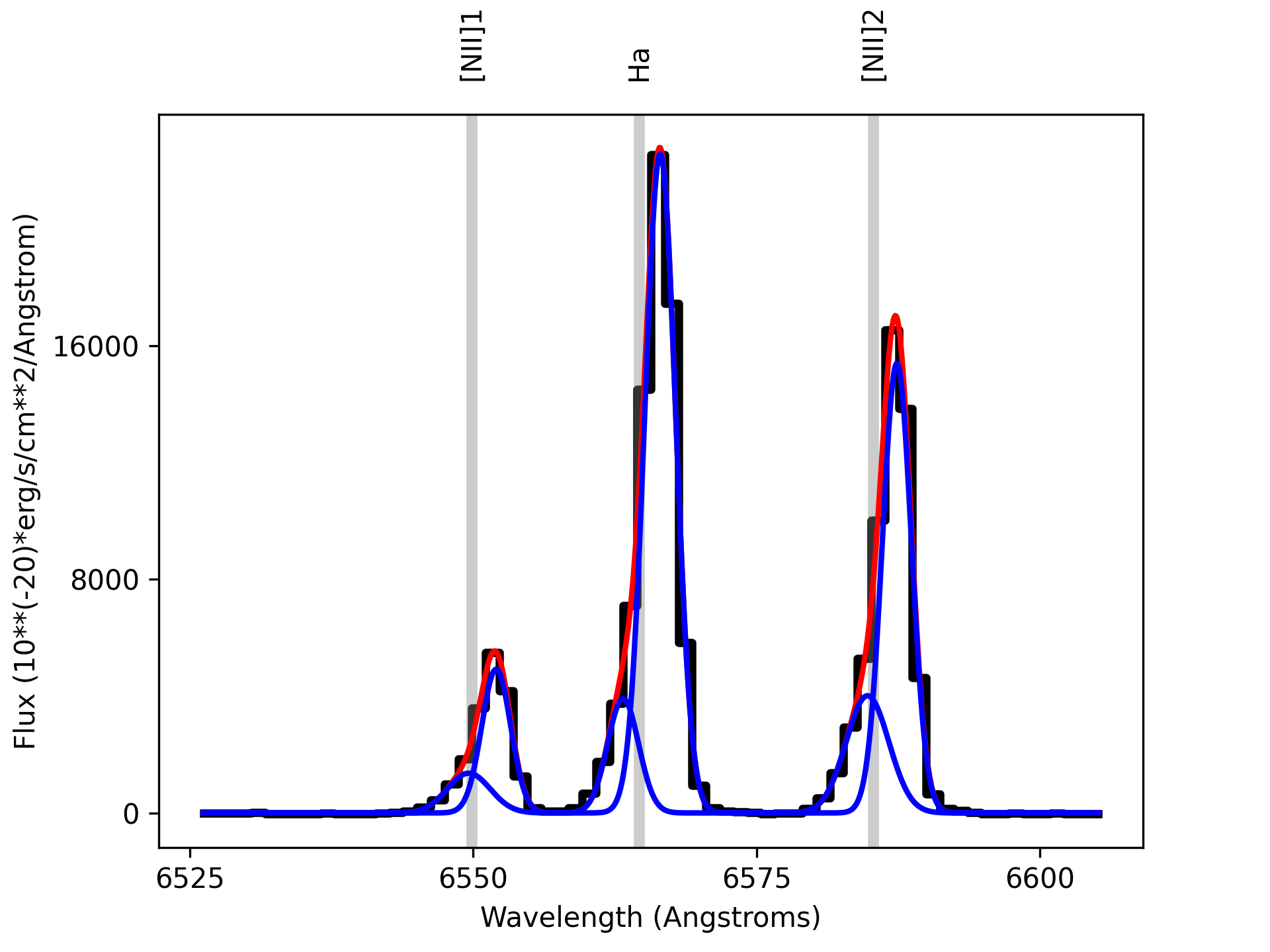}{0.25\textwidth}{}
          \fig{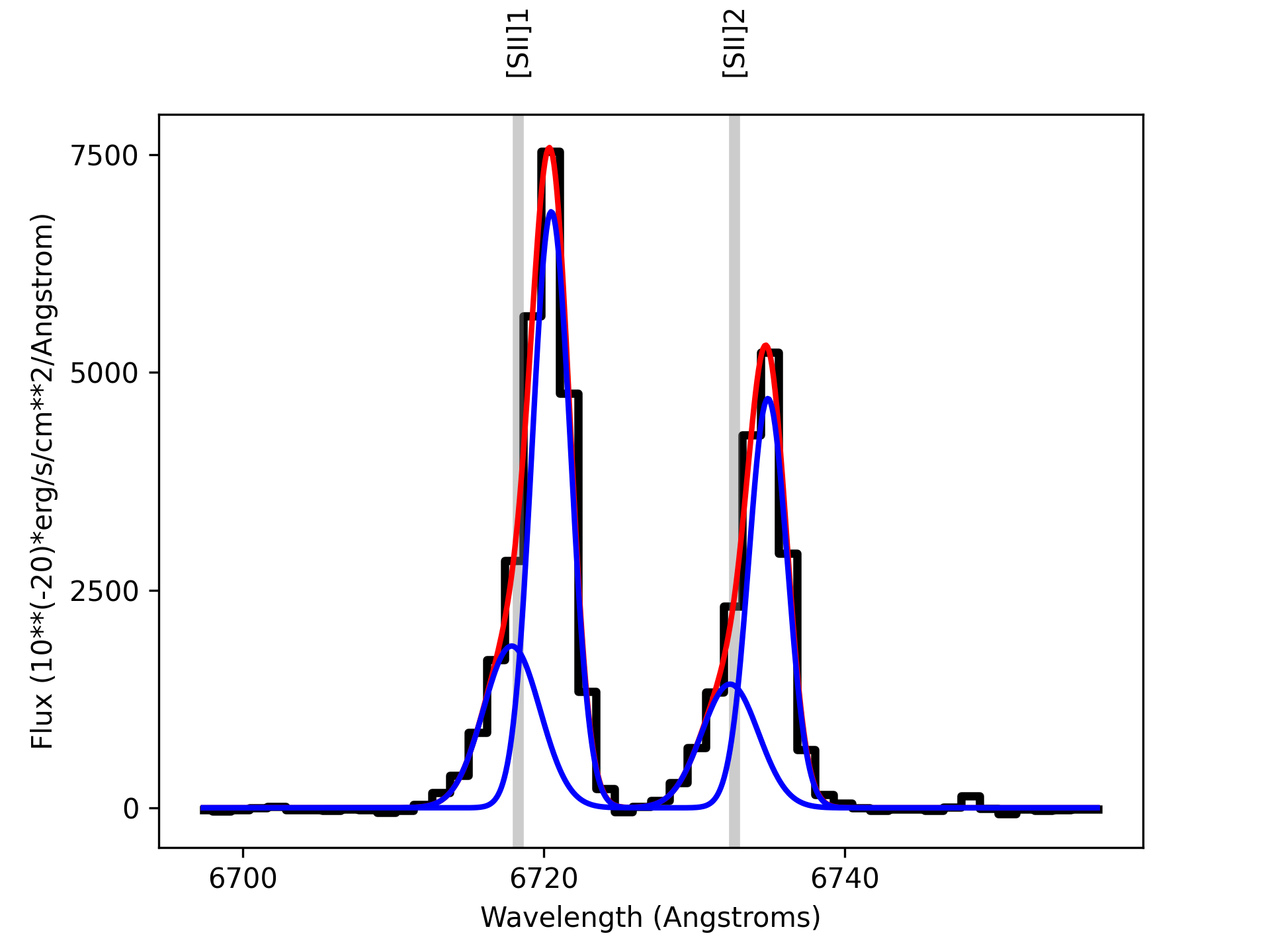}{0.25\textwidth}{}
          }
\gridline{\fig{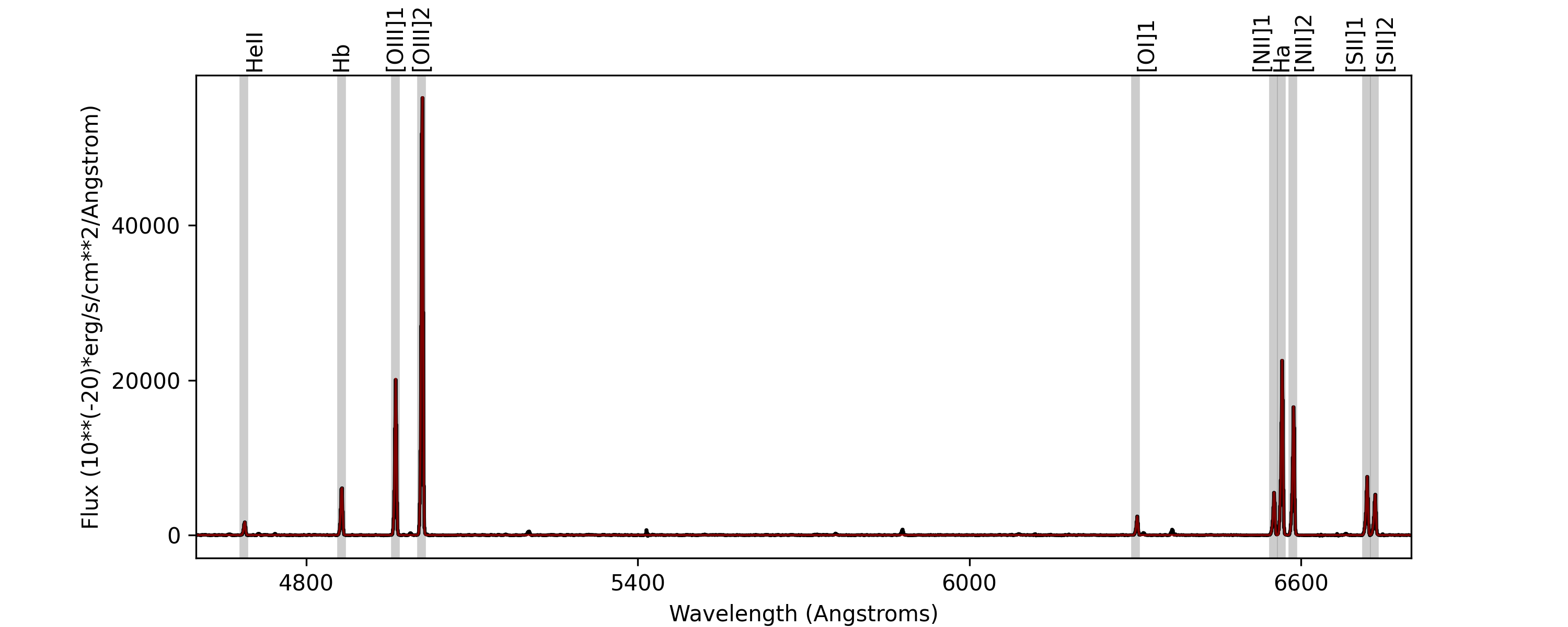}{\textwidth}{}
          }
\caption{Example of double Gaussian component fit to an aperture in the 'C' arm. {\it Upper panels:} Zoom into spectral windows, containing the emission lines, from left to right, H$\beta$, [OIII], [NII] and H$\alpha$, and [SII]. In blue are the two Gaussian components, and in red the sum of both components. {\it Bottom panel:} Fit to the complete spectra, trimmed at 7400 \AA. Gaussian components in blue and final fit of the summed components is represented by the dash red line.  
\label{fig:fit_example}}
\end{figure*}

\begin{figure}
	\includegraphics[width=1.0\columnwidth]{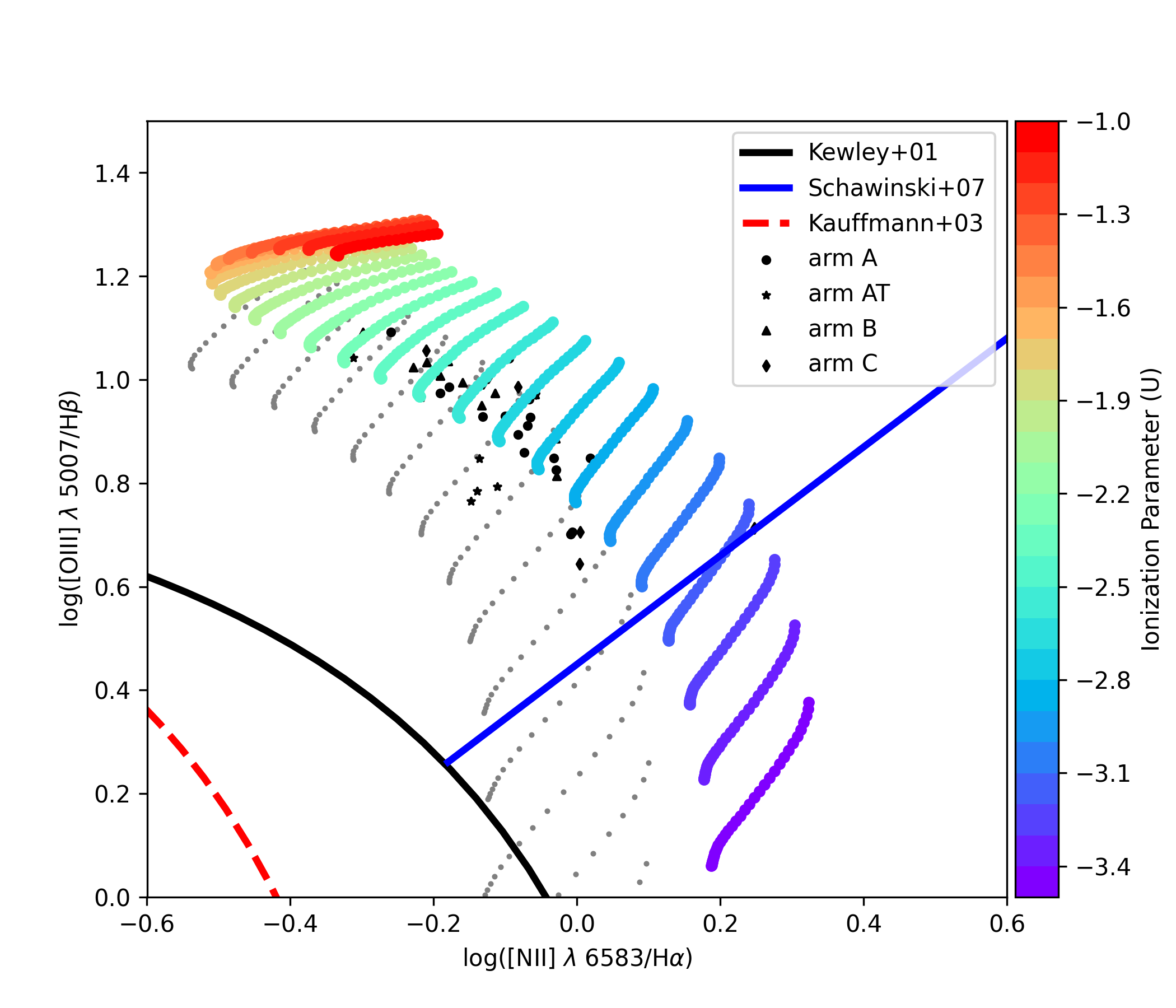}
    \caption{BPT diagram, black dots are derived from the fluxes coming from the apertures to each arm. Color dots come from the CLOUDY simulations, where the color indicates different ionization parameter values (U), and from left to right covers 1 to 4.5 dex of hydrogen density. The gray dots correspond to the CLOUDY simulations assuming solar metallicity}
    \label{fig:cloudy_sims}
\end{figure}

To obtain a model that fits the observations, it should be able to reproduce all the observed emission line ratios relative to H$\beta$. For each aperture we compare the following observed line ratios scaled to H$\beta$: [OIII], [NII], [SII], H$\alpha$, [OI], to those obtained from the simulations and choose the combination of ionization parameter and hydrogen density that delivers the best fit. We consider an acceptable fit when the difference between the emission line ratios is less than a factor of two for every line ratio. For [OIII]$\lambda 5007$ \AA\ we require the model to match the observations within 50$\%$. From the models that meet these criteria, we choose the one with the smallest reduced $\chi^2$. Examples of these results along three different apertures, one per arm, are shown in Fig. \ref{fig:cloudy_chisqr}.

Following previous EELRs analysis \citep{keel+2012a} we assume a fully ionized gas, and thus we can estimate the atomic hydrogen column density as being the same as the electron density (n$_{H}$ = n$_{e}$). The electron density can be constrained from the [SII]$\lambda6716/6730$ \AA\ emission line ratio (hereafter referred to as [SII] ratio). We use the Python package \texttt{PyNeb} which computes emission line emissivities \citep{luridiana+2012}, to obtain the n$_{e}$ from the observed [SII] ratio, assuming a $10^4$ K temperature \citep{osterbrock+2006}.
The $\chi^{2}$ maps for the U and $n_{H}$ parameters (Fig. \ref{fig:cloudy_chisqr}) show that for a given U the $n_{H}$ remains roughly constant. Considering that the required bolometric luminosity of the AGN is proportional to $U \times n_{H}$, and since the $n_{H}$ is not well constrained by the Cloudy fit, we adopt the electron density from the [SII] ratio as $n_{H}$. However, for completeness and given that not for every aperture the $n_{H}$ value for the Cloudy fit and the [SII] line ratio fall in the same U value, we calculate two different bolometric luminosities: one assuming the density from the [SII] ratio (hereafter referred to as 'model 1') and one assuming the hydrogen density from the best-fit Cloudy model (hereafter 'model 2').

Using the obtained ionization parameter and hydrogen density values, and assuming that the distance to the cloud is traced by the projected distance from the central source to the aperture, we can estimate $Q(H)$ for each aperture. Then, integrating our SED we can use the obtained $Q(H)$ to further estimate the bolometric luminosity required to ionize the gas at each point. \\

\begin{figure*}
	\includegraphics[width=1.0\linewidth]{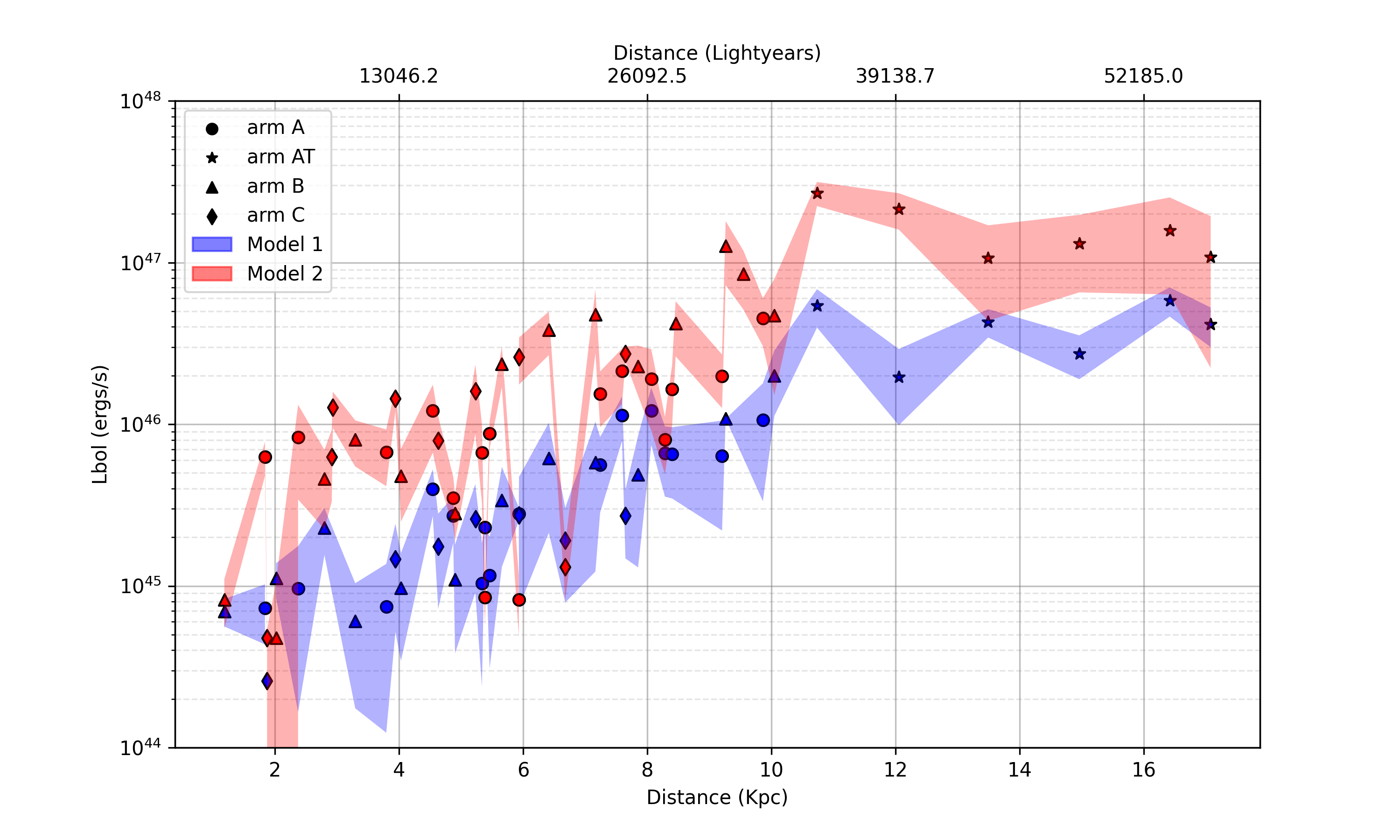}
    \caption{Bolometric luminosity vs projected distance from the center. The different symbols represent apertures over different arms. The shaded areas represent the error for each measurement from MC simulations to the data. The blue shaded area corresponds to the data obtained from model 1. Where model 1 corresponds to the bolometric luminosity obtained assuming the gas density is equivalent to the electron density derived from the [SII] line ratio, fixing this parameter in the fitting to the prhotoionization simulations, using the code Cloudy. In model 2 we leave as free parameters the ionization parameter and the gas density in the fitting to our photoionization simulations, which results in different L$_{bol}$ values for both approaches. There is a clear tendency of increasing L$_{bol}$ with projected distance from the center for both models. This implies a change of $\sim 120$ (for model 1) and $\sim 160$ (for model 2) times in luminosity over the past $5\times 10^4$ yr. }
    \label{fig:lboldist}
\end{figure*}

In figure \ref{fig:lboldist} we show the bolometric luminosity as a function of distance to the nucleus, using the electron density from model 1 (blue shaded area), and the electron density for model 2 (red shaded area).

The errors are derived from MonteCarlo (MC) simulations, where random noise is added to the observed emission line ratios. These simulations are then fitted with the described technique, and the errors are considered as the standard deviation (stdv). For model 2, we consider the stdv for the U and n$_{H}$ parameters. For model 1, the noise is applied first to the [SII] ratio, and we obtain the error as the stdv in electron density. With this value, we fix the density and fit the ionization parameter in the same manner as before. 
The results on both models show a clear trend of increasing luminosity with distance from the center. Both L$_{bol}$ derived from the different density calculations follow the same trend and overlap at some radii. For model 1 we see a change in luminosity of $\sim 124$ times between 1 kpc and 17 kpc, which considering time travel distance, corresponds to $5 \times 10^4$ years. While for model 2 the luminosity change is $\sim 160$ times.

The maximum bolometric luminosity is observed in the 'AT' arm, and reaches $4 \times 10^{46}$ erg s$^{-1}$ at $\sim 17$ kpc from the center for model 1, and $10^{47}$ erg s$^{-1}$ for model 2. 
 
The AGN in NGC 5972 has been reported to have a L(15-55 keV) = $1.0 \times 10^{43}$ erg s$^{-1}$ \citep{marchesi+2017}, a L(FIR)  $< 5.5 \times 10^{43}$ \citep{keel+2012a}, corresponding to a ionizing luminosity of L$_{ion} > 7.8\times10^{43}$ erg s$^{-1}$.
We estimate the present-day bolometric luminosity from the [OIII] emission line in the apertures closest to the center to be $\sim 2 \times 10^{44}$ erg s$^{-1}$, applying the correction factor of \citet[][]{heckman+2004}. Our results are in concordance with the values reported in \cite{keel+2012a} for this object, who estimate a L$_{ion} < 9.8 \times 10^{46}$ erg s$^{-1}$ at 15-18 kpc, in contrast with present-day L$_{ion} > 7.8 \times 10^{43}$ erg s$^{-1}$ for the AGN in its current accretion state.

\section{Discussion}
\label{sec:discussion}

\subsection{Morphology and kinematics}
The spatial resolution provided by the MUSE observations allows us to spatially resolve and characterize the EELR.
The most striking features of this EELR are the arms seen to the North and South of the center. Additionally, faint streams of gas reminiscent of tidal tails are observed towards the NE and SE. The spectra of these arms show that they are dominated by AGN-photoionized gas. The kinematic analysis shows complex structure with a component that follows a usual galaxy disk rotation reaching 120 km/s. Additionally, the NW and SE arms are a clearly distinct independent component. While the kinematics of the arms follow the sense of rotation in the disk, with blueshifted (redshifted) emission on the SE (NW), they show larger offsets from a simple rotation model. Given the inclination of the disk, this feature could represent an outflow if the blueshifted emission is on the near-side of the galaxy. The velocities reached by this feature ($\sim 300$ km/s) are consistent with an outflow in a low or moderate power source. However, another possibility is that this feature is another component in the line-of-sight, due to extraplanar gas. 
Finally, we identify a kinematically decoupled inner disk (radius 6\arcsec, peak velocity $\sim 200$ km/s). Inside the disk, in the inner 1\arcsec~ we observe a 50 km/s feature along PA 100\degree\ that could be an inflow if the NE side of the galaxy is the near side.  
Given the presence of tidal features, and the inner disk which seems kinematically decoupled from the large-scale disk, it is possible that this galaxy has experienced a merger or close-encounter event in the past. This event could have tidally disrupted the arms into the currently observed shape. However the presence of a undisturbed distribution of stellar populations strongly argues against this event being a major merger.

\subsection{Luminosity history and variability}

Previous luminosity history analyses of other sources have typically compared the luminosity of the current AGN to distant gas clouds $\sim 10^{4-5}$ light years from the center. In the case of \citet{gagne+2014} the analysis was made using a long-slit for the "teacup" galaxy, covering the change from center to the EELR along the slit.
In the case of NGC 5972 the EELR extends from the center to  $\sim 10^{4.7}$ ly. This allows us to carry out a tomographic analysis and thus trace the continuous change in luminosity over the past $\sim 10^{4}$ yr.
Using the photoionization code \texttt{CLOUDY}, we have created models that cover a range of ionization parameter, metallicity and hydrogen density, which we matched to the observed spectra from different apertures along the arms of ionized gas. To obtain a bolometric luminosity based on these models, some caveats must be taken into consideration:

a) \textit{Geometry:} we have assumed that the distance from the center to each ionized gas cloud corresponds to the projected distance in the plane of the sky between the two points (r$_{proj}$). However, it is possible that the ionized gas is not in the plane of the galaxy disk. In this case the true distance will correspond to r$_{proj}/sin(i)$, where $i$ is the inclination of the cloud with respect to the galaxy plane. Considering this inclination, the difference in epochs between the currently observed L$_{bol}$ and that inferred from the cloud emission, corresponds to:

$$ \Delta t = \frac{r_{proj}}{c sin(i)} (1-cos(i))$$

b) \textit{Density:} In the case of model 1, where the gas density was derived from the [SII] emission line ratio, it is important to consider that for electron density values between $\sim 100-1000$ cm$^{-3}$ the slope in the conversion curve is steep \citep{osterbrock+2006} and thus small changes in the [SII] ratios can translate into largely different electron density values.

c) \textit{Metallicity:}  In this analysis we assume a solar abundances for arms 'A', 'B' and 'C', which are sufficient to model the physical conditions of NLRs \citep[as suggested by][]{kraemer+2000}. However, the line ratios observed on the 'AT' arm do not fit within our grid of simulations while assuming solar metallicity. Thus, we vary the metallicity until it matches the observed values. 

\begin{figure}
     \centering
     \includegraphics[width=\columnwidth]{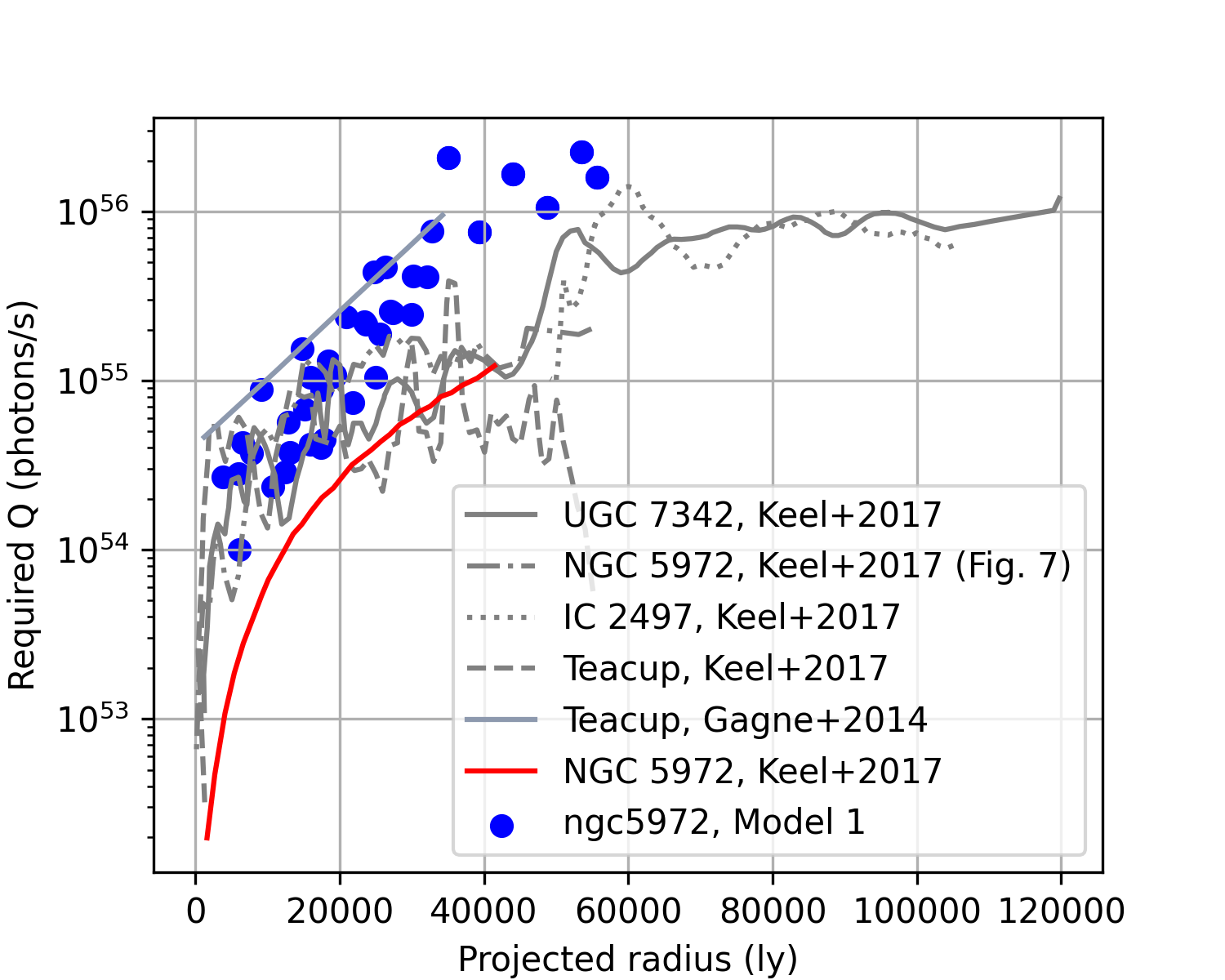}
	\caption{ Required Q in photons/s vs light time travel. This represents the luminosity required to ionize the gas at each radii, while the projected distance from the center is converted into time considering the light time travel. In black circles we show the values derived from model 1 for every arm. In red we show the lower limit of Q as shown in Fig. 5 of \cite{keel+2017}, and the solid light-blue line represents the data obtained by \citep[][]{gagne+2014}. The gray lines some from Fig. 7 in \cite{keel+2017}, and they represent the peak Q values in every 1000 ly bins for different objects. The Q values from \cite{keel+2017} are derived from a recombination balance method which provides a lower limit to the ionizing flux at different radii of the EELR. They use the highest surface brightness observed in H$\alpha$ as a function of projected distance to estimate the required (isotropic) emission rate of ionizing photons (Q$_{ion}$). Our results are consistent with the results from \cite{keel+2017}, as they show a similar tendency of increasing Q$_{ion}$ but with higher values which is expected as \cite{keel+2017} values represent a lower limit. }
	\label{fig:Q_plot}
\end{figure}

Taking this into consideration, we obtain a bolometric luminosity change vs light travel time along the cloud, based on the best-fitted CLOUDY model. To test the consistency with previous analysis in Figure \ref{fig:Q_plot}, we show a comparison of the required rate of ionizing photons (Q) from our CLOUDY models versus the values obtained by \cite{keel+2017} from recombination balance, which represents a lower limit. There is good agreement between both methods. 

For model 1, the apertures located in the bright arms ('A'-'C') show a decrease of $\sim 40$ times in bolometric luminosity between 10 to 1.2 kpc. However, the fainter tidal tail that covers between 11 and 17 kpc shows that this increase can reach a 125 times difference, reaching bolometric luminosities of $\sim 5\times 10^{46}$ erg/s at $\sim 17$ kpc. This external region was not covered previously in \cite{keel+2017}.
We compare the largest bolometric luminosities obtained from our analysis with the AGN luminosity as derived from the WISE MIR and FIR luminosity. This corresponds to a current AGN bolometric luminosity of $2\times 10^{44}$ erg/s \citep{keel+2017}. This implies a decrease of 85 times when considering only the 'A' - 'C' arms, over the past $\sim 3\times 10^{4}$ years, or $\sim 250$ times when considering the fainter 'AT' arm, over the past $5\times 10^{4}$ years. For model 2, we see a difference of 80 between 1.2 and 10 kpc and 160 between 1.2 and 17 kpc. While the difference between the largest bolometric luminosity with the current bolometric luminosity for this model is 790.

Support for order-of-magnitude variations in the AGN accretion state on $10^{5-6}$ yr timescales include simulations \citep[][]{novak+2011,gabor+2013,yuan+2018}, observational arguments \citep[e.g.,][]{schawinski+2015,sartori+2018}, and theoretical models \citep[][]{martini+2003,king+2015}. It is possible that the observed short timescale variability is caused by rapid AGN duty cycles \citep[{\it "flickering"};][]{schawinski+2015}. Possible scenarios that could explain the short timescales for these cycles include:

a) Simulations of {\it feedback-regulated BH accretion} which suggest changes in the character of the accretion over time, from well-separated sharp bursts, to chaotic, stochastic accretion. These bursts are needed to prevent gas pile-up, and can be caused by interactions between radiation pressure and winds with the galactic gas. The bursts of activity are followed by a rapid shutdown on timescales of $\sim 10^{5}$ \citep[][]{novak+2011,ciotti+2010}. 

b) {\it Chaotic cold accretion:}, whereby the AGN flickering can be caused by cold infalling gas that tends to fragment and fall as large discrete clumps. These clouds can condense from a hot halo due to thermal instabilities, losing angular momentum and falling into the SMBH on timescales of $\sim 10^{5}$ yr \citep[][]{gaspari+2013,king+2015}

c) {\it Sharply truncated accretion disk:} \cite{inayoshi+2016} suggests, based on nuclear starburst (SB) models \citep{thompson+2005}, that the growth of a SMBH over a few $10^{10}$ M$_{\odot}$ is stunted by small-scale physical processes. If a high accretion rate is achieved, vigorous star formation in the inner $\sim 10-100$ pc would be able to deplete most of the gas, causing the accretion rate to decrease rapidly (factor of $100-1000$).

In general, it remains unclear whether the main driver of the variability is the fueling mechanism at large scales or instabilities in the accretion disk.\\

We further compare our results to the framework for AGN variability presented by \cite{sartori+2018}, which links the variability over a wide range of timescales. We calculate the total variability for the AGN in NGC 5972 as the difference between the apertures closest and farther from the center, which translates into a magnitude difference ($\Delta$m $= 9.8$) at a given time lag \citep[$\tau = 5\times 10^{4}$ yr; eq. 1 in ][]{sartori+2018}. We compare this value with the structure function \citep[SF; Fig. 2 in][]{sartori+2018} and find that our results fall within the Voorwerpjes region in the SF. The Voorwerpjes cover a range of $\Delta$m from $\sim 6\times 10^{-1} - 6\times 10^{1}$ and of time lags of $\sim 10^{4} - 10^{5}$ yr. The total variability of NGC 5972 falls well into this region with $\Delta$m$ = 9.8$ and time lag$ = 5\times 10^{4}$ yr.

\begin{figure}
\gridline{\fig{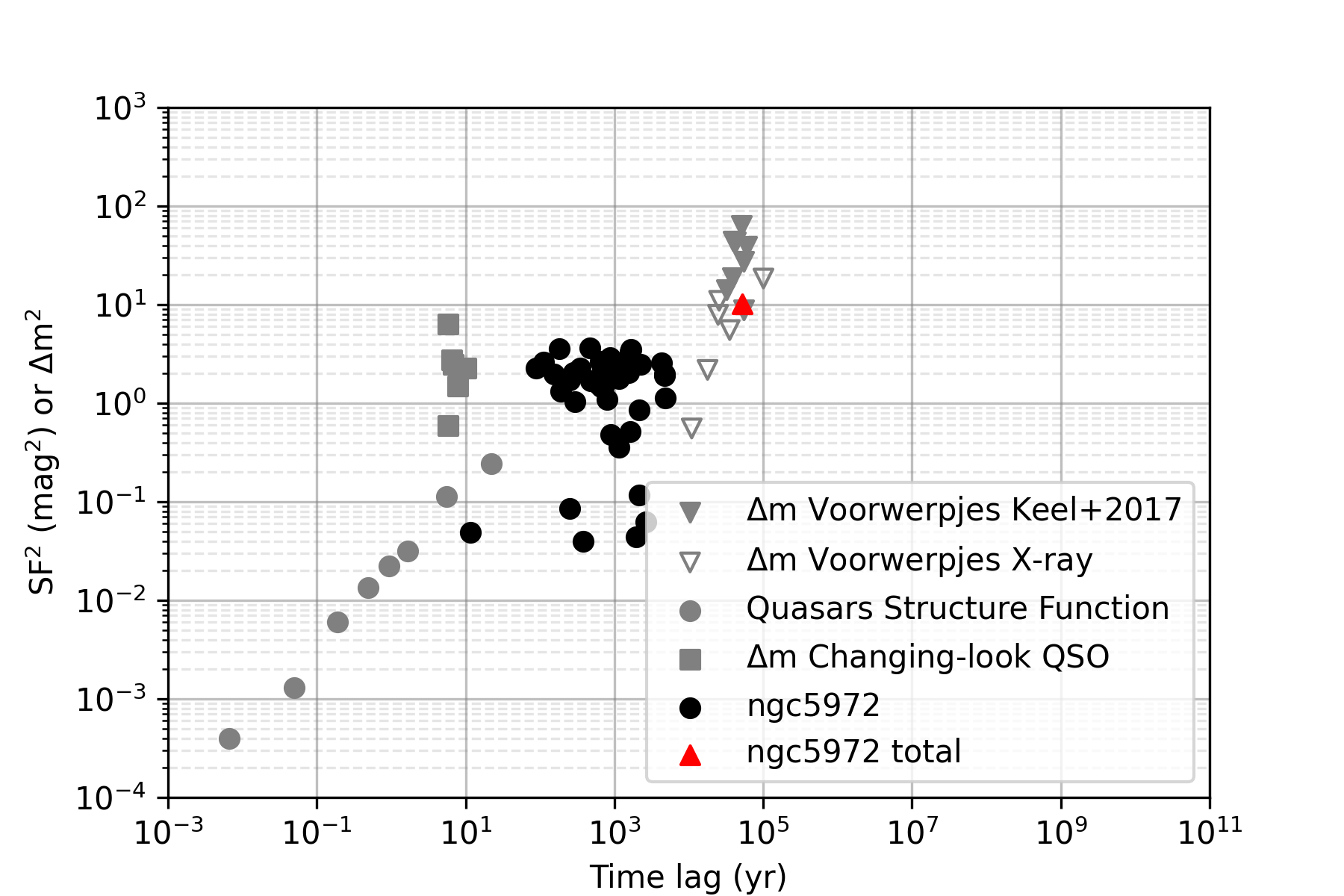}{\columnwidth}{}
          }
\gridline{\fig{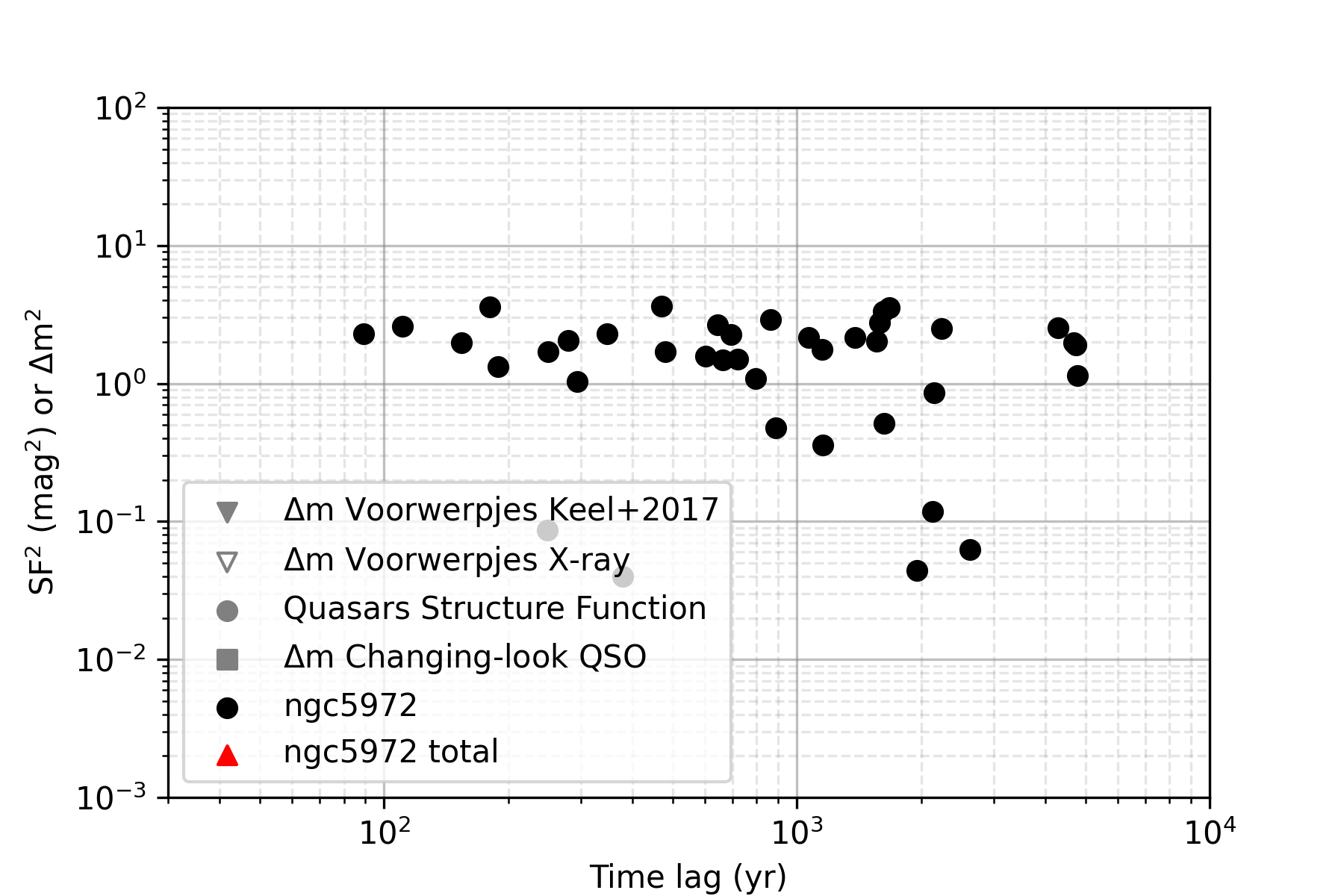}{\columnwidth}{}
          }
\caption{{\it Top panel:} SF for the magnitude difference of every radius (black circles) and the total magnitude difference (red triangle). We add in gray the variability data summarized in Fig. 2 of \cite{sartori+2018} for comparison. Gray circles represent the quasars structure function. Squares represent the $\Delta$m for changing-look quasars. Downward triangles show the data for Voorwerpjes from \cite{keel+2017} (filled triangles) and from X-ray data (empty triangles). For more details see \cite{sartori+2018}.	
	 {\it Bottom panel:} zoomed-in version of top panel. }
    \label{fig:SF}
\end{figure}

Furthermore, we calculate the $\Delta$m for the entire light travel time along the cloud, calculating the magnitude difference between each radius (r$_{i}$) with the following (r$_{i+1}$), and obtain a distribution that mostly follows the SF from light curve simulations in \cite{sartori+2018}, as can be observed in Fig. \ref{fig:SF}. Noticeably, this distribution fills a gap between data points from optical changing-look quasars, quasar structure function and Voorwerpjes derived from different methods.  

\section{Summary and conclusions}
\label{sec:summary}

In this work we present integral field spectroscopic VLT/MUSE observations for the nearby active galaxy NGC 5972 which shows a prominent EELR. The combination of the uninterrupted presence of ionized gas from the nucleus out to $\sim 17$ kpc and the spatial resolution and coverage of the MUSE observations has allowed us to study in detail the characteristics of this object. Our findings can be summarized as follows:

\begin{enumerate}
	
	\item We detect an EELR that extends over $\sim 11$ kpc with a fainter tidal tail that extends between $11-17$ kpc. The morphology of this region resembles a double-helix shape with highly filamentary structure. The analysis of the gas excitation though BPT diagnostic diagrams of this emission line region shows that it consistent with AGN photoionization. 
	
	\item The kinematics as disclosed by the emission lines shows a complex scenario, with multiple components as evidenced by the PV-diagrams and the broad, often double-peaked spectral profiles. We find evidence for a component that indicates disk rotation. The offsets from the systemic velocity are higher in the arms, reaching 300 km/s on the SE and NW regions, as compared to the disk rotation that reaches velocities of only $\sim$120 km/s. These features can be interpreted as extraplanar gas connected to the tidal debris. A final component is observed as a kinematically decoupled inner disk, in the inner 6\arcsec, which contains an outflow along the minor axis reaching $\sim 180$ km/s and appears to be dragging gas rotating in the large-scale disk. 
	
	\item  The faint tidal tails of ionized gas observed on the NE and SE can also be a hint of a past merger event. EELRs are often associated to events of this type, as a merger can create tidal tails that are then illuminated and ionized by the AGN. 
	
	\item We use the photoionization code \texttt{CLOUDY} to generate a grid of models covering a range of ionization parameter and hydrogen density values, which we then fit to each aperture of an array that covers between 1-17 kpc of the extended ionized gas. The bolometric luminosities derived from this analysis for each radii along the arms show a systematic decrease with radius from the center.
 This suggests a decrease in AGN luminosity for model 1 (gas density derived from [SII] line ratio) between 40 to 120. For model 2 (gas density fitted with Cloudy) the difference is between 80-160 times over the past $3-5\times 10^{4}$ years. These variability amplitudes and timescales are in good agreement with previous luminosity history analyses \citep[e.g.,][]{lintott+2009,keel+2012a,gagne+2014} and AGN variability models \cite[e.g.,][]{sartori+2018}.

\end{enumerate}

The extension of the ionized cloud in NGC 5972 makes it an ideal laboratory to carry out a comprehensive tomographic analysis of its EELR. It allow us to probe the AGN variability over a  continuous $\sim 10^{4}$ yr timescale, due to the light travel time. This timescale is significantly larger than human timescales, and therefore it provides a unique opportunity to fill in the timescale gap for studies of AGN variability at different timescales \citep[e.g.,][]{sartori+2018}.
The dramatic change in luminosity observed in the AGN of NGC 5972, as well as in other similar objects \citep[e.g.,][]{lintott+2009,gagne+2014,keel+2017},
suggests a connection to similarly dramatic changes in the AGN accretion state. 
The results presented are consistent with the scenario described in \cite{schawinski+2015} where AGN duty cycles ($10^{7-9}$ yr) can be broken down in shorter ($10^{4-5}$ yr) phases. To probe for longer timescales larger extensions of ionized gas, and larger FOV coverage with MUSE, would be required.

\section*{Acknowledgements}

We acknowledge support from FONDECYT through Postdoctoral grants 3220751 (CF) and 3200802 (GV), Regular 1190818 (ET, FEB) and 1200495 (ET., FEB); ANID grants CATA-Basal AFB-
170002 (ET, FEB, CF), FB210003 (ET, FEB, CF) and ACE210002 (CF, ET and FEB); Millennium Nucleus NCN19\_058 (TITANs; ET, CF); and Millennium Science Initiative Program – ICN12\_009 (FEB).
WPM acknowledges support by Chandra grants GO8-19096X, GO5-16101X, GO7- 18112X, GO8-19099X, and Hubble grant HST-GO-15350.001-A. D.T. acknowledges support by DLR grants FKZ 50 OR 2203.


\bibliography{refs}{}
\bibliographystyle{aasjournal}

\appendix

\section{Nuclear Outflow}  
\label{sec:outflow}

\begin{figure}
\gridline{\fig{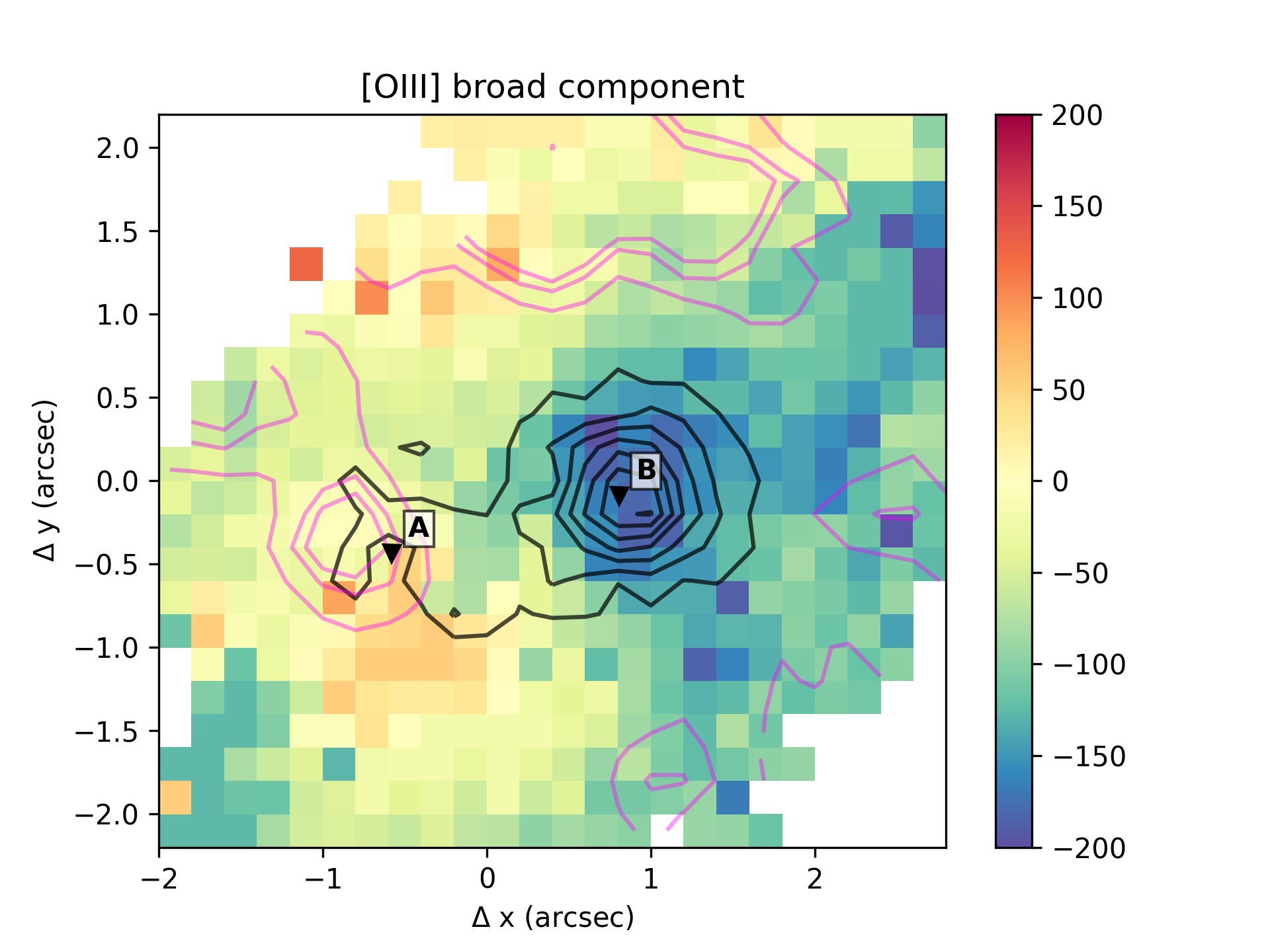}{0.49\columnwidth}{}
          \fig{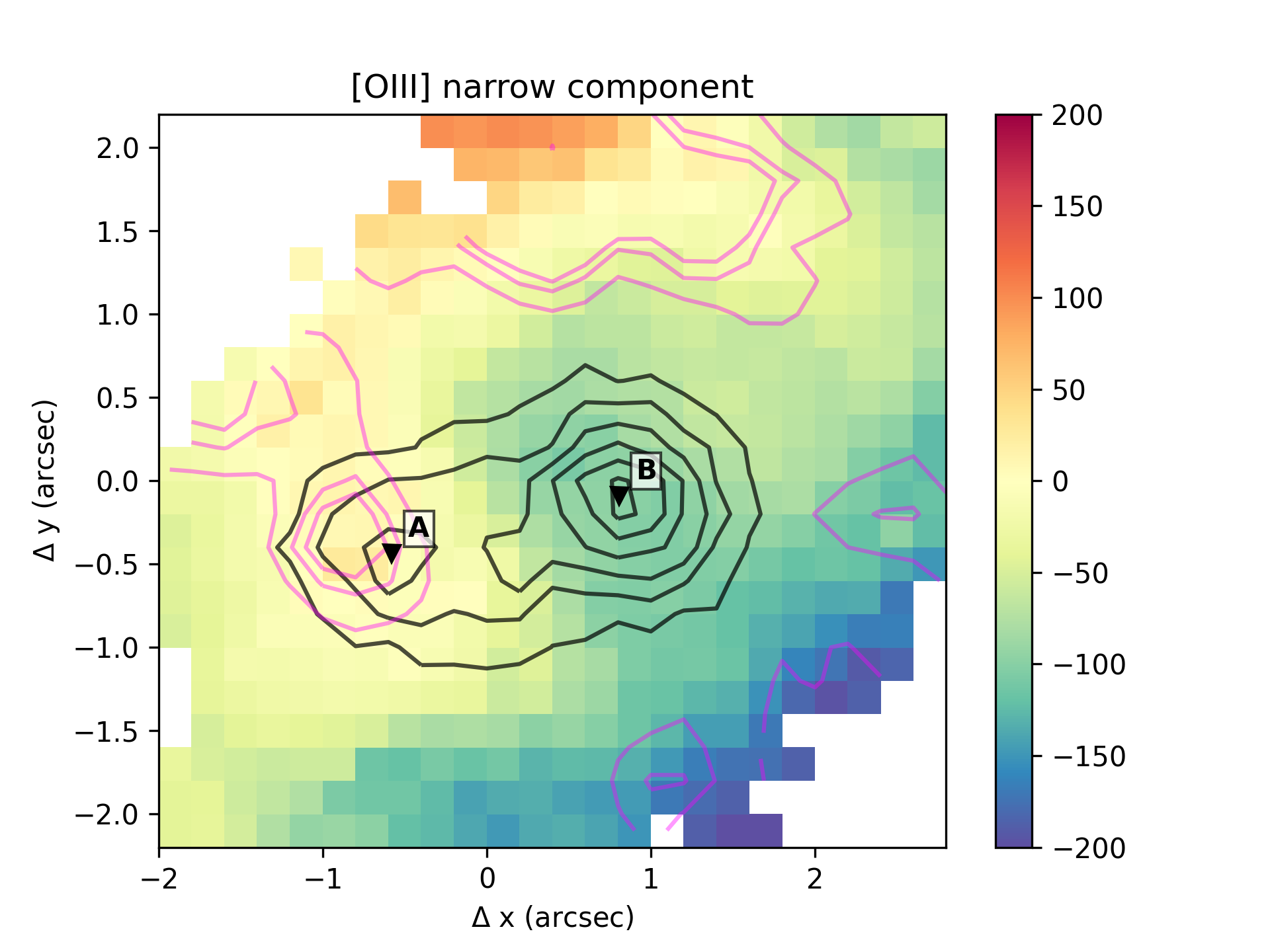}{0.49\columnwidth}{}
          }
\gridline{\fig{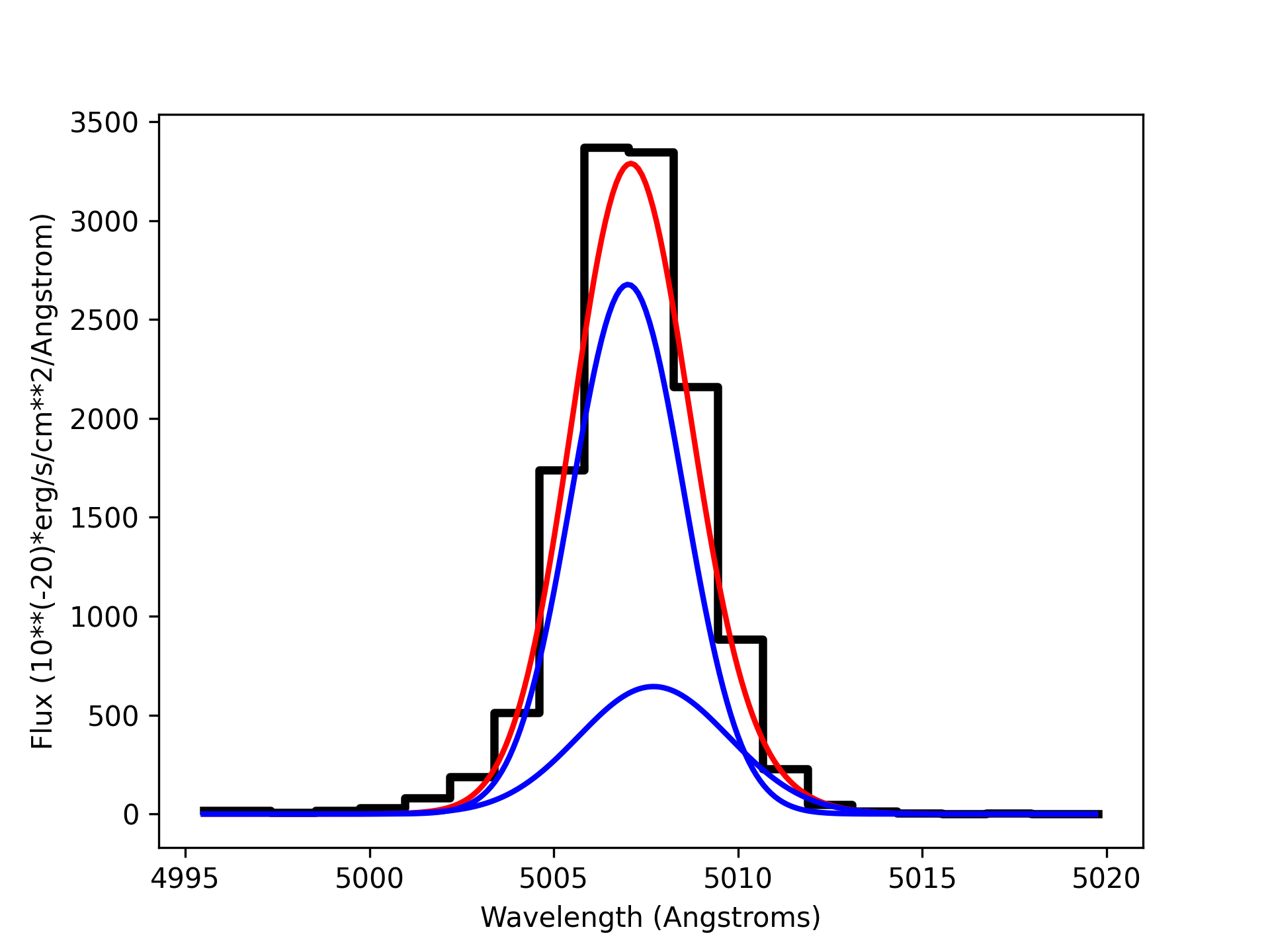}{0.49\columnwidth}{}
          \fig{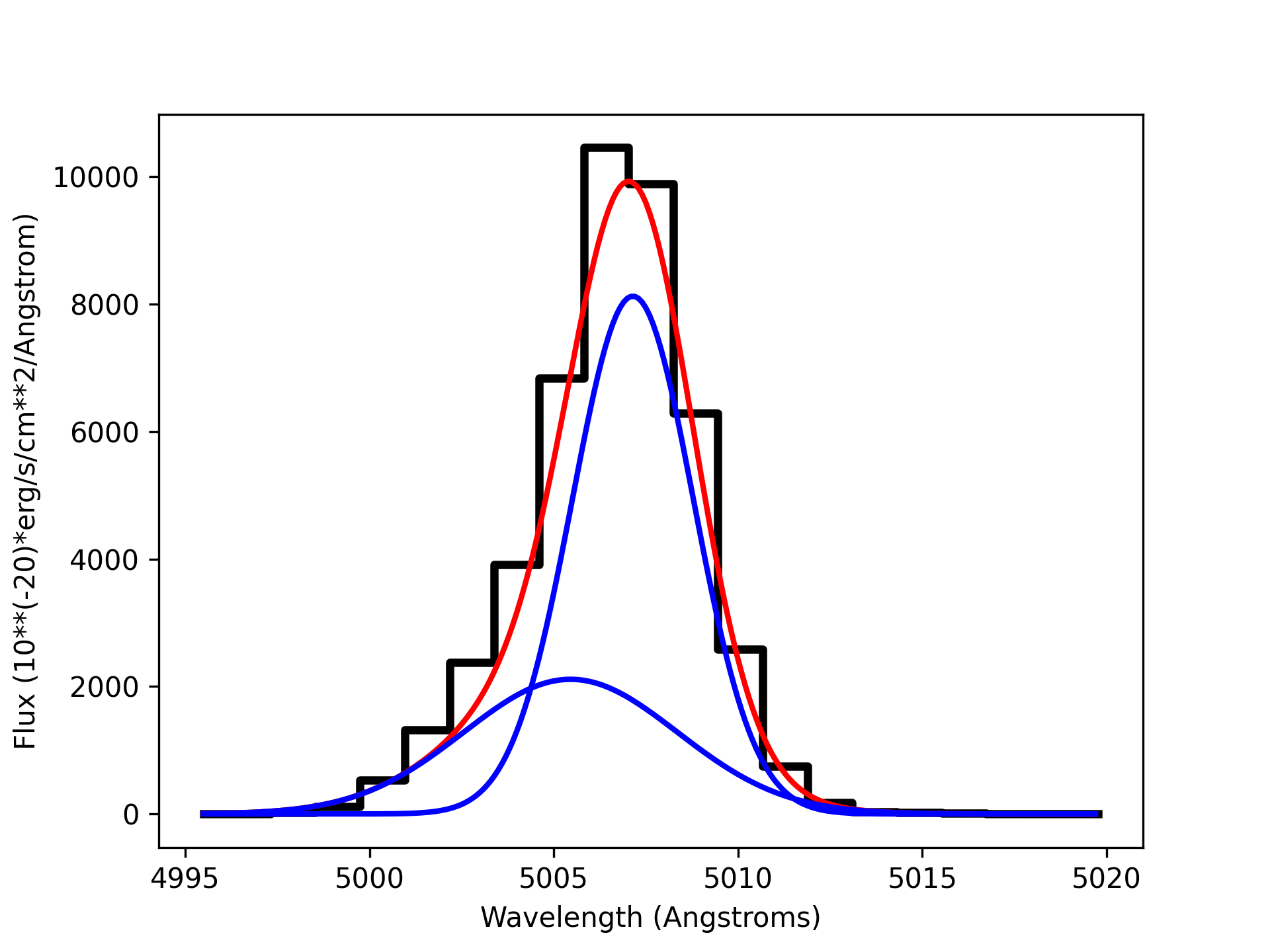}{0.49\columnwidth}{}
          }
\caption{{\it Top:} Velocity maps from 2 Gaussian component fit to the inner region, where the left corresponds to the broad component and the right to the narrow component. The black contours correspond to the respective flux for each component, while the magenta contours correspond to the extinction from a single component fit. The two points marked as 'A' and 'B' show the spaxels where example spectra are extracted from.
{\it Bottom:} Spectra profiles extracted from points 'A' (left) and 'B' (right), overplotted in red is the total fitted profile with the two components marked in blue.}
\label{fig:double_comp_outflow}
\end{figure}

\begin{figure}
	\includegraphics[width=\linewidth]{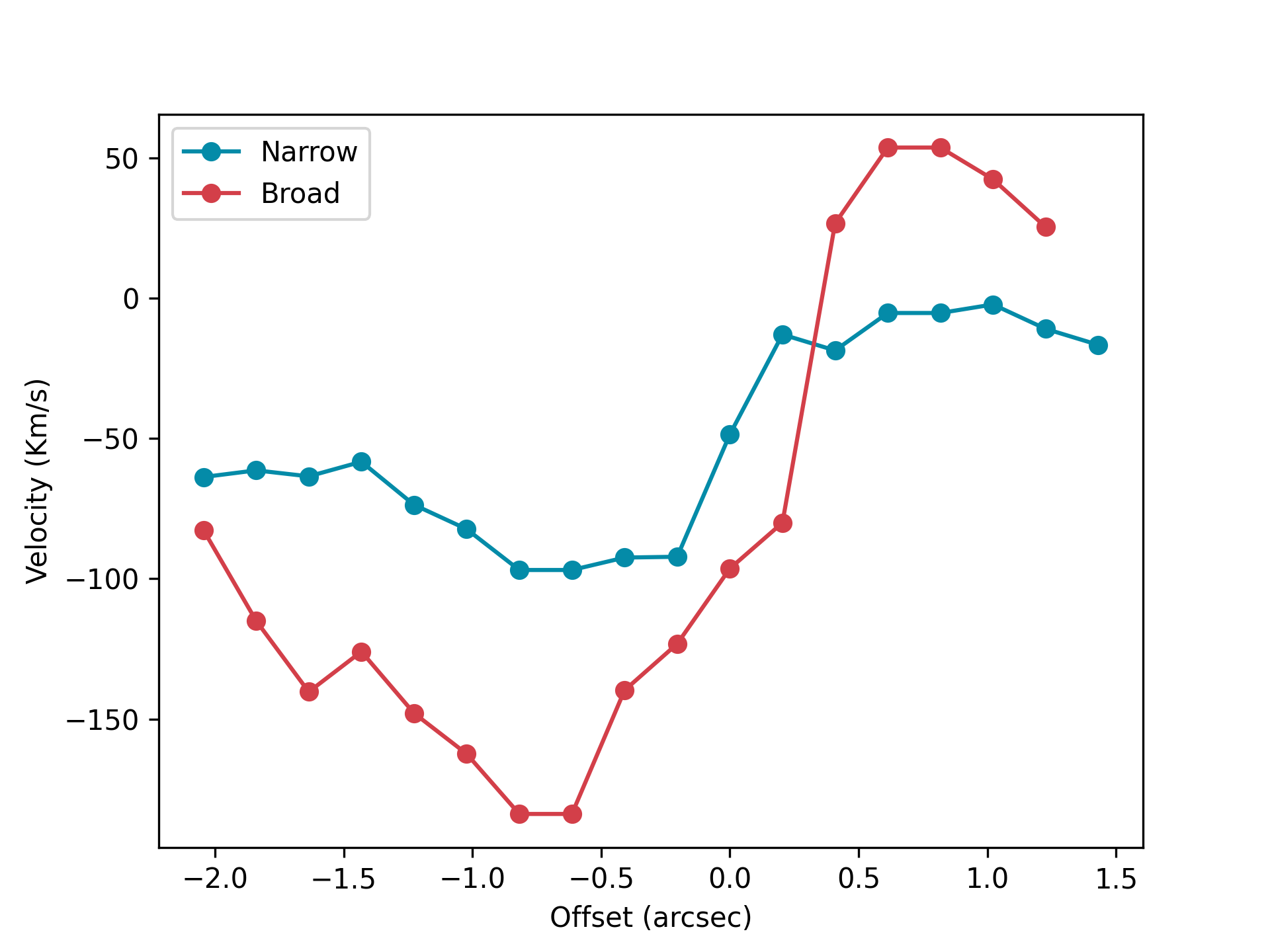}
    \caption{Velocity profiles extracted from the 2 Gaussian component fit velocity maps along PA 110\degree. The points in red represent the broad component while the points in blue represent the narrow component.}
    \label{fig:outflow_vel_profile}
\end{figure}

We perform a two-component Gaussian fit to the spaxels in the central 2\arcsec\ where we observe an outflow. The velocity maps and example spectra for each component (narrow and broad) are shown in Fig. \ref{fig:double_comp_outflow}. The outflow is extended along PA $\sim 100$\degree\ which corresponds to the minor axis of the large-scale rotation. The flux is obscured to the E side as shown by the magenta contours that mark the extinction. Red and blueshifted velocities are observed in both the narrow and the broad component, with the broad component reaching larger velocities. The observed velocities however are asymmetric, with the blueshifted region reaching velocities of 90 km/s and 180 km/s for the narrow and broad components, respectively. While the redshifted region shows velocities near systemic and  50 km/s for the narrow and broad components.
The outflow seems to extend along the same axis of the radio lobes \citep[][]{condon+1988}. It is possible that, at least partially, the receding outflow lies behind the galactic disk. Therefore, it could be affected by the extinction on the disk. In this scenario, the approaching section of the outflow would be, at least partially, above the disk in our line of sight. The velocity profiles extracted along PA 110\degree\ for both components (Fig. \ref{fig:outflow_vel_profile}) show that both follow similar patterns with the broad component reaching larger velocities, therefore it is possible that the main component of the outflow, represented by the broad component, is dragging along at least part of the gas that follows the main-disk rotation.

\section{Metallicity} \label{sec:metal}

\begin{figure}
\gridline{\fig{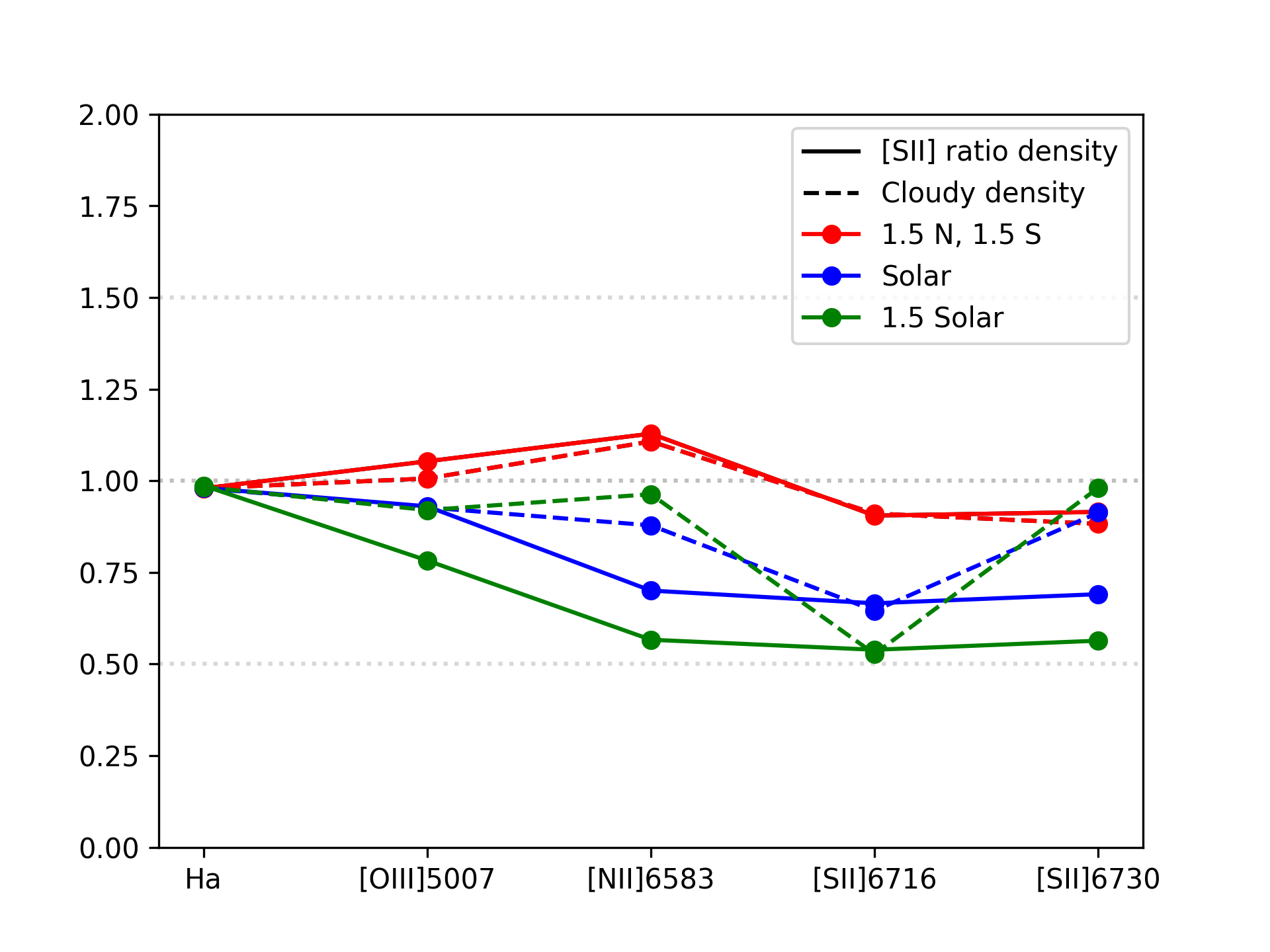}{\columnwidth}{} }
\gridline{ \fig{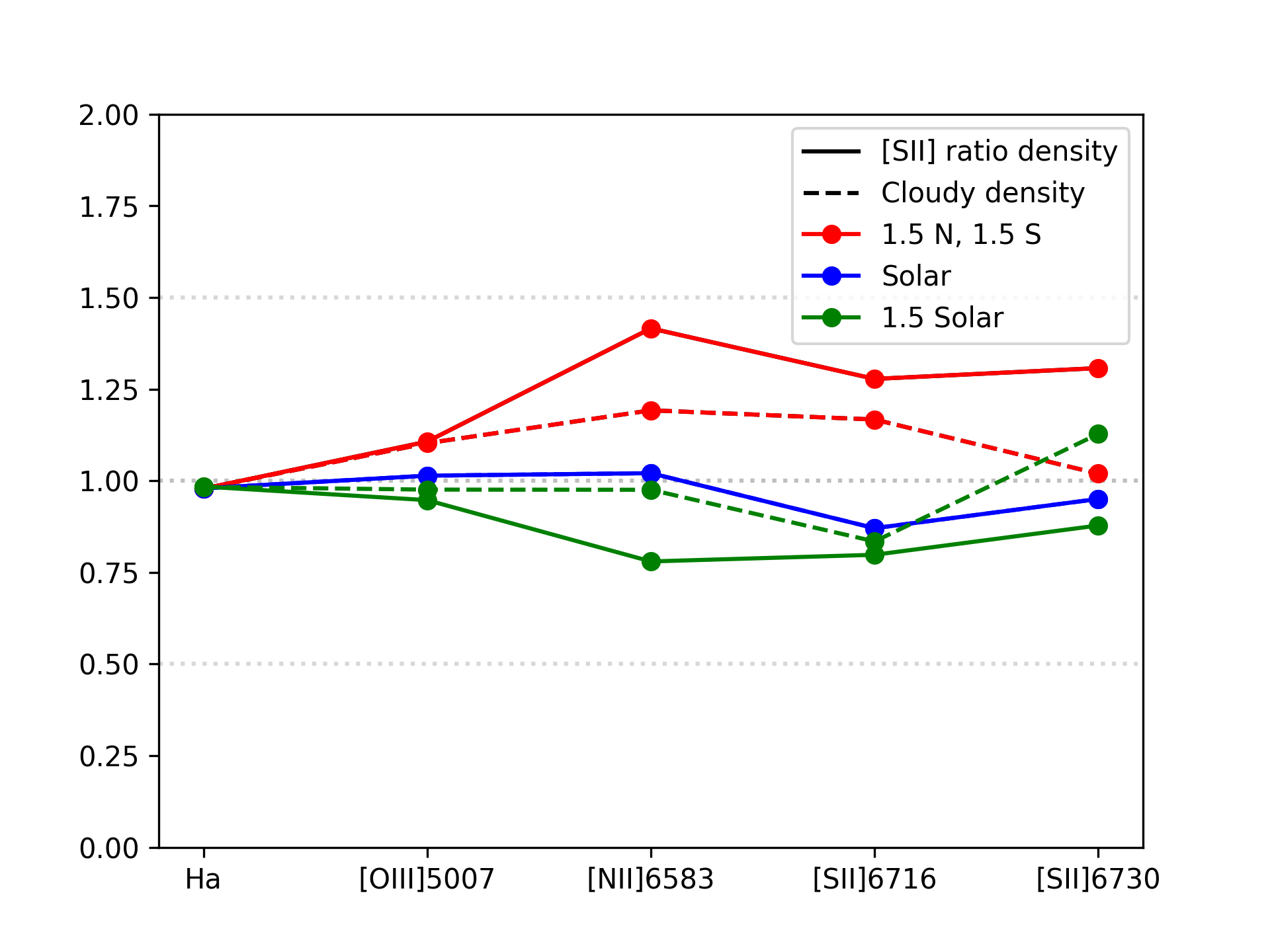}{\columnwidth}{}
         }
\caption{ CLOUDY model over data ratios for the emission lines, solid lines show the model obtained assuming a density from the [SII] emission line ratio, while dashed lines show a model assuming the density as a free parameter. The different colors represent the different metallicity scenarios. The top plot corresponds to an aperture over the 'A' arm while the bottom plot shows an aperture over the 'AT' arm. }
\label{fig:metallicity_bpt}
\end{figure}

The metallicity chosen for our final CLOUDY models come from creating a grid of models with different solar metallicities. In Fig. \ref{fig:metallicity_bpt} we show examples for an example aperture along arm 'A' and 'AT', we show solar and 1.5 times solar metallicities, and our custom metallicity where we only changed N and S elements based on their positions on the BPT diagram, these two elements were scaled 1.5 and 1.2 times their solar value. As it can be observed in Fig. \ref{fig:metallicity_bpt}, for apertures in arm 'A' (and the same is true for arms 'B' and 'C'), a simple increase of decrease of solar metallicity was not enough to achieve a good enough fit, which is why we opted for scaling N and S. However, for arm 'AT' this change resulted in a worse fitting, which is why we decided to maintain the metallicity as solar. Is important to remark that this change in metallicity kept very similar results for the U value obtained but changed the density (as can be observed in Fig. \ref{fig:cloudy_sims}), a problem we avoid by assuming the density obtained directly from the [SII] ratio in model 1.

\end{document}